\documentclass[footinbib,a4paper,aps,prx,superscriptaddress,reprint,twocolumn,preprintnumbers,amsmath,amssymb,nobalancelastpage,10pt]{revtex4-1}
\usepackage{amsmath}
\usepackage{verbatim}
\usepackage{graphicx}
\usepackage{color}
\usepackage{tikz}
\usepackage{subfigure}
\usepackage{float}
\usepackage{slashed}
\usepackage{mathptmx}
\usepackage{amssymb}
\usepackage{amsmath}
\usepackage{amsfonts}
\usepackage{array}

\usepackage{bm}
\usepackage{xcolor}
\graphicspath{{figures/}}

\usepackage{braket}

\definecolor{mygrey}{gray}{0.35}
\definecolor{myblue}{rgb}{0.2,0.2,0.8}
\definecolor{mygreen}{rgb}{0.2,0.8,0.5}
\definecolor{myzard}{cmyk}{0,0,0.05,0}
\definecolor{mywhite}{rgb}{1,1,1}
\definecolor{myred}{rgb}{1,0.,0.3}

\usepackage[colorlinks=true,citecolor=myblue,linkcolor=myred]{hyperref}

 \def\ee{\mathord{\rm e}}
 
 \def\ii{\mathord{\rm i}}
\def\min{\mathord{\rm min}}

\def\half{\textstyle\frac{1}{2}}

\renewcommand{\ii}{{\rm i}}
\renewcommand{\ee}{{\rm e}}
\def\beq{\begin{equation}}
\def\eeq{\end{equation}}
\def\barray{\begin{eqnarray}}
\def\earray{\end{eqnarray}}

\begin{document}

\title{ Large-$N$  Chern insulators:   lattice field theory  and quantum simulation approaches \\  to correlation effects in the  quantum anomalous Hall effect }

\author{L. Ziegler}
\affiliation{Departamento de F\'{i}sica Te\'{o}rica, Universidad Complutense, 28040 Madrid, Spain}
\author{E. Tirrito}
\affiliation{SISSA, Via Bonomea 265, 34136 Trieste, Italy}
\affiliation{ICFO - Institut de Ci\`encies Fot\`oniques, The Barcelona Institute of Science and Technology, Av. Carl Friedrich Gauss 3, 08860 Castelldefels (Barcelona), Spain}

\author{M. Lewenstein}
\affiliation{ICFO - Institut de Ci\`encies Fot\`oniques, The Barcelona Institute of Science and Technology, Av. Carl Friedrich Gauss 3, 08860 Castelldefels (Barcelona), Spain} 
\affiliation{ICREA, Lluis Companys 23, 08010 Barcelona, Spain}

\author{S. Hands}
\affiliation{Department of Mathematical Sciences, University of Liverpool, Liverpool L69 3BX, United Kingdom}

\author{A. Bermudez}
\affiliation{Departamento de F\'{i}sica Te\'{o}rica, Universidad Complutense, 28040 Madrid, Spain}

\begin{abstract}
Four-Fermi quantum field theories  in (2+1) dimensions  lie among the simplest models in high-energy physics,  the understanding of which requires a non-perturbative lattice  formulation  addressing their strongly-coupled fixed points. These lattice models are also  relevant in condensed matter, as they offer a neat playground to explore   strong correlations in the quantum anomalous Hall (QAH) effect. We give a detailed description of our multidisciplinary approach  to understand the fate of the QAH phases as the four-Fermi interactions are increased, which  combines strong-coupling and effective-potential techniques,  unveiling a rich phase diagram with large-$N$ Chern insulators and Lorentz-breaking fermion condensates. Moreover,  this  toolbox can be enlarged with recent advances in quantum information science, as we show that tensor-network algorithms based on projected entangled pairs can be used to improve our understanding of the strong-coupling limit. We also present a  detailed scheme that uses   ultra-cold atoms in optical lattices with synthetic spin-orbit coupling to build quantum simulators of these four-Fermi models. This yields a promising alternative to characterise the strongly-coupled fixed points and, moreover, could also explore real-time dynamics and finite-fermion densities.

\end{abstract}

\maketitle

\setcounter{tocdepth}{2}
\begingroup
\hypersetup{linkcolor=black}
\tableofcontents
\endgroup

\section{\bf Introduction}

One of the main tasks in condensed-matter physics is the prediction and classification of new phases of matter, understanding  how they can transform into each other via phase transitions~\cite{sachdev_2011}.  Symmetry has played a key role in this endeavour,~\cite{landau_37,ginzburg_50}, since the way in which it breaks can be used to understand many ordering patterns  at the microscopic level. Despite decades of intense research, observing how different orderings arise from the same microscopic model, and how they can be described by emerging effective theories that cannot be predicted  by simply looking  at the individual microscopic constituents~\cite{Anderson393},  is still a source of much fascination. This emergence is typically a consequence of the interplay of symmetry and strong inter-particle correlations, the latter being induced by interactions among the microscopic constituents. In condensed matter, these constituents typically correspond to the     ions forming the crystal structure and the valence electrons, all of them interacting via Coulomb forces~\cite{altland_simons_2006,doi:10.1142/2945}.

Narrow-band metals have turned out to be a particularly-rich playground for this emergence. Here,  the  Coulomb interaction between the electrons is screened, and can be approximated by a local four-Fermi term that leads to the  Hubbard model, the archetype of strongly-correlated fermions in condensed matter~\cite{PRSLSA_276_238}. The apparent simplicity of this model is deceptive, as it can host an interaction-induced metal-insulator  transition at partial band fillings~\cite{RevModPhys.70.1039}, which defies  the  naive  band-theory distinction between insulators and conductors. In fact, in the  half-filled case, these so-called Mott insulators display anti-ferromagnetic ordering as a consequence of magnetic super-exchange  interactions and a phase transition where  the  spin rotational symmetry is spontaneous broken~\cite{PhysRev.79.350,ANDERSON196399}. Moreover, as one dopes this Mott insulator away from half filling,  the interplay of anti-ferromagnetism with the  dynamics of the holes has been studied as a fully-electronic mechanism that may account for high-$T_{\rm c}$ superconductivity~\cite{1987Sci...235.1196A,RevModPhys.78.17}, as the mobility of holes can release the anti-ferromagnetic singlets, allowing them to condense into the superconductor.

Four-Fermi models have also played an important role at much higher energies, starting with the pioneering work of Enrico Fermi on  $\beta$-decay in nuclei~\cite{Fermi1934,doi:10.1119/1.1974382}; a precursor  to the theory of  electroweak interactions in the standard model of particle physics. Four-Fermi terms also appear in the so-called Nambu-Jona-Lasinio (NJL)  models~\cite{PhysRev.122.345,PhysRev.124.246}, introduced as a description of interactions between nucleons and predating the modern theory of the strong force in the quantum-chromodynamics (QCD) sector of the standard model. NJL models are nowadays considered as effective theories that capture essential properties of QCD, such as  dynamical mass generation by the spontaneous breakdown of chiral symmetry~\cite{RevModPhys.64.649}. Lowering the spacetime dimension, one finds the Gross-Neveu model in 1+1 dimensions~\cite{PhysRevD.10.3235}, where the specific form of the  Four-Fermi terms allows to study asymptotic freedom in a renormalizable framework, capturing in this way another essential feature of QCD. Note that, although all these four-Fermi models are defined by continuum quantum field theories (QFTs) of self-interacting Dirac fermions, one can always discretize spacetime in a lattice~\cite{PhysRevD.10.2445,gattringer_lang_2010}, which leads to lattice field theories (LFTs) that are closer in spirit to the aforementioned Hubbard models. These LFTs yield a non-perturbative approach to understand the strong-coupling nature of the fixed points  governing the chiral-symmetry-breaking  transitions~\cite{hep-lat/9706018}, around which one can perform a long-wavelength approximation  and connect to the  QFTs~\cite{creutz_1997}.

This  analogy between condensed matter and high-energy physics becomes more quantitative in the so-called {\it Dirac matter} including, as paradigmatic  examples, graphene~\cite{RevModPhys.81.109}, Weyl semi-metals~\cite{RevModPhys.90.015001}, and topological insulators and superconductors~\cite{RevModPhys.83.1057}. These materials have band structures that can be ultimately understood as specific lattice discretizations of Dirac-type Hamiltonian QFTs. The perspective, however, is rather different. Whereas the lattice is an artificial scaffolding in LFTs, and the focus lies on the critical points around which one  recovers the long-wavelength physics independent of lattice artifacts, the lattices of Dirac matter are physical and play a crucial role in determining the special properties of the phases.   In the case of topological insulators and superconductors, these properties are a consequence of a different  driving mechanism for the ordering of matter at the microscopic scale; a mechanism where topology and symmetry intertwine to determine the groundstate properties. Conceptually, topological phases of matter differ from those formed by spontaneous symmetry breaking (SSB), since they cannot be characterized in terms of local order parameters dictated by an underlying SSB process. Instead, the groundstates are characterized by quantized  topological invariants, the values of which cannot change unless the energy gap to the lowest-lying excitations closes, and a quantum phase transition takes place~\cite{RevModPhys.82.3045}. Note that these so-called topological phase transitions do not require any symmetry breaking. On the contrary, it might actually be the preservation of certain symmetries~\cite{PhysRevB.78.195125,doi:10.1063/1.3149495,classification_spt}, which determines the form of the topological invariant, and it becomes important to understand the robustness of these phases with respect to external perturbations. Besides their fundamental interest,  these topological phenomena can  lead to   novel functionalities and promising technological  applications~\cite{Soumyanarayanan2016}.

The epitome of the aforementioned robustness to external perturbations, such as disorder, occurs in  the quantum Hall  effect~\cite{PhysRevLett.45.494}. Remarkably, this phase of matter has an underlying topological invariant~\cite{PhysRevLett.49.405} that corresponds to the first Chern number of a fibre bundle associated to the electronic band structure~\cite{nakahara_2017}. Moreover, this description is not limited to band theory, as the Chern  numbers can be adiabatically connected to a many-body topological invariant when electron-electron interactions or disorder are switched on~\cite{niu1985}. In fact, this invariant is responsible for the
 robust quantization of the transverse conductivity under generic weak perturbations that do not close the gap of the system. In this case, regardless of their symmetry properties, these perturbations cannot induce back-scattering in current-carrying edge states that propagate along the boundaries of the system~\cite{PhysRevB.25.2185,PhysRevLett.71.3697}, and can thus  transport  charge robustly even in the presence of an insulating gap.    In this way, the naive band-theory distinction between insulators and conductors is again defied; only this time by the introduction of topology rather than correlations. 
 
 Coming back to the context of Dirac matter, the so-called
Chern insulators~\cite{PhysRevLett.61.2015, PhysRevB.74.085308,PhysRevB.78.195424} can feature the above quantum Hall effect even in absence of any external magnetic field, leading to what is currently known as the quantum anomalous Hall (QAH) effect~\cite{doi:10.1146/annurev-conmatphys-031115-011417}. In a seminal work~\cite{PhysRevLett.61.2015}, which stimulated subsequent contributions~\cite{PhysRevLett.95.146802, PhysRevLett.95.226801} that laid the foundations  of   topological insulators and superconductors~\cite{RevModPhys.83.1057,RevModPhys.82.3045}, Haldane showed that a time-reversal-breaking discretization of the Dirac QFT on a honeycomb graphene-type lattice can support a non-zero Chern number  that is  related to the parity anomaly in (2+1)-dimensional quantum electrodynamics ~\cite{PhysRevLett.51.2077,PhysRevLett.52.18,PhysRevD.29.2366,PhysRevLett.53.2449}. The associated transverse conductivity can again be associated with current-carrying edge states which circulate in a single direction along the boundaries of the system in spite of the absence of a net magnetic field. 
 
From a LFT perspective, the continuum QFT that describes the Haldane model is that of a pair of  Dirac fields  with a different mass. Here, the doubling  is related to a fundamental  theorem for the lattice discretization of a Dirac field~\cite{NIELSEN198120,NIELSEN1981173}, whereas the mass difference is an instance of Wilson's prescription to deal with such a doubling~\cite{Wilson1977}. The parallelism becomes clearer using a square-lattice discretization~\cite{PhysRevB.74.085308,PhysRevB.78.195424}, sometimes refereed to as the Qi-Wu-Zhang model of the QAH effect. This model can be readily  understood as  a Hamiltonian formulation  of LFTs~\cite{PhysRevD.11.395}, in which time remains continuous, and it is only the spatial coordinates that get discretized in a square lattice following Wilson's prescription~\cite{Wilson1977}. From this perspective,  the topological edge states that appear at the boundaries of these Chern insulators, and carry the transverse current in the QAH effect~\cite{PhysRevB.74.085308,PhysRevB.78.195424}, can be understood  as lower-dimensional versions of the so-called domain-wall
fermions~\cite{KAPLAN1992342}. In this context, the aforementioned topological invariants also appear in other dimensions, and control  a
Chern-Simons-type response to additional external gauge
fields~\cite{PhysRevB.78.195424,10.1143/PTP.73.528,GOLTERMAN1993219}. We note that the  QAH effect has been realized  with  thin films of semiconducting  tetradymite  compounds doped with magnetic atoms~\cite{Chang167,Chang2015}, as first proposed theoretically~\cite{doi:10.1126/science.1187485},  or in layered   compounds with intrinsic magnetic order~\cite{doi:10.1126/science.aax8156}. More recently, the QAH effect has also been  observed in magnetically-doped multi-layer systems, which have higher Chern numbers and display  plateau-to-plateau topological phase transitions~\cite{Zhao2020,zhao2021zero}.

A fundamental question that has generated considerable interest in recent years is how topological insulators and superconductors get  modified in the presence of interactions, ultimately seeking for new phases of matter driven by the interplay of symmetry, correlations and topology~\cite{Hohenadler_2013,doi:10.1142/S021797921330017X,Neupert_2015,Rachel_2018}. First of all,  starting from initial studies~\cite{niu1985}, we note that  topological invariants can be  generalised to the many-body case~\cite{doi:10.1143/JPSJ.75.123601,PhysRevLett.122.146601},   and formulated in terms of single-particle Green's functions that unveil the special role played by the self energy~\cite{PhysRevLett.105.256803,PhysRevB.83.085426,PhysRevX.2.031008,PhysRevB.86.165116,Wang_2013}. Additionally,  there are entanglement-related quantities that give an alternative route to explore topology~\cite{PhysRevLett.101.010504,PhysRevB.81.064439,ent_spectrum_berry}. Equipped with these many-body tools for the characterization of   topological phases, let us now briefly discuss some of the possible effects brought up by the inclusion of interactions. Due to the non-zero bulk gap, one expects that topological insulators will be generally robust to weak interactions. As first discussed in~\cite{PhysRevLett.95.226801}, electron-electron interactions can renormalize this bulk gap, and stabilize  the topological phase further. In other situations, this renormalisation of the band structure can even induce transitions from a trivial phase into a  topological one~\cite{PhysRevB.86.201407,PhysRevB.85.235135}. Further interaction-induced effects can occur by means of SSB, as time-reversal symmetry can be spontaneously broken when the interactions increase, such that a new topological phase  that cannot be understood from the renormalisation of the free-particle parameters arises~\cite{PhysRevLett.100.156401,PhysRevLett.103.046811,PhysRevLett.117.096402}. 
Note that SSB of unitary symmetries can also take place, leading to order parameters that coexist with non-zero topological invariants, leading to the concept of symmetry-breaking topological insulators induced by interactions~\cite{PhysRevLett.113.216404,z2_bhm_2,z2_bhm_3}. A final possibility brought up by interactions is that new topological phases arise without any free-particle counterpart, such as topological Mott insulators~\cite{Pesin2010} or fractional Chern insulators~\cite{PhysRevLett.106.236804,PhysRevX.1.021014,Sheng2011}. 

Despite this large body of theoretical work, the vast majority of materials that have been found to host topological phenomena in the absence of strong external magnetic fields, only display  weak electron-electron interactions. A notable exception is bilayer graphene, either  twisted at a specific magic angle~\cite{doi:10.1126/science.aay5533,Nuckolls2020,Saito2021,Choi2021,xie2021fractional}, or misaligned with respect to a substrate~\cite{doi:10.1126/science.aan8458}.
Let us note, however, that the microscopic description of these materials differs markedly from the paradigmatic lattice models that have been thoroughly studied in the presence of Hubbard-type interactions~\cite{Rachel_2018}, and where most of the interaction-induced effects discussed above have been identified. This  also occurs for the  experimental realizations of the   QAH effect in thin-film materials~\cite{Chang167,Chang2015,doi:10.1126/science.aax8156,doi:10.1126/science.aay5533}, in which a Haldane-type  model~\cite{PhysRevB.74.085308,PhysRevB.78.195424}  serves as a  guide to build a qualitative understanding, but where many of the microscopic details are clearly different~\cite{doi:10.1146/annurev-conmatphys-031115-011417}. Moreover, since the effective spinor degrees of freedom are related to bonding/anti-bonding states in the two opposite surfaces of the thin film, interactions will not be described by a simple contact Hubbard-type term.

Moving away from condensed matter into the realm of atomic, molecular and optical (AMO) physics, the microscopic tunability of gases of ultracold atoms trapped in optical lattices~\cite{Bloch_2008} allows one to face the quantum many-body problem from a different perspective, that of quantum simulations (QSs)~\cite{Feynman_1982,Cirac2012,Goldman2016}. The idea here is that one can control these dilute and highly non-relativistic gases of atoms, making them behave according to a specific model of interest at  a widely-different scale. For instance, one may use these AMO quantum simulators to explore  much denser systems in condensed matter, or much more energetic ones in particle physics.  Although research on QSs focused for some time on condensed-matter models~\cite{RevModPhys.86.153,doi:10.1080/00018730701223200,Bloch2012,doi:10.1146/annurev-conmatphys-070909-104059}, applications to high-energy physics are becoming more popular in recent years~\cite{doi:10.1002/andp.201300104,Zohar_2015, doi:10.1080/00107514.2016.1151199,Banuls2020,Carmen_Ba_uls_2020,2006.01258, aidelsburger2021cold,zohar2021quantum,klco2021standard}.   The present context of topological Dirac matter actually touches both of these directions, as recent experiments with ultra-cold fermions in shaken  optical lattices have allowed to realize Haldane's model for the first time~\cite{Jotzu2014}. In contrast to the previous condensed-matter realisations, where the archetype lattice models  serve to build  a qualitative understanding,  the Haldane model is an accurate  description of this AMO experiment and, moreover, one can control the fermion filling and microscopic parameters, and even modify the latter  dynamically and measure real-time evolution.

Beyond the interest in observing interaction-induced effects in the QAH effect, we note that the related (2+1)-dimensional four-Fermi LFTs are typically controlled by strongly-coupled fixed points, the properties of which can only be accessed by non-perturbative methods, and are still a subject of active research~\cite{HANDS199329,hep-lat/9706018,Braun_2012}. One can thus exploit the interdisciplinary character of this field, and combine the different tools developed by these communities to advance our understanding further.  For instance, in order to understand non-perturbative phenomena, one may employ a large-$N$ expansion through which  QFTs simplify significantly and become solvable in the limit of a large number of  flavours $N$~\cite{coleman_1985}. Within this framework, it has been  possible to predict some of the fundamental features of QCD in the simpler four-Fermi Gross-Neveu model in 1+1$d$, including asymptotic freedom and dynamical symmetry breakdown~\cite{PhysRevD.10.3235}. The large-$N$ expansion is also very important in four-Fermi models in (2+1) dimensions, as it offers a renormalizable framework that contrast the situation of a perturbative approach. In this article, we shall exploit large-$N$ tools to understand correlation effects in a QAH effect, and combine them with strong-coupling predictions based on the condensed-matter concept of super-exchange to understand the full phase diagram. Finally, exploiting the third facet, we give a detailed account of a  cold-atom realization  of the model for a single flavour based on a scheme for synthetic spin-orbit coupling.

From the perspective of the QS of a correlated QAH effect, the most direct route would be to explore a spinful version of  the  experiment with ultra-cold fermions in shaken honeycomb  lattices~\cite{Jotzu2014}. In this case,  the Hubbard interaction is directly implemented by the low-temperature scattering dominated by two-particle $s$-wave collisions~\cite{Bloch_2008}, which can actually be tuned by additional Feshbach resonances~\cite{RevModPhys.82.1225}. However, the combination of  Hubbard interactions with  periodic shaking has been hampered, in several experimental realizations, by a larger heating mechanism due  to a denser   spectrum of excitations. Here, the atoms  get excited by resonantly  absorbing quanta from the periodic drive  that shakes the lattice, a process that  causes an effective heating~\cite{PhysRevLett.119.200402,PhysRevX.10.011030,RevModPhys.89.011004}. In a recent manuscript~\cite{ziegler2020correlated}, we have explored a different route to explore correlated QAH insulators and their connection to strongly-coupled four-Fermi QFTs. This alternative route considers Fermi gases with synthetic spin-orbit coupling~\cite{Galitski2013,zhai_2015,book_soc} induced by a so-called Raman optical lattice~\cite{PhysRevLett.112.086401,PhysRevLett.113.059901}, which has  been recently demonstrated in   experiments~\cite{Wu83,PhysRevLett.121.150401,Songeaao4748,liang2021realization}. Although residual photon scattering from the Raman beams can also induce some heating, it should not be as severe as in the shaken optical lattice if sufficient laser power is available, allowing one to work at sufficiently large Raman detunings. In presence of interactions, correlated phenomena may need slower experimental time scales so this heating may again become a practical limitation.  In this case, one may consider realizations based on alkaline-earth or lanthanide atoms, where it  can be further minimised.  Below, we  describe this scheme in  detail.
    
In this work, we  start from a discretized anisotropic variant of a Gross-Neveu model with $N$ flavours of Wilson fermion in 2+1 dimensions. In contrast to common LFT approaches to the (2+1)$d$ Gross-Neveu model, we dispense with the notion of chirality from the outset by restricting our attention to two-component spinors instead of the four-component ones that permit a definition of a non-trivial $\gamma^5$ matrix~\cite{HANDS199329,hep-lat/9706018}. Since chiral symmetry and its breakdown is a fundamental ingredient of Gross-Neveu-type models, but is lacking in our case,  we prefer to refer to the current  discretization  as a {\it four-Fermi-Wilson model}. In the single-flavour limit, this lattice model has a neat connection  to a spin-orbit-coupled Hubbard model, bringing a direct connection to the aforementioned QSs of spin-orbit-coupled Fermi gases in optical lattices. Although the large-$N$ approximation is only exact in the limit $N\rightarrow \infty$, we provide a detailed analysis that shows that our large-$N$ calculations predict the same  structure of the  phase diagram as the $N=1$ limit, which can be explored using a super-exchange strong-coupling approach and various variational ans\"{a}tze.

This article is organized as follows. In section \ref{first}, we introduce our four-Fermi-Wilson model, and describe the topological properties of the non-interacting limit, showing that the model can host  $N$-flavoured Chern insulators and allow for various topological phase transitions as the microscopic bare parameters are modified. In Sec.~\ref{sec:strong_coupling},   we explore the opposite limit of very strong four-Fermi interactions and a single fermion flavour $N=1$, which we approach  by deriving an effective Hamiltonian using the concept of super-exchange interactions. We show that this effective model corresponds to a quantum compass model,  where symmetry-breaking orbital ferromagnets can be identified using a variational mean-field method, and in the framework of tensor networks using projected entangled pairs (PEPs).  In section \ref{gapequations}, we derive and solve the large-$N$ gap equations, both in continuous time and discretized time, showing that the aforementioned orbital ferromagnets can be interpreted as fermion condensates that spontaneously break a $\mathbb{Z}_2$ inversion symmetry and forbid recovering a QFT invariant under Lorentz boosts in the continuum limit.   We discuss   how an additive renormalisation and a suitable re-scaling must be considered in order connect both continuum and discrete-time approaches. In section \ref{effectivepotential}, we use the discrete-time approach to obtain an effective potential, which allows us to explore not only the symmetry-broken condensates, but also  to delimit the topological QAH phase at intermediate interactions. Using the large-$N$ self-energy in the symmetry-preserved region, we show that this QAH regions are characterised by large-$N$ Chern insulators that display  a non-zero value of a non-perturbative many-body topological invariant. We also discuss how the effective potential allows for a neat account of first- or second-order nature that separate these phases from the Lorentz-breaking condensates. Finally, in section \ref{optical}, an experimental scheme for the QS of the single-favour  four-Fermi-Wilson model with ultra-cold atoms in optical lattices is presented, where we also discuss in detail how the available measurement protocols could be used to explore this model in the laboratory.

\section{\bf Four-Fermi-Wilson  model in 2+1 dimensions} \label{first}

\subsection{Chiral and non-chiral four-Fermi field theories in 2+1 dimensions}

The original Gross-Neveu model is a chiral-invariant QFT describing $N$ flavours of massless Dirac fermions, which live in one spatial and one time dimension, and interact via four-Fermi terms~\cite{PhysRevD.10.3235}.  This model gained considerable attention in the early days of QCD, since it is a renormalizable field theory that shares  important features with QCD such as asymptotic freedom, chiral symmetry breaking by dynamical mass generation, and dimensional transmutation, all of which can be explored in a much simpler setup. In a Hamiltonian field-theory formulation, this model can be easily  extended to 2+1 dimensions, where the Hamiltonian density $\mathcal{H}(\boldsymbol{x})$ corresponding to the Hamiltonian $H= \int {\rm d}^2x\,  {:}\mathcal{H}(\boldsymbol{x})\colon$ reads
\begin{align}
\begin{split}
\mathcal{H}\!=\! -\!\sum_{f=1}^{N} \!
\overline{\psi}_\mathsf{f}(\boldsymbol{x}) (\ii\gamma^j \partial_j )\psi_\mathsf{f}(\boldsymbol{x})- \frac{g^2}{2N}\!\!\left(\!\sum_{\mathsf{f}=1}^{N}\!\overline{\psi}_\mathsf{f}(\boldsymbol{x})\psi_\mathsf{f}(\boldsymbol{x})\!\!\right)^{\!\!\!2}\!\!\!,
 \label{continuum}
\end{split}
\end{align}
where the repeated-index summation  only runs over the spatial indexes $j\in\{1,2\}$. Here, we have introduced the  field operators  $\psi_f(\boldsymbol{x})$, which annihilate a fermion  of a given flavor $\mathsf{f}\in\{1,2,\cdots,N\}$ at  a specific spatial location $\boldsymbol{x}=(x,y)^t$, 
and the corresponding adjoints  $\overline{\psi}_\mathsf{f}(\boldsymbol{x})=\psi^\dagger_\mathsf{f}(\boldsymbol{x}) \gamma_0$, which are  defined in terms of the analogous creation  operators $\psi^\dagger_\mathsf{f}(\boldsymbol{x})$.
While the first term in Eq.~\eqref{continuum} describes the kinetic energy of $N$ massless Dirac fermions, the second one represents the four-Fermi term responsible for intra- and inter-flavour interactions.  Note that, in contrast to the (1+1)-dimensional case~\cite{PhysRevD.10.3235}, the coupling strength $g^2$ is not dimensionless, but  has  units of inverse mass $[g^2]=\mathsf{M}^{-1}$ when considering natural units $\hbar=c=1$. By power counting, the four-Fermi term is no longer marginal as in (1+1) dimensions~\cite{PhysRevD.10.3235}, but becomes irrelevant thus forbidding a perturbative renormalization. Remarkably, a large-$N$ expansion provides such a renormalizable framework, and one can explore the nature of the strongly-coupled fixed point to different orders of $\mathcal{O}(1/N^{\alpha})$~\cite{HANDS199329}. 

In order to guarantee Lorentz invariance, the above Hamiltonian is expressed in terms of gamma matrices $\gamma^\mu$, such that  fermionic fields become spinors with well-defined transformations under the Lorentz group. These matrices must satisfy  Clifford's algebra $\{\gamma^\mu,\gamma^\nu\}=2 \eta^{\mu \nu}$, where $\eta={\rm diag}(1,-1,-1)$ is Minkowski's metric, and the  spacetime indexes for a (2+1)-dimensional spacetime are $\mu,\nu \in \{0,1,2\}$. In 2+1 dimensions, there are two possible strategies to satisfy this algebra. The simplest one  is to choose  a set of adapted Pauli matrices, such as 
\beq
\label{eq:gammas}
\gamma^0=\sigma^z,\hspace{1ex} \gamma^1=\ii\sigma^y,\hspace{1ex}\gamma^2=-\ii\sigma^x,
\eeq
which leads to  two-component spinor fields for each of the flavors  $\psi_\mathsf{f}(\boldsymbol{x})=(\psi_{\mathsf{f},1}(\boldsymbol{x}),\psi_{\mathsf{f},2}(\boldsymbol{x}))^{\rm t}$. The would-be chiral gamma matrix, which must be Hermitian and anti-commute with the rest, is typically defined as $\gamma^5=\ii\gamma^0\gamma^1\gamma^2$, which can be easily checked to be trivial in this case $\gamma^5=\mathbb{I}_2$. Accordingly, it does not satisfy the anti-commutation requirement $\{\gamma_5,\gamma_\mu\}=0$, and one speaks of the absence of chirality. This is actually a generic feature in all odd spacetime dimensions with an irreducible representation of gamma matrices, as one already exhausts all possibilities  with the gamma matrices assigned to the spacetime indexes.  
An alternative  is to consider 
  higher-dimensional representations of the gamma matrices, such as 
 \beq
 \label{eq:gammas_4}
\gamma^0=\sigma^z\otimes\sigma^z,\hspace{1ex} \gamma^1=\mathbb{I}_2\otimes(\ii\sigma^y),\hspace{1ex}\gamma^2=\mathbb{I}_2\otimes(-\ii\sigma^x),
\eeq
for $(2+1)$ dimensions. This representation allows for a pair of  choices  $\gamma^5\in\{\sigma^x\otimes\sigma^z,\sigma^y\otimes\sigma^z\}$ fulfilling the Hermiticity and anti-commutation constraints. For either choice, the four-Fermi QFT~\eqref{continuum} is invariant under a discrete axial rotation 
\beq
\label{eq:chiral_symmetry}
\psi_{\mathsf{f}}(x)\mapsto\ee^{\ii\frac{\pi}{2}\gamma^5}\psi_{\mathsf{f}}(x)=\ii\gamma^5\psi_{\mathsf{f}}(x),
\eeq
and 
one speaks of a discrete chiral symmetry that prevents the fermions from having a mass~\cite{HANDS199329,hep-lat/9706018}. In this work, however, we stick to the lowest-dimensional representation~\eqref{eq:gammas}, and thus dispense with the notion of chirality and axial rotations from the outset. As discussed below, however, there will still be remnants of dynamical mass generation and strongly-coupled phenomena, although one cannot relate them to chiral symmetry breaking. We note that our model exhibits a global $U(N)$ flavour symmetry, which becomes apparent by expressing the fields as a vector $\Psi(\boldsymbol{x})=(\psi_1(\boldsymbol{x}),...,\psi_N(\boldsymbol{x}))^{ t}$ and applying the transformation $\Psi(\boldsymbol{x})\mapsto (u \otimes \mathbb{I}_2)\Psi(\boldsymbol{x})$ on both components of the $N$-flavored Dirac spinors, where  $u\in U(N)$. 

\subsection{Wilson-type discretizations of four-Fermi  field theories and Hubbard bilayers}
\label{sec:discret}

Let us now describe the lattice  scheme chosen to regularize the QFT~\eqref{continuum}. First, spatial coordinates are discretized by introducing an anisotropic lattice spacing, leading to a simple rectangular Bravais lattice described by
\begin{equation}
\label{eq:lattice_an}
\Lambda_s=\{n_1 a_1 \boldsymbol{e}_1+n_2 a_2 \boldsymbol{e}_2:\hspace{1ex} \boldsymbol{n}=(n_1,n_2)\in\mathbb{Z}_{N_{1}}\times\mathbb{Z}_{N_{2}}\}.
\end{equation}
 This lattice is determined by the lattice spacings $a_1,a_2$, the unit vectors $\boldsymbol{e}_1,\boldsymbol{e}_2$,    the number of columns (rows) $N_{1}$ ($N_{2}$) along the $x$- and $y$-directions, and the corresponding spatial area $A_s=\prod_jN_{j}a_j$.  In the Hamiltonian version~\cite{PhysRevD.16.3031} of the naive-fermion prescription~\cite{gattringer_lang_2010}, the  Hamiltonian density $\mathcal{H}_N(\boldsymbol{x})$ in vector representation  includes the following tunnelling processes in the spatial lattice
 \begin{equation}
 \label{eq:naive}
\mathcal{H}_N(\boldsymbol{x})=-\sum_j\left(\overline{\Psi}(\boldsymbol{x})\frac{\ii(\mathbb{I}_N\otimes\gamma^j)}{2a_j}\Psi(\boldsymbol{x}+a_j \boldsymbol{e}_j)+\text{H.c.}\right),
\end{equation}
which follow from the substitution of the spatial derivatives by  forward finite differences.
The corresponding Hamiltonian is given by $H_N=a_1a_2\sum_{\boldsymbol{x}\in\Lambda_s}{:}\mathcal{H}_N(\boldsymbol{x})\!:$, where the integrals have been replaced by sums. We note that the lattice anisotropy $a_1\neq a_2$ should not lead to an anisotropic speed of light, such that Lorentz invariance will still be recovered in the continuum limit. Note that, from a condensed-matter perspective, the discretization anisotropy translates into anisotropic nearest-neighbor tunnellings~\eqref{eq:naive}, and has thus a natural microscopic interpretation. As discussed at length in the following sections, this anisotropy will play an important role in the correlation effects of a QAH effect.

\begin{figure*}[t]
	\centering
	\includegraphics[width=0.65\textwidth]{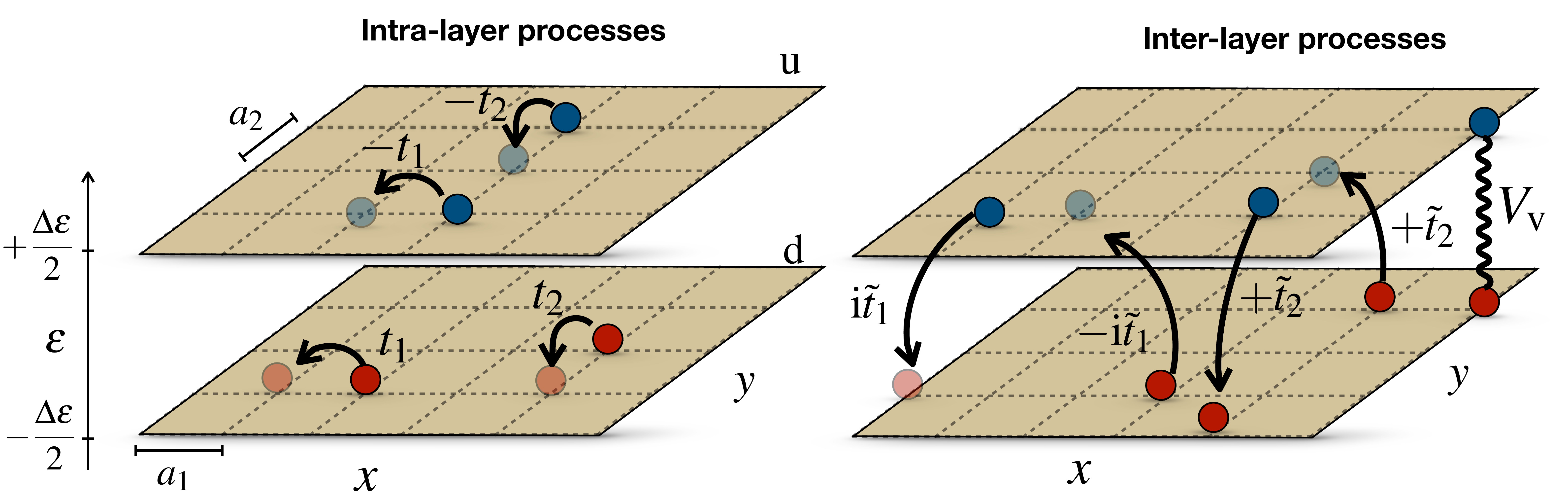}
	\caption{{\bf Hubbard bilayer for the single-flavour four-Fermi-Wilson model:} Visualization of hopping and interaction terms in the 2+1 dimensional Hamiltonian field theory  of Eq.~\eqref{eq:total_H} for $N=1$. The two-component Dirac spinor $\psi_\mathsf{f}(\boldsymbol{x})=(\psi_{\mathsf{f},1}(\boldsymbol{x}),\psi_{\mathsf{f},2}(\boldsymbol{x}))^{\rm t}$ can be mapped onto fermionic operators for the up/down layers $c_{\boldsymbol{n},{\rm u}},c_{\boldsymbol{n},{\rm d}}$. On the left panel, the intra-layer terms $t_1,t_2$ describe the tunnelling of of fermions along the $x$ and $y$ directions of the rectangular lattice~\eqref{eq:gammas_4}, such that the different signs of the tunnelling strengths allow one to implement the Wilson term~\eqref{wilson}. For our choice of gamma matrices,  the bare mass~\eqref{eq:bare_mass} corresponds to an energy imbalance $\pm\Delta\epsilon/2$ between the two layers. On the right panel, we depict the inter-layer terms $\tilde{t}_1,\tilde{t}_2$ corresponding to the tunnelling of of fermions along the $x$ and $y$ directions with the simultaneous change of layer. Finally, the four -Fermi term~\eqref{interaction} can be described as an inter-layer Hubbard interaction  of strength $V_{\rm v}$ along the vertical direction. The correspondence of the microscopic  parameters is derived in Sec.~\ref{sec:strong_coupling}  }
	\label{hoppings}
\end{figure*}

As is well-known in LFT~\cite{NIELSEN198120,NIELSEN1981173}, this naive-fermion approach  gives rise to additional  Dirac fermions in  the continuum limit, known as fermion doublers,  which actually appear as long-wavelength excitations around the boundaries of the Brillouin zone. 
As advanced in the introduction, we consider  Wilson's prescription~\cite{Wilson1977} to deal with the doublers, which can be  sent to the UV-cutoff by introducing an additional mass term that is proportional to the inverse lattice spacings $1/a_j$.   This is achieved through the following Wilson term 
\begin{equation}
\mathcal{H}_W(\boldsymbol{x})=\sum_{j} \left(	\overline{\Psi}(\boldsymbol{x})  \frac{r_j}{2a_j}\big(\Psi(\boldsymbol{x})-\Psi(\boldsymbol{x}+a_j\boldsymbol{e}_j)\big)+ \text{H.c.}\, \right) \label{wilson}
\end{equation}
where $r_j$ is an anisotropic version  the dimensionless Wilson parameter. This term is usually supplemented by an additional bare mass $m$ via
\begin{align}
\label{eq:bare_mass}
\mathcal{H}_M(\boldsymbol{x})= m\,\overline{\Psi}(\boldsymbol{x})\Psi(\boldsymbol{x}).
\end{align}

 The Wilson term~\eqref{wilson} stems from the finite-difference discretization of a second-derivative on the fields, which would be an irrelevant perturbation in the continuum QFT. Accordingly, one may  expect  that the doublers  shall not influence the universal low-energy properties in the continuum limit, which can be recovered around certain critical points of the lattice model. Let us emphasize, however, that this expectation can be a drastic oversimplification, especially in the presence of additional four-Fermi terms
\begin{align}
\mathcal{V}_g(\boldsymbol{x})=-\frac{g^2}{2N}\left(\overline{\Psi}(\boldsymbol{x})\Psi(\boldsymbol{x})\right)^2 \label{interaction}.
\end{align}
As already noted in the introduction, the critical points of the Wilson-discretized model  can actually separate topological from trivial phases, where the relative signs of the doubler masses play a crucial role. Such phases have  widely different low-energy behaviors, as is the case of  the QAH effect. This phase of matter can support a non-zero topological invariant~\cite{PhysRevB.78.195424,10.1143/PTP.73.528,GOLTERMAN1993219}, and a transverse current in spite of the bulk gap~\cite{PhysRevLett.61.2015,PhysRevB.74.085308,PhysRevB.78.195424}, which is  carried by mid-gap current-carrying edge states that may be understood as (1+1)-dimensional versions of  domain-wall fermions~\cite{KAPLAN1992342}. In fact, from  this perspective~\cite{PhysRevLett.105.190404,PhysRevLett.108.181807,PhysRevX.7.031057,BERMUDEZ2018149,PhysRevB.99.064105,PhysRevB.99.125106,Roose2021,PhysRevD.102.094520}, topological insulators in different
symmetry classes and dimensions~\cite{Ryu_2010} correspond to lower-dimensional versions of these domain-wall
fermions~\cite{KAPLAN1992342} with different representations of the Clifford algebra.  In the trivial phase, on the other hand, these edge states are absent, and the long-wavelength description of the model is that of a trivially gapped phase. These differences can be formalised quantitatively in the presence of interactions by means of the renormalisation group~\cite{WILSON197475,RevModPhys.66.129}, as explicitly shown for lower-dimensional instances in~\cite{PhysRevB.99.125106}.
The other possibility, also found in lower-dimensional models~\cite{PhysRevX.7.031057,BERMUDEZ2018149},  is that the interplay of interactions and topology  leads to critical lines in the phase diagram that contain  continuum QFTs defining different universality classes.

The full Hamiltonian of the  four-Fermi-Wilson model~\cite{ziegler2020correlated} can be defined by summing up the four terms defined above
\beq
\label{eq:total_H}
H=a_1 a_2\sum_{\boldsymbol{x}\in\Lambda_{\rm s}} {:}\mathcal{H}_N(\boldsymbol{x})+\mathcal{H}_W(\boldsymbol{x})+\mathcal{H}_M(\boldsymbol{x})+\mathcal{V}_g(\boldsymbol{x}):.
\eeq
As displayed in Fig.~\ref{hoppings} for the single-flavour limit $N=1$, and discussed in more detail in the following section, the discretized model can be regarded as a bilayer model, where the upper (lower) plane belongs to the first (second) component of the Dirac spinor. Within this picture, and in light of our choice of gamma matrices~\eqref{eq:gammas}, the first  contribution~\eqref{eq:naive} describes  inter-layer tunnellings. The Wilson term~\eqref{wilson} contributes to intra-layer tunnellings  of opposite  strengths for each of the layers and, together with the bare mass~\eqref{eq:bare_mass}, determines the energy imbalance between the upper and lower layers. Finally,  the four-Fermi coupling~\eqref{interaction} is depicted as spring that describes  the interactions of two fermions that occupy simultaneously  the same site of the upper and lower layers, and can be  seen to correspond to a Hubbard-type density-density interaction. 

This  representation can be interpreted as a (2+1)-dimensional generalization of Creutz's cross-link ladder~\cite{PhysRevLett.83.2636}, a paradigmatic model of topological insulators in the $\mathsf{AIII}$ class that has been the subject of intensive studies in recent years~\cite{PhysRevLett.102.135702,PhysRevLett.112.130401,PhysRevB.94.245149,PhysRevA.88.063613,PhysRevX.7.031057,PhysRevB.96.035139,PhysRevLett.121.150403,PhysRevB.98.155142,PhysRevX.10.041007}. If we now introduce more flavours $N>1$, the picture would be that of stacked bilayers that get coupled among each other via the four-Fermi term. This perspective can be understood in light of the synthetic dimensions~\cite{PhysRevLett.108.133001,PhysRevLett.112.043001}, and  will offer valuable insight when exploring the phase diagram of the model in the strong-coupling limit. Let us note, however, that we can interpret this model as a  Fermi-Hubbard model with a generalized spin-orbit coupling, which also yields valuable insight as it underlies the proposal for the QS cold-atom scheme, briefly presented in~\cite{ziegler2020correlated}, and described in detail below.  

\subsection{ $N$-flavored Chern insulators and  the quantum anomalous Hall effect}
\label{sec: chern}

Due to the discrete translation symmetry introduced by the lattice~\eqref{eq:lattice_an}, it is natural to Fourier transform the fields 
\beq
\label{eq:fourier_transform}
\Psi(\boldsymbol{x})=\frac{1}{\sqrt{A_s}}\sum_{\boldsymbol{k}\in {\rm BZ}}\ee^{\ii\boldsymbol{k}\cdot\boldsymbol{x}}\Psi(\boldsymbol{k})
\eeq
where the momenta  $\boldsymbol{k}=(k_1,k_2)^t$ are restricted to the first Brillouin zone ${\rm BZ}= \{k_{j}=-\pi/a_j+2\pi n_j/N_{j}a_j:\hspace{1ex} n_j\in\mathbb{Z}_{N_{j}},\hspace{1ex} j\in\{1,2\}\}$, and we assume periodic boundary conditions such that the ${\rm BZ}$ is a 2-torus.
 After this Fourier transform, and in the absence of the four-Fermi term~\eqref{interaction}, the  quadratic free Hamiltonian obtained by restricting the sum in Eq.~\eqref{eq:total_H} to the naive, Wilson and bare mass terms, can be rewritten as
\begin{equation}
H_F= \sum_{\boldsymbol{k} \in {\rm BZ}} \!\Psi^\dagger \!(\boldsymbol{k}) h_{\boldsymbol{k}}(m) \Psi(\boldsymbol{k}).
\end{equation}
 Here, the single-particle Hamiltonian  reads
\beq
h_{\boldsymbol{k}}(m)=\boldsymbol{d}_{\boldsymbol{k}}(m)\cdot(\mathbb{I}_N\otimes\boldsymbol{\sigma}),
\label{freehamiltonian}
\eeq
where we have introduced the vector of Pauli matrices $\boldsymbol{\sigma}$, and the  following mapping of the BZ to a real vector
\beq
\boldsymbol{d}_{\boldsymbol{k}}(m)\!=\!\!\!\left(\!\frac{\sin(k_1 a_1)}{a_1},\frac{\sin(k_2 a_2)}{a_2},m+\sum\limits_{j}\frac{r_j}{a_j}\bigg(1-\cos(k_j a_j)\bigg)\!\! \right)\!\!.
\label{freehamiltonian_d_vector}
\eeq

 This  can be readily diagonalized  yielding  $2N$  energy bands, a pair  $\epsilon_{\pm}(\boldsymbol{k})=\pm\epsilon(\boldsymbol{k})=\pm\parallel\boldsymbol{d}_{\boldsymbol{k}}(m)\parallel$ for each   flavour. These bands display a relativistic dispersion at small momenta  $\delta k_j\ll 1/a_j$, namely $\epsilon(\delta \boldsymbol{k})\approx({\delta\boldsymbol{k}^2+m^2})^{1/2}$, where   we find an effective speed of light $c=1$ in accordance with natural units and the emergence of Lorentz invariance. Note that there are additional points in $\boldsymbol{k}$-space, $\boldsymbol{K}_{\boldsymbol{n}_d}=(\pi n_{d,1}/a_1,\pi n_{d,2}/a_2)$ for $\boldsymbol{n}_d=(n_{d,1},n_{d,2}) \in \{0,1\}\times \{0,1\}$,  around which the energy dispersion corresponds again to that of a continuum Dirac equation with a non-zero mass. These fermion doublers can thus be described by a collection of   Dirac  spinors $\{\Psi_{\boldsymbol{n}_d}(\boldsymbol{x})\}_{\boldsymbol{n}_d}$ with $N$ flavours, which are governed by the following long-wavelength quantum field theory in the free case
 \beq
 \label{eq:continuum_theory}
 {H}_F=\!\!\int\!\!{\rm d}^2x\sum_{\boldsymbol{n}_d} 
\overline{\Psi}_{\boldsymbol{n_d}}(\boldsymbol{x})\! \left(-\ii\big(\mathbb{I}_N\otimes\gamma_{\boldsymbol{n_d}}^j\big) \partial_j+m_{\boldsymbol{n_d}} \right)\!\Psi_{\boldsymbol{n_d}}(\boldsymbol{x}).
 \eeq
 Here, depending on the particular doubler, the corresponding representation of the gamma matrices may involve different signs with respect to that of the original QFT~\eqref{eq:gammas}, namely
 \beq
\label{eq:gammas_doubler}
\gamma^0_{\boldsymbol{n}_d}=\gamma^0,\hspace{2ex} \gamma^1_{\boldsymbol{n}_d}=(-1)^{n_{d,1}}\gamma^1,\hspace{2ex}\gamma^2_{\boldsymbol{n}_d}=(-1)^{n_{d,2}}\gamma^2.
\eeq
 Likewise, the  effective mass depends on the particular doubler, and is usually referred to as a Wilson mass
\begin{equation}
\label{eq:WIlson_masses}
m_{\boldsymbol{n}_d}=m+\frac{2r_1}{a_1}n_{d,1}+\frac{2r_2}{a_2}n_{d,2}.
\end{equation}

We thus see that, in addition to the Dirac fermion of mass $m_{(0,0)}=m$ around the centre of the Brillouin zone, the so-called $\Gamma$ point for $\boldsymbol{n}_d=(0,0)$; there are three additional fermions at high-symmetry points, such as the corner  $R$ for $\boldsymbol{n}_d=(1,1)$, and the edge centers $M$ for $\boldsymbol{n}_d\in\{(1,0),(0,1)\}$, which have Wilson masses that scale with the inverse lattice spacings. If the bare mass is small enough $ma_j\ll 1$, and one takes the continuum limit $a_j\to 0$, these doublers become infinitely heavy and are naively expected to decouple from the universal long-wavelength physics. Note, however, that the bare mass is simply a lattice parameter and could also take large negative values $m a_j\approx -2r_j$, such that some of the Wilson masses~\eqref{eq:WIlson_masses} flip their sign. This connects to the band-inversion that leads to topological insulators~\cite{RevModPhys.83.1057,RevModPhys.82.3045}, and is responsible for the non-zero quantisation of the transverse conductivity in the QAH effect~\cite{PhysRevLett.61.2015, PhysRevB.74.085308,PhysRevB.78.195424}.

In the non-interacting regime, the Thouless-Kohmoto-Nightingale-den-Nijs formula~\cite{PhysRevLett.49.405} provides a direct link of   such a transverse conductivity $\sigma_{xy}$ with the Chern numbers $\{N_{{\rm Ch},\mathsf{b}}\}$ that characterise  the band structure. Mathematically~\cite{nakahara_2017},  these  correspond  to the topological invariants associated to the fibre sub-bundles for each occupied energy band. In the present context, $N$ bands get fully occupied in the  half-filled groundstate, such that the transverse conductivity reads
\begin{equation}
\label{eq:TKNN}
\sigma_{xy}	= \frac{e^2}{2\pi \hbar} N_{{\rm Ch}},\hspace{2ex}N_{{\rm Ch}}=\sum_{\mathsf{b}=1}^N N_{{\rm Ch},\mathsf{b}}.
\end{equation}
These Chern numbers turn out to be proportional to  the Berry phase $\gamma_{\mathsf{b}}$ defined as the integral of the Berry curvature in momentum space $\mathcal{F}^{ij}_{\mathsf{b}}\!(\boldsymbol{k})=\partial_{k_i}\mathcal{A}_{\mathsf{b}}^j(\boldsymbol{k})-\partial_{k_j}\mathcal{A}_{\!\mathsf{b}}^i(\boldsymbol{k})$, where $\boldsymbol{\mathcal{A}}_{\!\!\mathsf{b}}(\boldsymbol{k})=\bra{\epsilon_{\mathsf{b}}(\boldsymbol{k})}\ii\boldsymbol{\nabla}_{\boldsymbol{k}}\ket{\epsilon_{\mathsf{b}}(\boldsymbol{k})}$ is the Berry connection~\cite{doi:10.1098/rspa.1984.0023}. In the present context,  and considering periodic boundary conditions to explore the bulk band structure, the Brillouin zone corresponds to a toroidal manifold acting as the base space of the principal $U(1)$ bundle associated to the eigenstates where the Chern invariants are defined
\begin{equation}
N_{{\rm Ch},\mathsf{b}}= -\frac{1}{2\pi} \gamma_\mathsf{b}=-\frac{1}{4\pi} \int_{\text{BZ}}dk_i\wedge dk_j \,\mathcal{F}^{ij}_{\mathsf{b}}\!(\boldsymbol{k}).
\end{equation}

As neatly discussed in~\cite{bernevig_hughes_2013}, this topological invariant can be rewritten as the Pontryagin index, the winding number of the mapping between the momentum-space 2-torus and  the unit 2-sphere $\hat{\boldsymbol{d}}:\mathbb{T}^2\to\mathbb{S}^2$, given by the unitary vector field $\hat{\boldsymbol{d}}_{\boldsymbol{k}}(m)=\boldsymbol{d}_{\boldsymbol{k}}(m)/\!\!\parallel\boldsymbol{d}_{\boldsymbol{k}}(m)\!\!\parallel$  obtained by normalizing Eq.~\eqref{freehamiltonian_d_vector}. This winding number counts how many times the mapping wraps around the unit sphere, and yields
\beq
\label{eq:chern_winding}
N_{\rm Ch,\mathsf{b}}=\frac{1}{4\pi}\int_{\rm BZ}\!{\rm d}^2k\,\hat{\boldsymbol{d}}_{\boldsymbol{k}}(m)\cdot(\partial_{k_1}\hat{\boldsymbol{d}}_{\boldsymbol{k}}(m)\wedge\partial_{k_2}\hat{\boldsymbol{d}}_{\boldsymbol{k}}(m)).
\eeq
As one can numerically check, this integral can be accurately  evaluated using the long-wavelength approximation~\eqref{eq:continuum_theory}. Each of the $N$-flavour massive Dirac fermions, including the doublers,   may be considered  as a monopole for the Berry curvature with a non-zero contribution to the winding number
 \begin{equation}
\label{eq:Chern_free}
N_{\rm Ch}=\frac{N}{2}\sum_{\boldsymbol{n}_d}(-1)^{(n_{d,1}+n_{d,2})}{\rm sign}(m_{\boldsymbol{n}_d}).
\end{equation}
Assuming that the Wilson parameters are  $r_j=1$, the lattice constants  are $a_1\leq a_2$, and introducing their anisotropy ratio
\beq
\label{eq:spatial_anisotropy}
 \xi_2=\frac{a_1}{a_2}
\eeq
fulfilling $ \xi_2<1$, we find that the Chern numbers are 
\beq
\begin{aligned}
\label{eq:critical_lines_chern}
N_{\rm Ch}=\left\{   \begin{matrix} 
      \hspace{-11ex} 0, \hspace{2.5ex} {\rm if}\,\,\,\,\,\,0<ma_1,\hspace{3ex} \phantom{,} \\
    \hspace{-10ex}   -N,   \hspace{2ex} {\rm if}\,\,\,-2\xi_2<ma_1<0,\\
             \hspace{-8ex} 0,   \hspace{2.ex} {\rm if}\,\,\,-2<ma_1<-2,\\
   +N,   \hspace{2ex} {\rm if}\,\,\,\,\,\,-2(1+\xi_2)<ma_1<-2\xi_2,\\
           \hspace{-6ex}    0,   \hspace{2.ex} {\rm if}\,\,\,\,\,\,ma_1<-2(1+\xi_2).\\
   \end{matrix}\right.
   \end{aligned}
\eeq

\begin{figure}[t]
	\centering
	\includegraphics[width=0.5\textwidth]{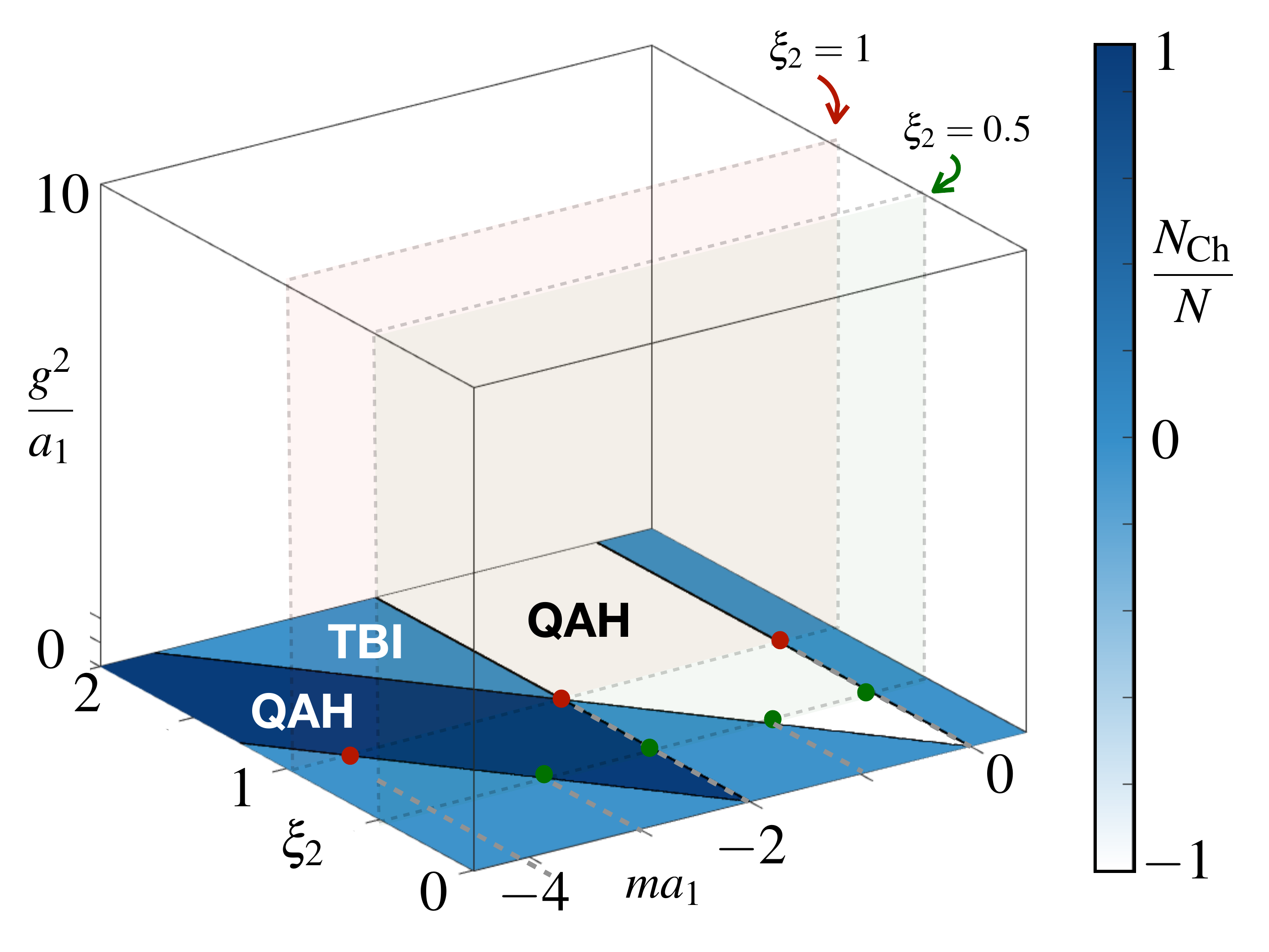}
	\caption{{\bf Phase diagram and non-interacting QAH phases:} We define parameter space as $\{ma_1,\xi_2,g^2/2a_1\}$, such that the free-fermion case $g^2=0$ corresponds to the base plane where one can depict the Chern numbers~\eqref{eq:Chern_free} in a contour plot. In this limit, one finds quantum anomalous Hall (QAH) and trivial band insulating (TBI) phases separated by topological quantum phase transitions. For spatial anisotropy $\xi_2=\half$, there are four critical points which, according to Eq.~\eqref{eq:critical_lines_chern}, lie  at $ma_1\in\{-3,-2,-1,0\}$ and are marked with green dots in the figure  where the Wilson mass~\eqref{eq:WIlson_masses} of a single fermion gets inverted by changing $m$. In the isotropic case $\xi_2=1$, there are only three critical points depicted by red dots. The central one at $ma_1=-2$ marks the mass inversion of a couple of fermions, and thus connects two different QAH phases with edge currents circulating in opposite directions. In the sections below, we shall explore the vertical direction as $g^2>0$ along the two planes $\xi_2\in\{\half,1\}$ shaded green/red. }
	\label{fig:chern}
\end{figure}

As advanced previously, as  the bare mass takes negative values proportional to the inverse lattice spacings, the groundstate of the system can support a non-zero Chern number~\eqref{eq:critical_lines_chern}, and thus transport current transversally~\eqref{eq:TKNN}. We shall refer to these states as $N$-flavored Chern insulators. The specific values of the dimensionless parameter $-ma_1$ where the Chern number undergoes an abrupt change are associated with the closure of the energy gap and, thus, to a second-order quantum phase transition that cannot be characterized by an order parameter or any spontaneous symmetry breaking. On the contrary, there is an observable~\eqref{eq:TKNN} that displays robust quantized values  depending on an underlying topological invariant~\eqref{eq:Chern_free} that changes across these critical points and leads to a topological quantum phase transition (TQPT). Each of the TQPTs in Eq.~\eqref{eq:critical_lines_chern} are marked by the mass inversion of one of the emerging Dirac fermions~\eqref{eq:continuum_theory}. Only when the number of negative Wilson masses~\eqref{eq:WIlson_masses} is odd, does the groundstate lead to a Chern insulator with a non-zero transverse conductivity.

For instance, in the single-flavour case,  there are  topological phases with ${N_{\rm Ch}\in\{1,-1\}}$ characterized by either 1 or 3 Dirac fermions with a negative Wilson mass, respectively. As depicted in Fig.~\ref{fig:chern}, for spatial anisotropies $\xi_2\neq 1$, these Chern insulators are separated in parameter space by an intermediate trivial phase with ${N_{\rm Ch}=0}$, which is characterized by the same number of Dirac fermions with negative and positive Wilson masses. In the isotropic limit $a_1=a_2$, the Wilson masses at the edge centers $M$  of the Brillouin zone become degenerate $m_{(1,0)}=m_{(0,1)}$,  such that there is no intermediate trivial phase separating the two QAH phases with $N_{\rm Ch}=\pm 1$. We note that the regime of opposite anisotropy $\xi_2>1$ has a completely analogous description~\eqref{eq:critical_lines_chern}, where one must only exchange $a_1\leftrightarrow a_2$ in the expressions above. Also, although we have set  $r_j=1$, we note that different values of the Wilson parameters would simply lead to rescalings of the bare mass axis to $ma_1\to ma_1/r_1$, and the spatial anisotropy $\xi_2\to\xi_2r_2/r_1$,   without changing the overall structure of the phase diagram. In the following, we will set $r_j=1$. The question to be answered below is to what extent these topological phases survive for  substantial non-zero interactions $g^2>0$, as one explores the parameter space in the vertical direction of Fig.~\ref{fig:chern}. 
\section{\bf  Strong couplings  and orbital magnetism}
\label{sec:strong_coupling}

In this section, we explore the strong-coupling limit where the four-Fermi term~\eqref{interaction} dominates. We note that,  as discussed below Eq.~\eqref{continuum}, the  four-Fermi  coupling strength $g^2$ is not dimensionless, as for the Gross-Neveu model in (1+1) dimensions where dynamical mass generation yields an example of dimensional transmutation~\cite{PhysRevD.10.3235}. In $(2+1)$ dimensions, the coupling strength has units of inverse mass, and we can define the strong-coupling regime of the four-Fermi-Wilson model by letting the bare parameters $g^2/a_j\gg1$, $\forall j$.

\subsection{Super-exchange  and   quantum  compass models}

Let us start by discussing the single-flavour limit $N=1$. To find the explicit connection with the Hubbard bilayer depicted in Fig.~\ref{hoppings}, let us note that the lattice field operators in Eq.~\eqref{eq:fourier_transform} have dimensions of mass, and can be thus rescaled to define the upper- and lower-layer $\ell\in\{\rm u,d\}$ fermionic operators
\beq
c_{\boldsymbol{n},{\rm u}}=\sqrt{a_1a_2}\psi_1(\boldsymbol{x}),\hspace{1ex} c_{\boldsymbol{n},{\rm d}}=\sqrt{a_1a_2}\psi_2(\boldsymbol{x}),
\eeq
for the anisotropic rectangular lattice~\eqref{eq:lattice_an}. These operators satisfy the standard anti-commutation relations used in condensed matter $\{c^{\phantom{\dagger}}_{\boldsymbol{n},\ell},c^{\dagger}_{\boldsymbol{n}',\ell'}\}=\delta_{\ell,\ell'}\delta_{\boldsymbol{n},\boldsymbol{n}'}$. This rescaling leads directly to the following free Hamiltonian 
\beq
\begin{split}
H_F\!=\!&-\!\sum_{\boldsymbol{n},j}\!\!\left(t_{\! j}^{\phantom{\dagger}}\!\!\!\left({c}^\dagger_{\boldsymbol{n},{\rm u}\phantom{d}\!\!\!\!}{c}^{\phantom{\dagger}}_{\boldsymbol{n}+{\bf e}_j,{\rm u}\phantom{d}\!\!\!\!}-{c}^\dagger_{\boldsymbol{n},{\rm d}}{c}^{\phantom{\dagger}}_{\boldsymbol{n}+{\bf e}_j,{\rm d}}\right)\!\!+{\rm H.c.}\!\right)\\
&\hspace{-4ex}-\sum_{\boldsymbol{n},j}\left( \ii^{j}\tilde{t}_{\! j}^{\phantom{\dagger}}\!\!\left({c}^\dagger_{\boldsymbol{n},{\rm u}\phantom{d}\!\!\!\!}{c}^{\phantom{\dagger}}_{\boldsymbol{n}+{\bf e}_j,{\rm d}}+(-1)^j{c}^\dagger_{\boldsymbol{n},{\rm d}\phantom{d}\!\!\!\!}{c}^{\phantom{\dagger}}_{\boldsymbol{n}+{\bf e}_j,{\rm u}\phantom{d}\!\!\!\!}\right)+{\rm H.c.}\right)\\
&\hspace{-4ex}+\sum_{\boldsymbol{n},j}\frac{\Delta\epsilon}{2}\!\!\left({c}^\dagger_{\boldsymbol{n},{\rm u}\phantom{d}\!\!\!\!}{c}^{\phantom{\dagger}}_{\boldsymbol{n},{\rm u}\phantom{d}\!\!\!\!}-{c}^\dagger_{\boldsymbol{n},{\rm d}}{c}^{\phantom{\dagger}}_{\boldsymbol{n},{\rm d}}\right)\!,
\end{split}
\eeq
 where we recall that {$j=1,2$. Here, we have introduced  the intra- $t_j$ and inter-layer $\tilde{t}_j$ tunnellings along the $j$-th axis in the first two lines, depicted by black arrows in   Fig.~\ref{hoppings}
  \beq
 t_1=\frac{1}{2a_1},\hspace{1.5ex} t_2=\frac{1}{2a_2},\hspace{1.5ex}  \tilde{t}_1=\frac{1}{2a_1},\hspace{1.5ex} \tilde{t}_2=\frac{1}{2a_2}.
 \eeq
  The third line of the above Hamiltonian contains an energy imbalance $\Delta\epsilon$ between the layers, which is depicted by the energy scale  in   Fig.~\ref{hoppings}, and corresponds to 
\beq
\frac{\Delta\epsilon}{2}=m+\frac{1}{a_1}+\frac{1}{a_2}.
\eeq

 In addition, up to irrelevant quadratic terms that only contribute with a constant shift of the energies for a fixed number of particles,  the single-flavour four-Fermi term~\eqref{interaction} can be rewritten as a density-density  Hubbard interaction between the two layers
\beq
V_g=\sum_{\boldsymbol{n}}V_{\mathsf{v}} c^{{\dagger}}_{\boldsymbol{n},{\rm u}\phantom{d}\!\!\!\!\!}c^{\phantom{\dagger}}_{\boldsymbol{n},{\rm u}\phantom{d}\!\!\!\!\!}c^{{\dagger}}_{\boldsymbol{n},{\rm d}}c^{\phantom{\dagger}}_{\boldsymbol{n},{\rm d}},
\eeq 
where the  Hubbard interaction strength is repulsive 
\beq
V_{\mathsf{v}}=\frac{g^2}{a_1a_2}.
\eeq 

According to the strong-coupling condition $g^2/a_j\gg1$, we see that the Hubbard interactions  are much larger than any of the intra- or inter-layer tunnellings in this particular regime,  since $V_{\mathsf{v}}\gg 1/a_j=|t_j|=|\tilde{t}_j|$. Accordingly, the half-filled groundstate corresponds to a Mott insulator without any pair of fermions occupying simultaneously both layers at the same site (see Fig.~\ref{super-exchange}). Following the condensed-matter nomenclature, we  call such high-lying excitations doublons, which necessarily appear as we dope the system away from half-filling, where a non-zero charge gap appears.   The remaining question to address is if this Mott insulator is featureless, or if there is some ordering with respect to the orbital/layer degrees of freedom and a non-zero orbital gap.

\begin{figure*}[t]
	\centering
	\includegraphics[width=0.75\textwidth]{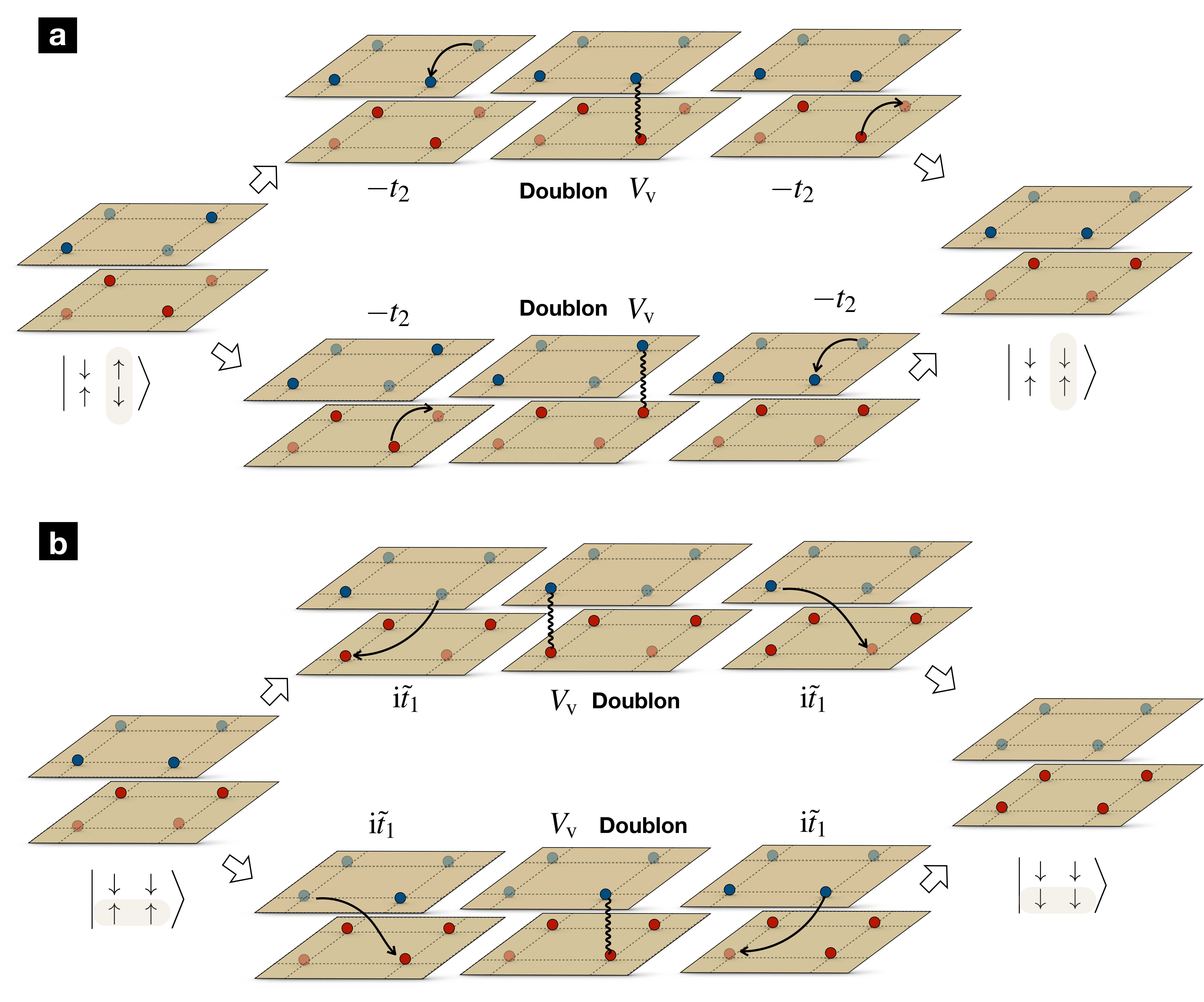}
	\caption{{\bf Super-exchange processes in the strong-coupling limit:} For $g^2/a_j\gg 1$, the half-filled groundstate corresponds to a Mott insulator where fermions occupy single sites of each layer, avoiding  double occupancies of neighboring sites along the vertical direction (i.e. doublons). This is can be achieved by different fermion configurations corresponding to individually half-filled layers, as depicted in the left schemes of {\bf (a)} and {\bf (b)}. Sequential  tunnelling processes can  virtually create doublons with an  energy cost  proportional to the Hubbard interaction strength $V_{\mathsf{v}}$ and, as depicted by the central panels, connect to a different groundstate configuration that is consistent with the Mott-insulating constraint. The depicted super-exchange processes correspond to {\bf (a)} single spin-flip along the $y$ axis, and {\bf (b)} double spin-flip along the $x$ axis.}
	\label{super-exchange}
\end{figure*}

 Although the tunnelling of fermions is inhibited in the half-filled Mott insulator, as it leads to the creation of  doublons with a very large energy cost, there can be second-order processes where such doublons are virtually created and annihilated. These virtual tunnellings, also known as super-exchange processes~\cite{PhysRev.79.350,ANDERSON196399}, are the leading perturbative corrections and become responsible for the antiferromagnetic ordering of Mott insulators in the strong-coupling limit of the standard Hubbard model, as discussed in the introduction. In the present context, analogous super-exchange mechanisms can be formalized using
 the language of  $SU(2)$ orbital spin operators
 \beq
 \label{eq:orbital_spins}
 \begin{split}
 \tau_{\boldsymbol{n}}^x&=c^{\dagger}_{\boldsymbol{n},{\rm u}\phantom{d}\!\!\!\!\!}c^{\phantom{\dagger}}_{\boldsymbol{n},{\rm d}}+c^{\dagger}_{\boldsymbol{n},{\rm d}}c^{\phantom{\dagger}}_{\boldsymbol{n},{\rm u}\phantom{d}\!\!\!\!\!},\\
 \tau_{\boldsymbol{n}}^y&=\ii c^{\dagger}_{\boldsymbol{n},{\rm d}}c^{\phantom{\dagger}}_{\boldsymbol{n},{\rm u}\phantom{d}\!\!\!\!\!}-\ii c^{\dagger}_{\boldsymbol{n},{\rm u}\phantom{d}\!\!\!\!\!}c^{\phantom{\dagger}}_{\boldsymbol{n},{\rm d}},\\
  \tau_{\boldsymbol{n}}^z&=c^{\dagger}_{\boldsymbol{n},{\rm u}\phantom{d}\!\!\!\!\!}c^{\phantom{\dagger}}_{\boldsymbol{n},{\rm u}\phantom{d}\!\!\!\!\!}-c^{\dagger}_{\boldsymbol{n},{\rm d}}c^{\phantom{\dagger}}_{\boldsymbol{n},{\rm d}},\\
  \end{split}
 \eeq
  and  a Schrieffer-Wolff transformation~\cite{PhysRevB.37.9753} with the following graphical interpretation.  Let us start from a common situation that is also found in the standard Hubbard model if we consider that  the spin up/down states of the electrons correspond to the two layers. In this process, a pair of fermions occupying neighbouring sites in different  layers   tunnel,  creating a virtual doublon, and then annihilate it through a second tunnelling event. This may lead to an effective swap of the fermions that can be interpreted as a spin-flip exchange $\ket{\uparrow_{\boldsymbol{n}},\downarrow_{\boldsymbol{n}+\boldsymbol{e}_j}}\to\ket{\downarrow_{\boldsymbol{n}},\uparrow_{\boldsymbol{n}+\boldsymbol{e}_j}}$ along any of the two spatial directions (see  Fig.~\ref{super-exchange}{\bf (a)} for such an spin-flip exchange along the $y$ axis). We note that, due to the anisotropic fermion tunnellings (see  Fig.~\ref{hoppings}), these spin-flip exchanges will have different strengths along the two axes, the specific value of which can be obtained through the Schrieffer-Wolff formalism $J_{\perp,1}=-1/{g^2\xi_2}$, $J_{\perp,2}=-{\xi_2}/{g^2}$. 
  
  A new virtual process allowed by the pattern of  inter-layer tunnellings, with no counterpart in the standard Hubbard model, occurs for a pair of fermions occupying neighbouring sites within the same layer. These fermions  can  tunnel to the adjacent site by simultaneously changing layer, virtually creating a doublon, and then annihilate it through a second inter-layer tunnelling event.  This can lead to an effective  pair-tunnelling between the different layers that is consistent with the Mott-insulating state, and  may be interpreted as a double-spin-flip exchange $\ket{\uparrow_{\boldsymbol{n}},\uparrow_{\boldsymbol{n}+\boldsymbol{e}_j}}\to\ket{\downarrow_{\boldsymbol{n}},\downarrow_{\boldsymbol{n}+\boldsymbol{e}_j}}$ along any of the two spatial directions (see  Fig.~\ref{super-exchange}{\bf (b)} for such a double-spin-flip exchange along the $x$ axis). In this case, the anisotropy and the different phases of the inter-layer tunnellings conspire to yield  spin flips  of  a different strength, but also of an opposite   sign depending on direction $\tilde{J}_{\perp,1}=+1/{g^2\xi_2}$, $\tilde{J}_{\perp,2}=-{\xi_2}/{g^2}$.  This sign difference is crucial, as the combination of the two types of spin-flip process can be rewritten as the following effective spin model with direction-dependent interaction:
\beq
\label{eq:compass}
\begin{split}
H_{\rm eff}= \sum_{\boldsymbol{n}} \left( J_x \tau^{x\phantom{d}\!\!\!\!\!}_{\boldsymbol{n}}\tau_{\boldsymbol{n}+\boldsymbol{e}_2}^{x\phantom{d}\!\!\!\!\!}+J_y \tau^y_{\boldsymbol{n}\phantom{e_1}\!\!\!\!\!\!\!\!}\tau_{\boldsymbol{n}+\boldsymbol{e}_1}^y-h\tau_{\boldsymbol{n}}^z\right), 
\end{split}
\eeq
where we have introduced the following coupling strengths
\beq
\label{eq:spin_couplings}
J_x=\frac{-a_1}{2g^2a_2 }, \hspace{1.5ex}J_y=\frac{-a_2}{2 g^2a_1},\hspace{1.5ex} h=-m-\frac{1}{a_1}-\frac{1}{a_2}.
\eeq

Let us note that, in contrast to the original lattice model~\eqref{eq:total_H} we started from,  the above model  only contains nearest-neighbour quartic terms, and must be supplemented by a so-called Gutzwiller projector $\mathcal{P}_G=\Pi_{\boldsymbol{n}}(1-c^{{\dagger}}_{\boldsymbol{n},{\rm u}\phantom{d}\!\!\!\!\!}c^{\phantom{\dagger}}_{\boldsymbol{n},{\rm u}\phantom{d}\!\!\!\!\!}c^{{\dagger}}_{\boldsymbol{n},{\rm d}}c^{\phantom{\dagger}}_{\boldsymbol{n},{\rm d}})$ onto the no-doublon  subspace. Accordingly, at half filling,  every site labelled by $\boldsymbol{n}$ is filled with a single  fermion that may occupy either layer, such that the projected orbital four-Fermi operators~\eqref{eq:orbital_spins} can be represented in a  tensor-product Hilbert space $\mathsf{H}_{\rm eff}=\otimes_{\boldsymbol{n}}\mathbb{C}^2=\otimes_{\boldsymbol{n}}{\rm span}\{\ket{{\uparrow_{\boldsymbol{n}}}},\ket{{\downarrow_{\boldsymbol{n}}}}
\}$ using the standard Pauli matrices  
\beq
\label{eq:spins}
 \tau_{\boldsymbol{n}}^\alpha=\mathbb{I}_2\stackrel{m-1}{\cdots}\otimes\mathbb{I}_2\otimes\sigma^\alpha\otimes\mathbb{I}_2\otimes\stackrel{N_{1}\!N_{2}-m}{\cdots}\otimes\mathbb{I}_2, 
 \eeq
 where $m=(n_2-1)N_{1}+n_1$ labels the sites of the rectangular lattice from east to west and south to north. In the rest of this section, we use this tensor-product description of the strong-coupling Hilbert space. This allows us to interpret Eq.~\eqref{eq:compass} as a spin model with ferromagnetic couplings~\eqref{eq:spin_couplings}, such that the original fermionic statistics no longer apply. We note that, by performing a unitary operation  flipping   the operators $\tau^{x\phantom{y}\!\!\!\!}_{\boldsymbol{n}},\tau^y_{\boldsymbol{n}}\to,(-1)^{(n_1+n_2)}\tau^{x\phantom{y}\!\!\!\!}_{\boldsymbol{n}},(-1)^{(n_1+n_2)}\tau^y_{\boldsymbol{n}}$ in a checkerboard pattern, the ferromagnetic couplings become anti-ferromagnetic $J_x,J_y\to-J_x,-J_y$, while the transverse field is preserved. There is thus no fundamental difference between the ferromagnetic or anti-ferromagnetic spin models.

Notably, this  spin model~\eqref{eq:compass} belongs to the family of \textit{quantum compass models}~\cite{RevModPhys.87.1}. In comparison to the Heisenberg model~\cite{Heisenberg1928}, which arises in the strong-coupling limit of the standard Hubbard model~\cite{assa},  there are no  $\tau^z\tau^z$ interactions. Moreover, $\tau^x\tau^x$ interactions only couple neighbouring sites  along the $y$-direction, whereas the $\tau^y\tau^y$ interactions do so along the $x$-direction, as depicted in the scheme of Fig.~\ref{fig:compass}. Up to an irrelevant  relabelling, this corresponds to the anisotropic 90$^{\rm o}$ compass model in a square lattice~\cite{PhysRevB.71.195120}, since $J_x\neq J_y$ in general~\eqref{eq:spin_couplings}. This  characteristic directionality of the spin-spin interactions, which are no longer invariant under $C_4$ rotations of the spatial lattice,  evokes  a compass that distinguishes  north/south from east/west directions. In  the honeycomb lattice~\cite{KITAEV20062}, a similar directionality is responsible for the appearance of intrinsic topological order in a spin-liquid groundstate~\cite{doi:10.1146/annurev-conmatphys-033117-053934,doi:10.1146/annurev-conmatphys-031218-013401}. Again, this ordering  cannot be understood by the spontaneous breakdown of symmetry or by the onset of a non-zero order parameter, all of which are absent in spin liquids even down to zero temperature. Instead, intrinsic topological order hosts long-range entanglement in the groundstate  and   anyonic excitations, which differs from the  QAH phases that have been discussed so far.

\begin{figure}[t]
	\centering
	\includegraphics[width=0.45\textwidth]{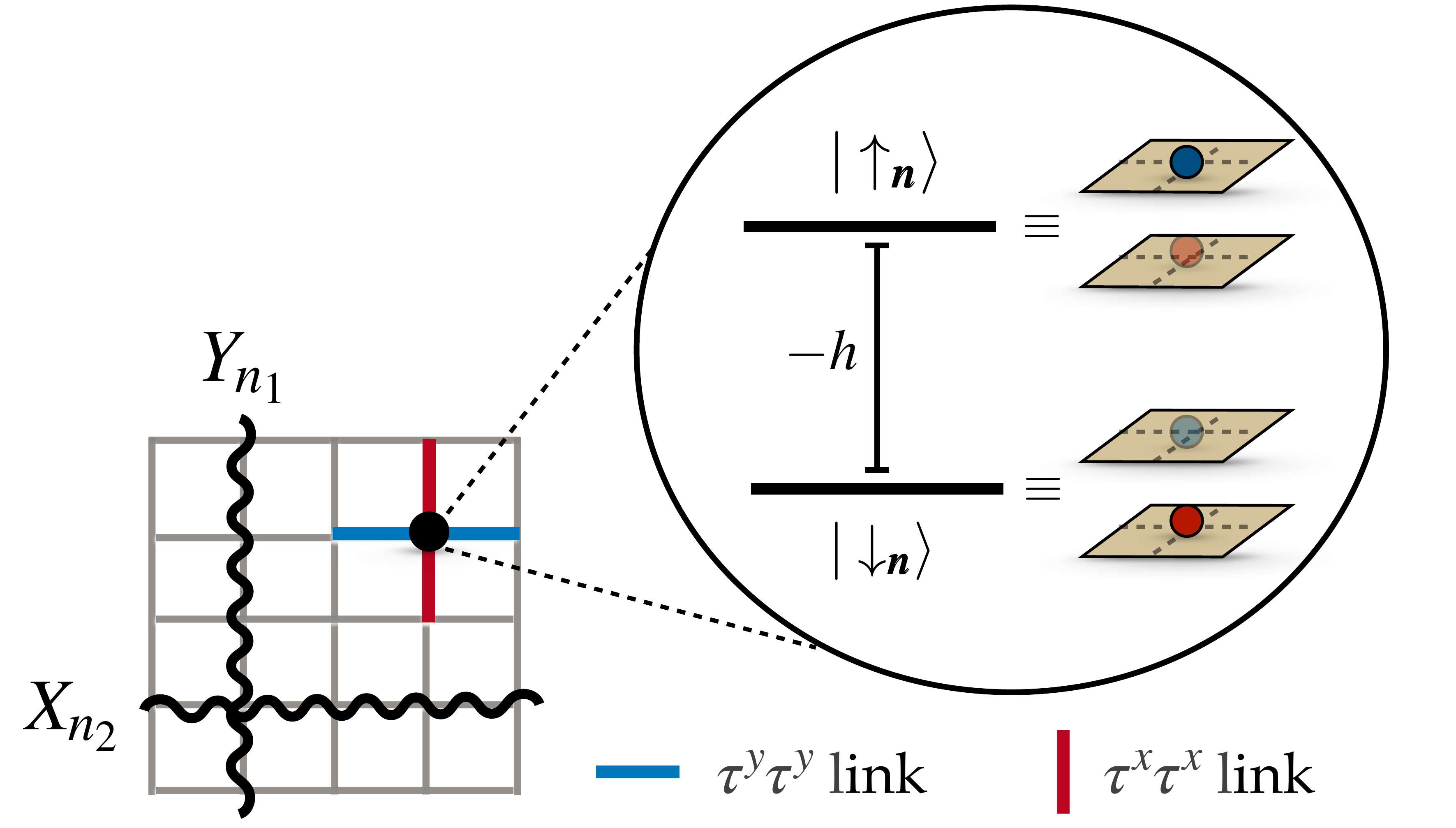}
	\caption{{\bf Scheme of the 90$^{\rm o}$ quantum compass model in a transverse field:} The orbital spins are arranged on the sites of  a rectangular lattice, where the inset depicts the mapping of the spin states to the two possible distribution of fermions avoiding doublons. The energy difference between these two spin states is proportional to the transverse field $h$. The spin-spin interactions can be depicted by blue (red) links forming 90$^{\rm o}$, which represent a $\tau^y\tau^y$ ($\tau^x\tau^x$) coupling of strength $J_x$ ($J_y$). The wavy lines represent the two types of sliding symmetry described in the text.}
	\label{fig:compass}
\end{figure}

 In the case of the square lattice, the anisotropic  90$^{\rm o}$ compass model for a vanishing transverse field ${h=0}$~\eqref{eq:compass} has been thoroughly studied from both condensed-matter and  quantum-information perspectives. This model is invariant under the so-called sliding symmetries~\cite{PhysRevB.71.024505,PhysRevB.71.195120}, which lie midway between  a  local gauge symmetry and a purely global one. These sliding symmetries consist of strings formed by the product of $\tau_{\boldsymbol{n}}^{x\phantom{y}\!\!\!\!\!} ,\tau^y_{\boldsymbol{n}}$ operators along  rows and columns $X_{n_2}={\displaystyle \prod_{n_1}}\tau^x_{(n_1,n_2)}, Y_{n_1}={\displaystyle \prod_{n_2}}\tau^y_{(n_1,n_2)}$ which, due to the interaction directionality, clearly commute with the  Hamiltonian $[X_{n_2},H_{\rm eff}]=[Y_{n_1},H_{\rm eff}]=0$ , but anti-commute with each other $\{X_{n_2},Y_{n_1}\}=0$, $\forall n_1,n_2$. In  Kitaev's toric~\cite{KITAEV20032} and surface~\cite{quant-ph/9811052,Freedman2001,doi:10.1063/1.1499754} codes for quantum error correction (QEC), where the spins are arranged  on the lattice links rather than the sites  and interact via four-body terms instead of two-body ones, these sliding symmetries also appear  along rows and  columns  of the real and reciprocal lattices, and can be related to Dirac's electric and 't Hooft's magnetic field lines of a $\mathbb{Z}_2$ lattice gauge theory~\cite{doi:10.1063/1.1665530,RevModPhys.51.659,fradkin_2013}. This underlying gauge symmetry allows one to identify an extensive set of local operators that commute  with the Hamiltonian and with each other, and generate the  so-called stabilizer group~\cite{quant-ph/9705052} under which the  groundstate manifold remains invariant.  Moreover, these stabilizers also allow one to connect the sliding symmetries acting on different rows (columns), such that they have the same effect on the groundstate manifold. In this way, one can use the sliding symmetries as   unitary operations on logical qubit(s) encoded in the groundstate manifold. Although these logical qubits are not inherently robust to noise, one can perform QEC by measuring the local stabilizers, inferring the most likely error and, subsequently, correcting it by applying simple unitaries to bring the system back to the groundstate manifold, withstanding a considerable  amount of physical errors~\cite{RevModPhys.87.307}. From a condensed-matter perspective, the groundstate of the toric/surface-code Hamiltonian  is a quantum spin liquid with intrinsic topological order rather than symmetry-breaking long-range order, such that the encoded logical qubits  and QEC codes are typically referred to as topological qubits and topological codes, respectively.

 \begin{figure*}[t]
	\centering
	\includegraphics[width=0.85\textwidth]{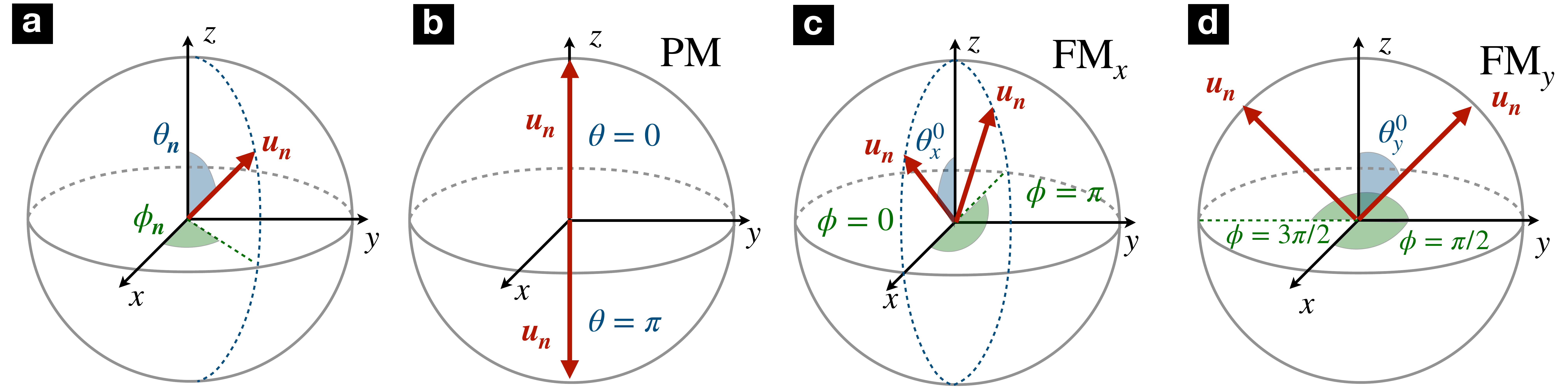}
	\caption{{\bf Spin coherent states and mean-field solutions:} {\bf (a)} The variational ansatz~\eqref{eq:spin_coherent_states} is built from a tensor product of spin coherent states characterized by a unit vector $\boldsymbol{u}_n$ with polar and azimuthal angles $\theta_{\boldsymbol{n}},\phi_{\boldsymbol{n}}$. For spin $S=1/2$, the unit sphere is commonly referred to as Bloch's sphere. {\bf (b)} Paramagnetic PM phase, where each of the spins points along the north or south pole depending on the sign of the transverse field $h$. {\bf (c, d)} Ferromagnetic  phases where all spins are parallel and point in two possible directions within the $xz$ (FM$_x$) or $yz$ (FM$_y$)  plane.  }
	\label{fig:bloch_sphere}
\end{figure*}

 For the 90$^{\rm o}$ compass model~\eqref{eq:compass}, on the contrary, there is no   $\mathbb{Z}_2$ gauge symmetry and the commuting set of stabilizers is no longer formed by local observables. Instead,  one defines the stabilizers by taking the product  of  two  sliding symmetries along a  neighboring pair of rows (columns) $X_{n_2}X_{n_2+1}$ $\big(Y_{n_1}Y_{n_1+1}\big)$, which leads to the so-called  subsystem Bacon-Shor  codes~\cite{PhysRevA.73.012340}. Although gauge symmetry is no longer present in these codes, the underlying notion of redundancy that underlies such local symmetries  still appears in a different guise.  These codes host   additional gauge qubits, the state of which is irrelevant, and  can be manipulated using certain gauge operations without affecting the encoded logical information~\cite{PhysRevLett.95.230504}. This can be exploited to improve the QEC routines. For instance,  QEC based on the Bacon-Shor  code can still be performed by the sequential measurements of a set of local gauge operators corresponding to some of the compass-model pairwise interactions~\eqref{eq:compass}, such that their product  yields the measurement outcome of the desired non-local stabilizer. This  approach lowers the  overhead and  complexity with respect to fault-tolerant protocols  for the toric/surface code~\cite{PhysRevLett.98.220502,yoder2017universal} but, on the other hand, can only attain a bounded  reduction of the logical error rate with respect to the physical one, which is achieved for intermediate optimal lattice sizes~\cite{napp2012optimal}. Interestingly, one  may also use these pairwise interactions to modify the stabilizers by a so-called gauge fixing, making the encoded logical qubits more robust to certain types of biased errors. This leads to the so-called two-dimensional compass codes which, in contrast to Bacon-Shor codes, have a non-zero error threshold, below which  the protection of the logical qubits  can be  improved  arbitrarily by steadily increasing  the lattice size~\cite{PhysRevX.9.021041,PhysRevA.101.042312}.

 Regardless of the relevance of the  90$^{\rm o}$ compass model for QEC, the limit of  zero transverse field ${h=0}$ has also received considerable attention from the condensed-matter community.
Note that the total number of sliding symmetries $X_{n_2},Y_{n_1}$ scales with half of the perimeter of the square lattice $N_1+N_2$, which could lead   to exponentially-large groundstate degeneracies as occurs for classically-frustrated magnets. Since these symmetries anti-commute with each other $\{X_{n_2},Y_{n_1}\}=0$,  while the aforementioned stabilisers satisfy $[X_{n_2}X_{n_2+1},Y_{n_1}]=0=[X_{n_2},Y_{n_1}Y_{n_1+1}]$, $\forall n_1,n_2$, one can indeed prove rigorously that all of the eigenstates are at least two-fold degenerate, independently of the specific system size~\cite{PhysRevB.71.024505}. This does not preclude, however, that higher degeneracies exist. In fact, it has been argued numerically that there are exponentially-many low-lying excitations that collapse exponentially fast to the groundstate  as the lattice size increases~\cite{PhysRevB.72.024448}, leading to the aforementioned analogue of frustrated magnets with an exponentially-large groundstate manifold. This connects to the quantum-mechanical version of the mechanism of order by disorder~\cite{villain_bidaux_carton_conte_1980} whereby  quantum fluctuations, which typically tend to destroy a possible long-range  order, instead  induce it by selecting a particular groundstate from the exponentially-large     manifold of possible candidates. In fact, instead of  a  quantum spin-liquid phase, one finds two different ferromagnetic orders  in the groundstate, which are connected by a  first-order phase transition~\cite{PhysRevB.72.024448,PhysRevB.75.144401,PhysRevLett.102.077203} as  the ratio of the exchange couplings is varied across  the self-dual point ${J_x=J_y}$~\cite{PhysRevB.71.195120}.

To the best of our knowledge,  the 90$^{\rm o}$ compass model  in a non-vanishing transverse field~\eqref{eq:compass} remains largely unexplored in comparison to the zero-field limit discussed above. The transverse field is a source of  additional quantum fluctuations, which favours a paramagnetic phase where all spins align in the direction of the transverse field. In our recent work~\cite{ziegler2020correlated}, we briefly discussed how the competition with the magnetic phases can give rise to second-order phase transitions  at finite ratios of the transverse-field to exchange couplings. In the two following subsections, we shall give a detailed exposition of our findings using  variational mean-field and tensor-network approaches. Since either magnetic or paramagnetic phases are ultimately different from the QAH phase discussed in the previous section, our results show that the fate of this topological phase is to disappear in the strong-coupling limit. Equipped with the lessons learned from the strong-coupling limit, and in particular the identification of the magnetic orders that compete with the topological phase, we will analyze in the next section  how large the interactions can be before the correlated QAH phase gives way to the (para)magnets. Before turning to this discussion, let us note that the effective spin models will be modified when other choices of the Wilson parameters $r_j\neq 1$ are explored. By simple inspection of Fig.~\ref{super-exchange}, it is clear that the intra-layer second-order processes~{\bf (a)}
 and the inter-layer ones~{\bf (b)} have different strengths even when acting along the same axis, which will lead to perturbations breaking the perfect directionality of the spin-spin interactions. However, these perturbations should preserve the underlying global symmetry, and we believe that unless one focuses on very large values of the Wilson parameters, they will not modify the nature of the critical lines. This question will be  studied in the future.

\subsection{Variational mean-field description}
\label{sec:var_mf}

Let us now describe a variational mean-field approach to the groundstate of the 90$^{\rm o}$ compass model  in a non-vanishing transverse field~\eqref{eq:compass}. There are a variety of methods for obtaining a mean-field approximation in a many-body problem~\cite{chaikin_lubensky_1995}, all of which  share the common aspect of addressing non-perturbative  phenomena, such as criticality and phase transitions, by underestimating the effect of inter-particle correlations. Although  mean-field methods have their own well-known limitations, they typically give a correct qualitative picture of effects that cannot be captured by perturbation theory. In this subsection, we focus on variational mean-field theory~\cite{chaikin_lubensky_1995}, where one constructs a variational ansatz  by a family of  fully-uncorrelated  tensor product states.

In the present context, we define the variational ansatz as $\ket{\Psi_{\rm MF}(\{\theta_{\boldsymbol{n}},\phi_{\boldsymbol{n}}\})}= \otimes_{\boldsymbol{n}}\ket{\psi_{\rm MF}(\theta_{\boldsymbol{n}},\phi_{\boldsymbol{n}})}$, where the state of each of the lattice spins $\ket{\psi_{\rm MF}(\theta_{\boldsymbol{n}},\phi_{\boldsymbol{n}})}$ is described by a spin coherent state pointing along the unit vector $\boldsymbol{u}_{\boldsymbol{n}}=(\sin\theta_{\boldsymbol{n}}\cos\phi_{\boldsymbol{n}},\sin\theta_{\boldsymbol{n}}\sin\phi_{\boldsymbol{n}}, \cos\theta_{\boldsymbol{n}})$ within  the 2-sphere~\cite{Radcliffe_1971}. Here, $\theta_{\boldsymbol{n}}\in[0,\pi]$ and $\phi_{\boldsymbol{n}}\in[0,2\pi)$ are the so-called polar and azimuthal angles depicted in Fig.~\ref{fig:bloch_sphere} {\bf (a)}, which can in principle be inhomogeneous across the lattice. However, given our previous discussion of the zero-field limit, where we remarked on the competition of two possible ferromagnetic orders, we shall assume that all spins point  in the same direction $\theta_{\boldsymbol{n}}=\theta,\phi_{\boldsymbol{n}}=\phi$, $\forall \boldsymbol{n}$, such that translational invariance is maintained in the groundstate. By selecting the state where all spins point to the south pole as a fiducial state, the spin coherent state can be written as
\begin{equation}
\label{eq:spin_coherent_states}
\ket{\Psi_{\rm MF}(\{\theta,\phi\})}= \otimes_{\boldsymbol{n}} \ee^{\frac{\ii}{2}\tan\left(\frac{\theta}{2}\right)\left(\cos\phi\tau_{\boldsymbol{n}}^x-\sin\phi\tau^y_{\boldsymbol{n}}\right)}\ket{\downarrow_{\boldsymbol{n}}},
\end{equation}
which leads to the following set of expectation values
\beq
\label{eq:exp_values}
\begin{split}
&\bra{\Psi_{\rm MF}(\{\theta,\phi\})}{\tau}^\alpha_{\boldsymbol{n}}\ket{\Psi_{\rm MF}(\{\theta,\phi\})}={u}^\alpha_{\boldsymbol{n}},\\
&\bra{\Psi_{\rm MF}(\{\theta,\phi\})}{\tau}^\alpha_{\boldsymbol{n}}\cdot{\tau}^\beta_{\boldsymbol{n}'}\ket{\Psi_{\rm MF}(\{\theta,\phi\})}={u}^\alpha_{\boldsymbol{n}}{u}^\beta_{\boldsymbol{n}'}.
\end{split}
\eeq	
We can now calculate the variational energy through the expectation value of the effective compass Hamiltonian on these coherent states $\epsilon(\theta, \phi)=\bra{\Psi_{\rm MF}}H_{\rm eff}\ket{\Psi_{\rm MF}}/{N_{1}N_{2}}$, yielding
\begin{equation}
\label{eq:mf_energy}
\epsilon(\theta, \phi)=J_x \sin^2\theta \cos^2\phi+J_y \sin^2\theta \sin^2\phi-h  \cos\theta.
\end{equation}

In a variational approach, the  groundstate can be found by calculating the partial derivatives of the variational energy  with respect to $\theta$, $\phi$ and setting them equal to zero,  which leads to the non-linear system of equations
\begin{flalign} \label{derivatives}
 \sin(2\theta)(J_x\cos^2\phi+J_y \sin^2\phi)+h  \sin\theta&=0,\\ 
 \sin(2\phi)\sin^2\theta(J_x-J_y)&=0.
\label{derivatives2}
\end{flalign}
There are three different types of solution. If $\theta=p\cdot \pi$ with ${p \in \mathbb{Z}_2}$, all the terms in (\ref{derivatives}) and (\ref{derivatives2}) are zero for arbitrary $\phi$. In this case all the spins point along the poles of the 2-sphere, as depicted in Fig.~\ref{fig:bloch_sphere} {\bf (b)}  such that the  energy  is 
\beq
\epsilon(p\pi,\phi)= \pm h.
\label{eq:energy_pm}
\eeq
The two possible states represent paramagnetic states with the spins either aligned or anti-aligned with respect to the transverse magnetic field. We shall refer to this phase as PM.

The second set of possible solutions is found for  $\phi=p\cdot \pi$ with ${p \in \mathbb{Z}_2}$, such that equation~(\ref{derivatives2}) is always zero. In this case the first equation reduces to
$
2 J_x \cos(\theta)+ h=0
$,
and the solution  is found for the groundstate polar angle 
\begin{equation}
\theta^0_{x}= \arccos\left(\left|\frac{h}{2 J_x}\right|\right).
\end{equation}
The particular ground state energy is given by
\begin{equation}
\label{eq:gs_energy_x}
\epsilon(\theta_{x}^0,p\pi)= -|J_x|\left(1-\frac{h^2}{4 J_x^2}\right).
\end{equation}
Considering the expectation values in Eq.~\eqref{eq:exp_values}, these solutions describe a ferromagnetic state where all spins have a non-zero magnetization along the $x$-$z$ plane (see  Fig.~\ref{fig:bloch_sphere} {\bf (c)}). We shall refer to this phase as FM$_x$.

 The third type of solution occurs for  $\phi=(2p+1) \frac{\pi}{2}$ with ${p \in \mathbb{Z}_2}$, where  equation~(\ref{derivatives2}) is again zero. Then, the first equation reduces to $2 J_y \cos(\theta)+h=0$. It follows that the groundstate polar angle  is given by
\begin{equation}
\label{eq:angle_MF_y}
\theta_{y}^0= \arccos\left(\left|\frac{h}{2J_y}\right|\right),
\end{equation}
and the corresponding energy per spin is
\begin{equation}
\label{eq:gs_energy_y}
\epsilon\left(\theta_{y}^0,(2p+1)\frac{\pi}{2}\right)= -|J_y|\left(1-\frac{h^2}{4 J_y^2}\right).
\end{equation}
Considering the expectation values in Eq.~\eqref{eq:exp_values}, these solutions describe a ferromagnetic state where all spins have a non-zero magnetization along the $y$-$z$ plane (see  Fig.~\ref{fig:bloch_sphere} {\bf (d)}). We shall refer to this phase as FM$_y$.

Let us now discuss the possibility of finding  also a  mean-field solution with mixed  magnetization along both the $x$ and $y$ directions. As one can see, the condition $J_x=J_y=:J$ solves directly the second equation~(\ref{derivatives2}) for all possible angles. Similarly, all the dependence of  equation~(\ref{derivatives}) on  $\phi$ disappears, such that one can find solutions for arbitrary  azimuthal angle provided that $2J\cos\theta+h=0$, all of which have the same energy~\eqref{eq:mf_energy}. These solutions contain the previous second- and third-type ferromagnetic solutions discussed previously, and will be referred to as  FM$_{\phi}$.  For zero transverse field, this is the mean-field account of the special isotropic point of the  90$^{\rm o}$ compass model,   where we recall that an exponentially-large groundstate degeneracy was predicted in the thermodynamic limit~\cite{PhysRevB.72.024448}. Although the variational mean-field does not capture the exponentially-fast clustering of the low-lying excitations with the system size, it identifies the isotropic regime $J_x=J_y$ as a special point. Let us note that, contrary to the standard situation in Heisenberg ($XY$) models, the independence on the azimuthal angle does not derive from a global $SU(2)$ ($U(1)$) symmetry, where spontaneous symmetry breaking would in principle select only a specific groundstate angle. For the compass model, this independence is instead related to the intermediate sliding symmetries.
By inspecting the two solutions in Eqs.~\eqref{eq:gs_energy_x}-\eqref{eq:gs_energy_y}, one can also see that the point $J_x=J_y$ corresponds to a level crossing, such that the mean-field analysis also captures the  first-order nature of the phase transition at zero transverse field. At this point, the spins change their orientation abruptly from $x$- to $y$-direction.

It is now a matter of comparing the energies of all the possible mean-field solutions in Eqs.~\eqref{eq:energy_pm},~\eqref{eq:gs_energy_x} and~\eqref{eq:gs_energy_y} to find the groundstate and possible quantum phase transitions for a non-vanishing transverse field. Note that the compass Hamiltonian~\eqref{eq:compass} in the limit $ |J_x|\gg |J_y|, |h|$ corresponds to a set of decoupled Ising columns with a ferromagnetic ground state where all spins point along the $x$-direction.  From our mean-field solution, this phase turns out to be the low-energy configuration for weaker exchange couplings $J_x$ whenever $|J_x|>|J_y|$, which requires $a_1>a_2$ in light of Eq.~\eqref{eq:spin_couplings}, and
\beq
\label{eq:x_mag}
\langle\tau^x_{\boldsymbol{n}}\rangle=\sin(\theta^0_x)=\sqrt{1-\frac{h^2}{4 J_x^2}},\hspace{2ex}{\rm if}\hspace{1ex}  |J_x|\geq \frac{|h|}{2}.
\eeq 
Note that this  leads to  a quantum phase transition towards the second type  of solution~\eqref{eq:x_mag} at $|h|=2|J_x|$, where the global $\mathbb{Z}_2 $   symmetry generated by $U=\Pi_{\boldsymbol{n}}\tau^z_{\boldsymbol{n}}$ is spontaneously broken. This symmetry can also be combined with inversion symmetry of the spins about the  center of the rectangular lattice $\boldsymbol{n}\to (N_1,N_2)-\boldsymbol{n}$. For smaller transverse fields, the order parameter~\eqref{eq:x_mag} becomes non-zero, and can be attained by two possible azimuthal angles $\phi=p\pi$ with $p\in\{0,1\}$, corresponding to the two possible ferromagnetic arrangements, which are degenerate due to the spontaneous  breaking of the $\mathbb{Z}_2$ symmetry. For larger transverse  fields, the spins align with the transverse field, pointing along the $\pm z$ axis for positive/negative values of $h$, and the spontaneous magnetization along the $x$ axis  vanishes exactly. Note that only one of the paramagnetic configurations is the low-energy state, and there is thus no degeneracy as the $\mathbb{Z}_2$ symmetry is preserved in this case. This critical point thus represents a second-order quantum phase transition  between a  ferromagnet along the $x$ axis (FM$_x$) and a paramagnet (PM), in contrast to the first-order nature of the self-dual critical point $J_x=J_y$ for  vanishing transverse field. We emphasize that the symmetry being broken is not related to the internal $U(N)$ symmetry, but is instead a combined spatial and spin  inversion symmetry in the  language of the quantum compass model. In the following sections, we shall re-interpret this discrete symmetry in the language of the Dirac spinors and the emerging continuum QFTs.

 The same type of argument holds in the opposite direction of the exchange-coupling anisotropy, where the model becomes a set of decoupled Ising rows in the limit of very large anisotropy $ |J_y|\gg |J_x|, |h|$ . We find that for $|J_y|>|J_x|$ and $|J_y|>|h|/2$,  the non-zero order parameter is
\beq
\label{eq:y_mag}
\langle\tau^y_{\boldsymbol{n}}\rangle=\sin(\theta^0_y)=\sqrt{1-\frac{h^2}{4 J_y^2}}, \hspace{2ex}{\rm if}\hspace{1ex}  |J_y|\geq \frac{|h|}{2}.
\eeq  
and corresponds to the third type of variational solutions discussed above. In this case, the critical point $|h|=2|J_y|$ describes a second-order quantum phase transition between an  ferromagnet along the y axis (FM$_y$) and a paramagnet (PM).

 At this point it is worth taking a step back to understand these phases from the perspective of the orbital spin operators~\eqref{eq:orbital_spins} and the Hubbard bilayer interpretation. The PM corresponds to a density-imbalanced Mott-insulating phase, where one finds   the fermions occupying all of the sites of a single layer. Depending on the sign of the effective transverse field~\eqref{eq:spin_couplings}, the upper ($h>0$) or lower ($h<0$) layers will be the preferred choice. The FM$_x$ (FM$_y$)  phases represent a long-range order whereby the fermions equally populate both layers, avoiding double occupancies, and   delocalising over the two vertically-neighboring sites of the bilayer. This establishes a real (complex) non-zero bond density $\langle\tau^x_{\boldsymbol{n}}\rangle\propto{\rm Re}\{\langle c^{\dagger}_{\boldsymbol{n},{\rm u}\phantom{d}\!\!\!\!\!}c^{\phantom{\dagger}}_{\boldsymbol{n},{\rm d}}\rangle\}$ $\left(\langle\tau^y_{\boldsymbol{n}}\rangle\propto{\rm Im}\{\langle c^{\dagger}_{\boldsymbol{n},{\rm u}\phantom{d}\!\!\!\!\!}c^{\phantom{\dagger}}_{\boldsymbol{n},{\rm d}}\rangle\}\right)$ that preserves translational invariance and long-range order $\langle \tau^x_{\boldsymbol{n}}\tau^x_{\boldsymbol{n}+\boldsymbol{r}}\rangle\to\langle \tau^x_{\boldsymbol{n}}\rangle^2$ $\left(\langle \tau^y_{\boldsymbol{n}}\tau^y_{\boldsymbol{n}+\boldsymbol{r}}\rangle\to\langle \tau^y_{\boldsymbol{n}}\rangle^2\right)$ as $|\boldsymbol{r}|\to\infty$. We will generally refer to these phases as orbital ferromagnets and orbital paramagnets.

   \begin{figure}[t]
	\centering
	\includegraphics[width=0.5\textwidth]{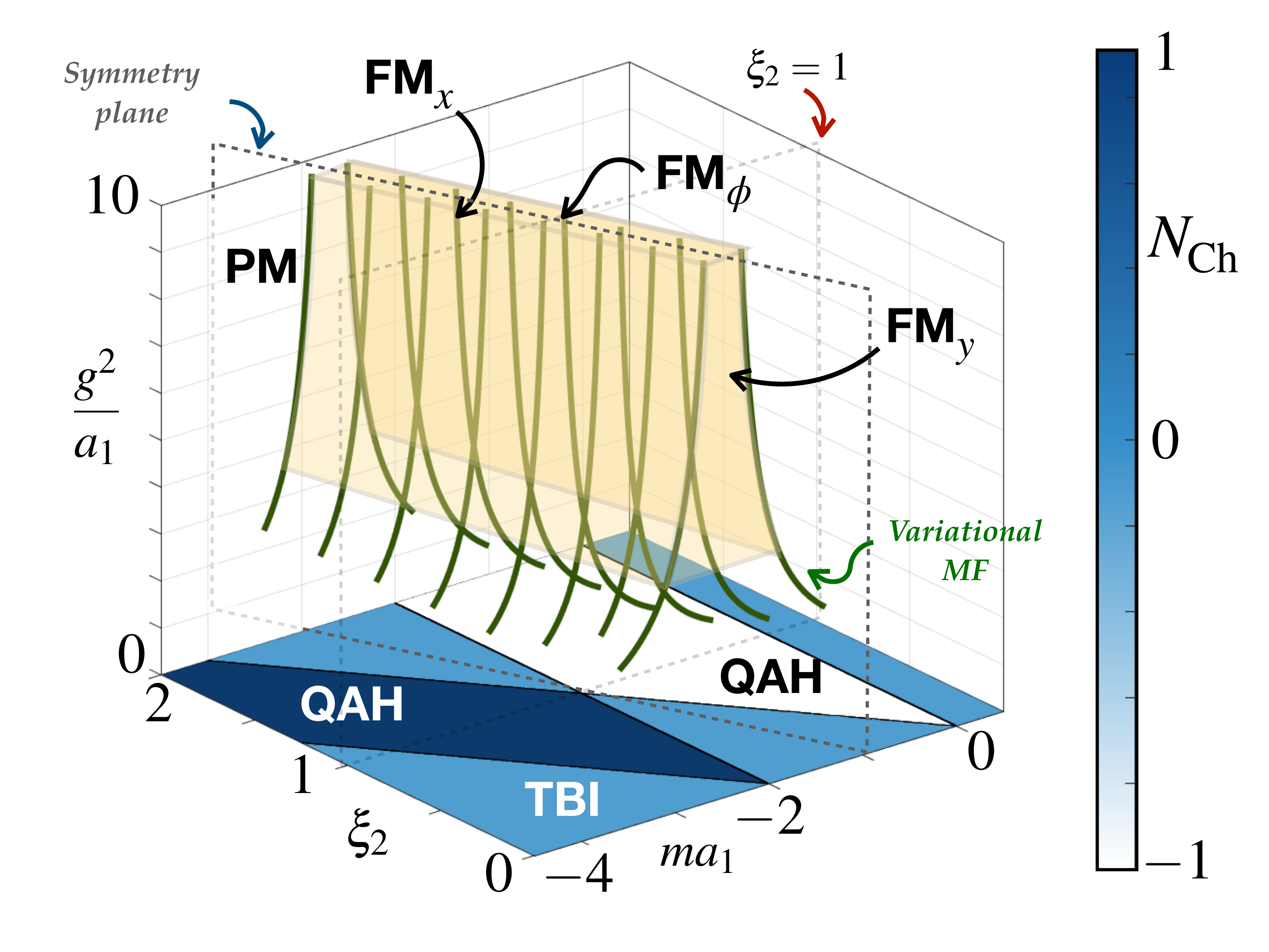}
	\caption{{\bf Orbital ferromagnetism for strong couplings and $N=1$:} In the limit of large interactions $g^2/a_j\gg 1$, we have identified orbital paramagnetic (PM) and ferromagnetic (FM$_{x}$,FM$_{y}$, separated by the critical lines in Eq.~\eqref{criticalcompass}. The inner volume, colored with a light orange, contained within these lines corresponds to a ferromagnetic phase where the orbital spins point along the $x$ axis for spatial anisotropies $\xi_2>1$, or along the $y$ axis for $\xi_2<1$. In the isotropic limit $\xi_2=1$, the spins van order along any azimuthal angle $\phi\in[0,2\pi)$, leading to the FM$_{\phi}$ phase contained in the area within the critical lines in Eq.~\eqref{criticalcompass}. All the surrounding region   in white for $g^2>0$ corresponds to the paramagnet. We also plot the symmetry plane that divides the ferromagnets into two equal halves, given by the condition of a vanishing transverse field $h=0$, i.e. $ma_1=1+\xi_2$.  }
	\label{fig:ferrophases}
\end{figure}

Let us now  rewrite the equations for the critical lines separating these orbital phases in terms of the original microscopic parameters of the four-Fermi-Wilson model~\eqref{eq:spin_couplings}.  We find two different critical lines depending on the anisotropy
\beq
\begin{split}
g^2=\frac{a_1}{a_2 \left|m+\frac{1}{a_1}+\frac{1}{a_2}\right|},\hspace{1ex}{\rm if} \hspace{1ex}a_1>a_2,\\
g^2=\frac{a_2}{a_1 \left|m+\frac{1}{a_1}+\frac{1}{a_2}\right|},\hspace{1ex}{\rm if} \hspace{1ex}a_1<a_2.\\
\end{split}
\label{criticalcompass}
\eeq
These critical lines are depicted as green solid lines in Fig.~\ref{fig:ferrophases}, and delimit a volume in parameter space where one expects to find the symmetry-broken orbital ferromagnets. We are thus certain that the QAH phases described in the previous section will disappear when the interactions $g^2$ are sufficiently strong. However, the variational mean-field approach for the strong-coupling limit has its own limitations. First of all, being a strong-coupling limit that is exact in the strict limit $g^2/a_j\to\infty$, we cannot predict what happens at intermediate interactions,  nor locate the critical lines separating the QAH phase from the ferromagnets and paramagnets. In addition, being a mean-field approximation, we expect that the exact location of the critical lines and the scaling of the order parameter will differ from the correct critical phenomena, which would require other methods that can better accommodate for inter-particle correlations. We will address both limitations in the following sections.

\subsection{Variational tensor-network description}
\label{sec:var_tn}

In this section, we benchmark the results of the  mean-field approximation for the 90$^{\rm o}$ compass model presented in the previous section by means of a variational algorithm based on a projected entangled-pair state (PEPS)~\cite{verstraete2004renormalization,verstraete2008matrix,orus2014practical,ran2020tensor}. The PEPS represents a natural generalization of the one-dimensional variational ansatz based on matrix product states (MPS)~\cite{verstraete2008matrix,schollwock2011density} to two, or even  higher, spatial dimensions. These variational states improve upon the separable mean-field ansatz~\eqref{eq:spin_coherent_states} by including inter-particle correlations, and can be understood in terms of pairs of maximally-entangled states describing  auxiliary degrees of freedom on  neighboring lattice sites, which are locally projected onto the  lower-dimensional subspace of  physical spins residing at each lattice site. This construction can be mathematically expressed as a  network of tensors with multiple indexes corresponding to the physical and auxiliary degrees of freedom, such that those  corresponding to the auxiliary ones are contracted. Accordingly,  the PEPS belongs to the family of tensor-network variational algorithms.

To study groundstate properties of  quantum lattice Hamiltonians in two spatial dimensions,
one can  {\it (i)} variationally optimize the PEPS tensors, so as to minimise the expectation value of the  corresponding Hamiltonian~\cite{corboz2016variational,lubasch2014algorithms}. Alternatively,  in analogy to spectroscopic methods that determine the particle spectrum via the imaginary-time evolution of correlators in Euclidean LFTs~\cite{gattringer_lang_2010}, {\it (ii)}  one may evolve the system in imaginary time until a stationary state corresponding to the groundstate is reached. This assumes  that this groundstate is unique, and  that the energy gap is non-zero, as done in the time-evolving block-decimation method (TEBD) for one-dimensional chains~\cite{vidal2007classical,orus2008infinite,jordan2008classical}. In the following, we will use this second method in the thermodynamic limit, for the infinite PEPS state (iPEPS), which we describe  briefly in the next section.

\begin{figure}[t]
\centering
\includegraphics[scale=0.8,width=0.5\textwidth]{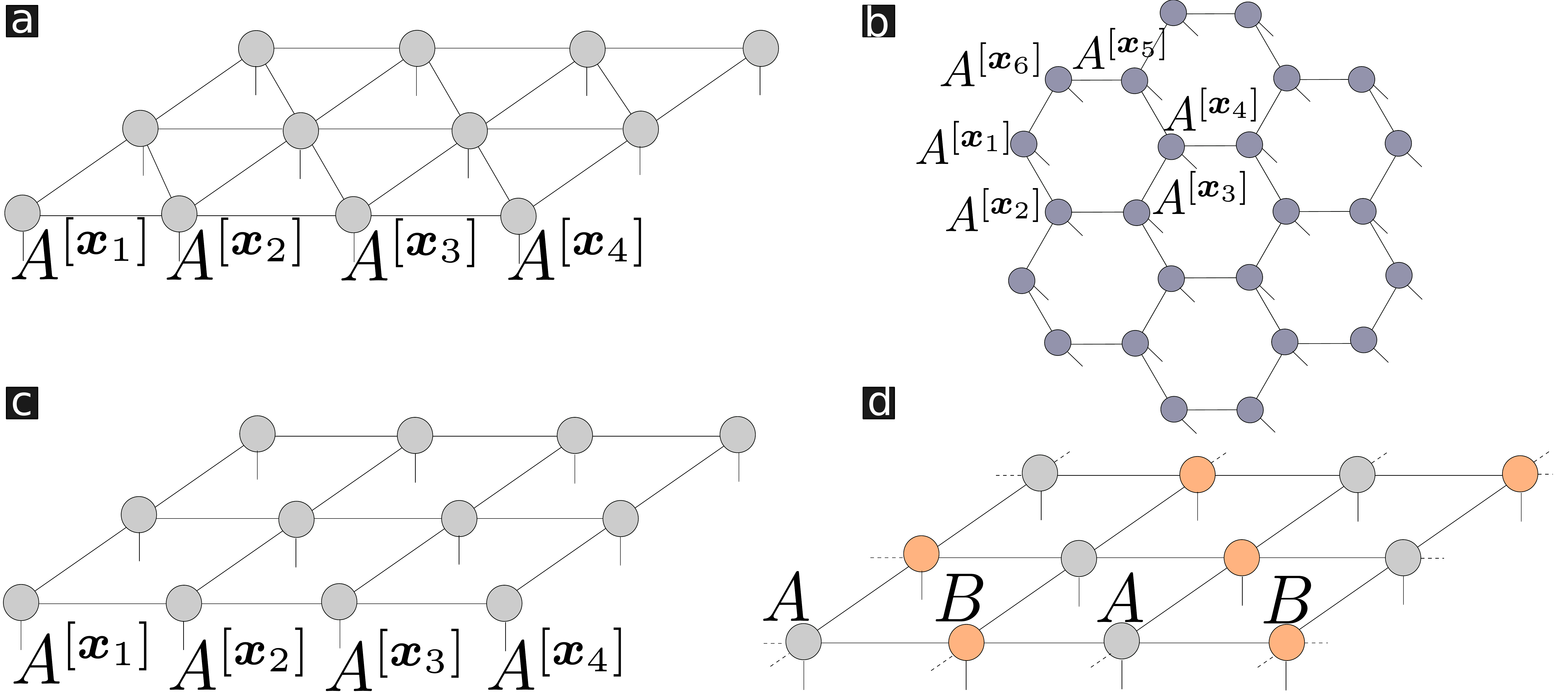}
\caption{{\bf PEPS representations:} Diagrammatic representations of PEPSs corresponding to different lattice patterns: {\bf (a)} a triangular lattice, {\bf (b)} a hexagonal lattice, {\bf (c)} a square lattice, and {\bf (d)} infinite square lattice PEPS with a two-site unit cell.
}
\label{fig:peps_sketch}
\end{figure}

\subsubsection{Projected entangled pair states (PEPS)}
Here, we briefly review the notation and fundamental properties of projected entangled pair states. Let us consider a two-dimensional  spatial lattice  
consisting of $N_1N_2$ sites, each of which hosts a quantum sub-system with a $d$-dimensional local Hilbert space $\mathbb{C}^d$, e.g. $d=2$ for a spin-$1/2$ lattice model. The full Hilbert space of the system is thus $\mathsf{H}= \mathbb{C}^d\otimes\stackrel{N_1N_2-2}{\cdots}\otimes\mathbb{C}^d$, such that   the dimension grows exponentially with the number of lattice sites, and the problem quickly becomes intractable, already for moderately low values of $N_1N_2$. In order to avoid this problem, we use a PEPS to represent a pure state.

\begin{figure*}[t]
\centering
\includegraphics[width=1.0\textwidth]{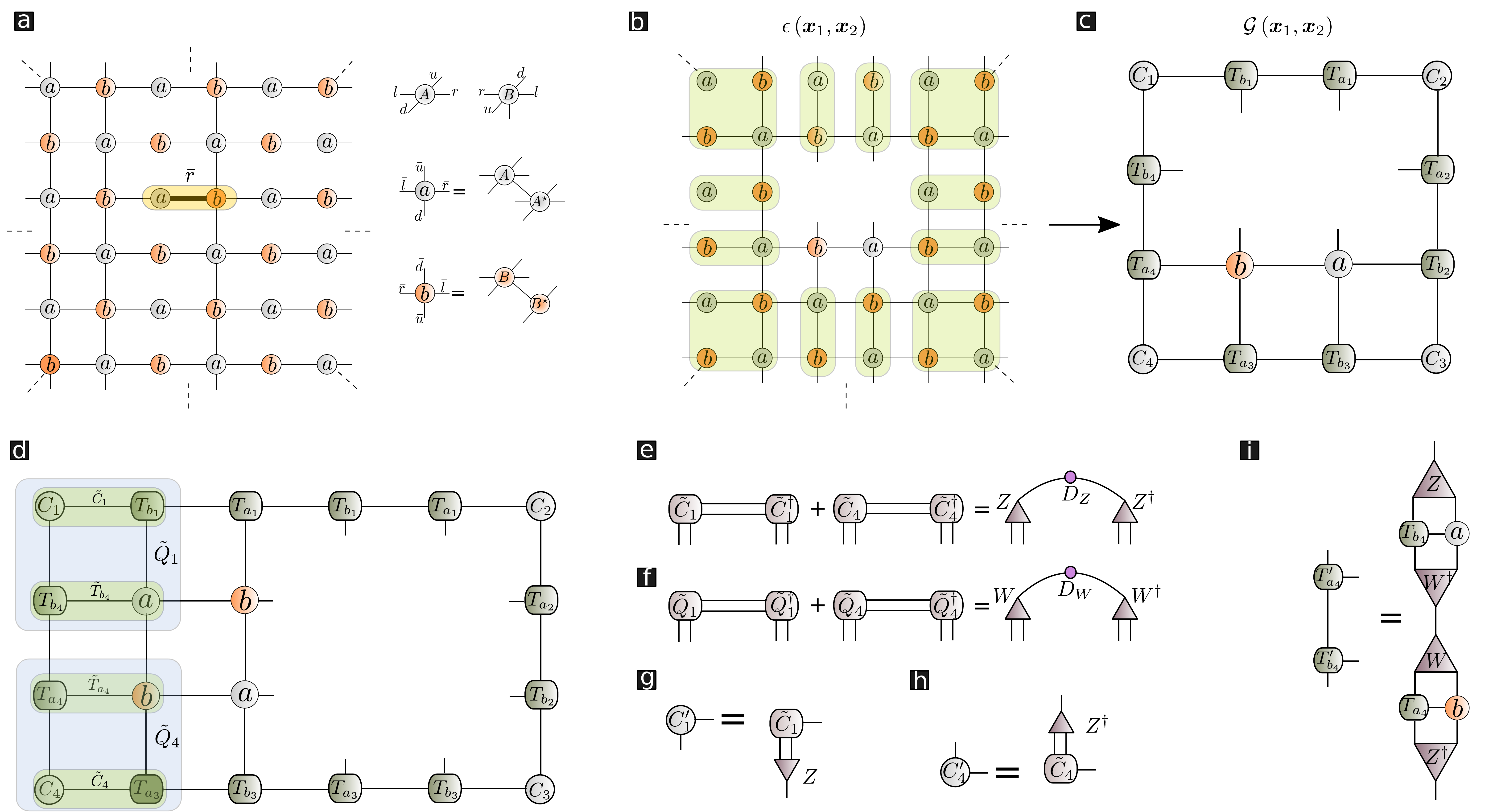}
\caption{{\bf Effective environment in the iPEPS algorithm:} {\bf (a)} Left: $2d$ lattice of tensors formed from $a$ and $b$; right: contractions to obtain tensors $a$ and $b$. {\bf (b)} Environment of the two-site unit cell for a specific $r$ link.  {\bf (c)} Effective environment of a given link on the lattice
(here an $r$ link). It is formed by twelve tensors.
{\bf (d)} two new columns are inserted,  absorbed towards the left, and renormalized individually. {\bf (e)} Eigenvalue decomposition for the sum of the
squares of CTMs. {\bf (f)} Eigenvalue decomposition for the sum of the
squares of $Q$ tensors. {\bf (g)}- {\bf (h)} The CTMs $\tilde{C}_1$, and $\tilde{C}_4$  are renormalized with isometry $Z$. {\bf (i)} 
Two isometries $Z$ and $W$ are used to obtain the renormalized half-row transfer
matrices $T^{\prime}_{a_4}$ and $T^{\prime}_{b_4}$.}
\label{fig:peps_ctmrg_sketch}
\end{figure*} 

The PEPS describes a state through interconnected tensors. As in Eq.~\eqref{eq:spins}, we use an integer $m\in\{1,\cdots,N_1N_2\}$ to label the lattice sites ordered from east to west and south to north $\boldsymbol{x}_m=(x_m,y_m)$, and define  the PEPS variational ansatz as
\beq \label{eq:def_peps}
|\psi \rangle = \!\!\!\sum_{ \left\lbrace s_{\boldsymbol{x}_m} \right\rbrace}\!\!\! F\!\!\left(\! A^{[\boldsymbol{x}_1]}_{s_{\boldsymbol{x}_1}},A^{[\boldsymbol{x}_2]}_{s_{\boldsymbol{x}_2}}, \cdots, A^{[\boldsymbol{x}_{N_1\!N_2}]}_{s_{\boldsymbol{x}_{N_1\!N_2}}} \right)
|s_{\boldsymbol{x}_1},s_{\boldsymbol{x}_2} \cdots, s_{\boldsymbol{x}_{N_1\!N_2}} \rangle ,
\eeq
where we sum over all possible states  in the basis of the local Hilbert space, labelled by $\{s_{\boldsymbol{x}_m} \}$. This PEPS is represented by a network of  $N_1N_2$ tensors $A^{[\boldsymbol{x}_m]}_{s_{\boldsymbol{x}_n}}$, some of which are 
connected  according to the geometry of the lattice and the notion of neighboring lattice sites.
Each tensor of the PEPS has $n$ so-called bond indices of dimension $D$, which describe the aforementioned auxiliary degrees of freedom, and a single physical 
index of dimension $d$. The choice of $n$ in the tensor network depends on the geometry of the lattice and can be chosen arbitrary.
The function $F$ contracts all the tensors $A^{[\boldsymbol{x}_m]}_{s_{\boldsymbol{x}_m}}$, according to this pattern, and then performs the trace to obtain a scalar quantity such that Eq.~\eqref{eq:def_peps} can be understood as a parametrization of a  particular  set of states in the exponentially-large physical Hilbert space $\mathsf{H}$. In Fig. \ref{fig:peps_sketch}, we show diagrammatically several PEPSs for systems corresponding to different geometries with open boundary conditions (OBC). The spheres represent the tensors, with solid lines depicting the auxiliary and physical indexes. The vertical ones stand for the physical indexes, whereas the planar ones that connect neighbouring tensors stand for the contracted auxiliary indexes.  In the case of the system geometry being a square lattice pattern with OBC, the PEPS consists of tensors in the bulk $A^{[\boldsymbol{x}_m]}$, which have four bond indices connecting neighboring tensors, and one physical index $s_m$ (Fig. \ref{fig:peps_sketch} {\bf (c)}). Overall, the PEPS depends on $O(N_1N_2 D^4 d)$ variational parameters.

If we assume a translationally-invariant state with bond dimension $D=1$ and spin-$1/2$ physical states $d=2$, we are left with only two variational parameters, corresponding to our previous mean-field ansatz. Accordingly, for any $D>1$, the PEPS captures inter-particle correlations and improves upon the mean-field variational family. The accuracy of the ansatz can be systematically controlled by the bond dimension $D$ of the auxiliary indices. This parameter is related to the
maximum entanglement content that can be handled by the simulations \cite{verstraete2004renormalization}. In practice, increasing the value of $D$ leads to a better description of the groundstate and, therefore, to more accurate estimations of the different observable quantities. It turns out that the PEPS parametrization can describe very well the entanglement structure of many
interesting two-dimensional quantum systems, including low-energy eigenstates of gapped Hamiltonians with local interactions. More specifically, PEPS satisfy the entanglement area law and the scaling of entanglement entropy of an $L \times L$ block within the larger $N_1\times N_2$ lattice of a PEPS scales with  $O(L \log D)$ .

This construction can be generalized to any lattice shape and dimension and one can show that any state can be written as a PEPS if we allow the bond dimension to become very large. In the thermodynamic limit, an efficient variational tensor network ansatz is the iPEPS. It consists of a rectangular unit cell of tensors with one tensor per lattice site, $A^{[x,y]}$ where $[x,y]$ label the coordinates of a tensor relative to the unit cell of size $L_x \times L_y = N_T$ , as shown in Fig \ref{fig:peps_sketch} {\bf (d)} for a 4-site  cell.

\subsubsection{Infinite PEPS ansatz and optimal update}
In order to get an approximate representation of the groundstate of a given
Hamiltonian $H$, the tensors of the PEPS in Eq. (\ref{eq:def_peps}) need to be optimized. This is typically obtained by minimizing the expectation value 
of $\langle \psi | H | \psi \rangle$, or by simulating an evolution in
imaginary time $|\psi \rangle \simeq e^{-\beta H} |\psi_0\rangle$, where 
$|\psi_0\rangle$ is some initial state. 
In either case, the tensors that define the iPEPS are optimized iteratively. 

In this work, this optimization has been performed based on the imaginary-time evolution of the initial state. In particular, we use the full update introduced in reference~\cite{orus2009simulation,phien2015infinite}. 
Indeed, for a given Hamiltonian $H$, and for a given initial state described by an iPEPS with a unit cell composed by two tensors $A$ and $B$, the ground state of the system can be obtained by evolving an initial state $|\psi_0\rangle$ in imaginary time $\beta$ as 
\beq
|\psi_{GS} \rangle = \lim_{\beta \rightarrow \infty} \frac{e^{-\beta H} |\psi_0 \rangle}{|| e^{-\beta H} |\psi_0 \rangle ||}
\eeq
This imaginary-time evolution is achieved in practice by breaking the evolution operator $e^{-\beta H}$ into a sequence 
of two-body gates, using a Suzuki-Trotter decomposition expansion. 
To this end, the Hamiltonian $H$ is rewritten as
\beq 
H=H_l+H_r+H_u+H_d,
\eeq
where each term $H_i$ is the sum of commuting Hamiltonian terms for links 
labelled as (left $l$, right $r$, up $u$, and down $d$). Using the first order Suzuki-Trotter decomposition, the  time-evolution operator is split into 
\beq \label{eq:time_evolution_PEPS}
e^{-\beta H }=\left( e^{-H \delta} \right)^M \simeq \left( e^{-H_l \delta} e^{-H_r \delta} e^{-H_u \delta} e^{-H_d \delta} \right)^m ,
\eeq
where we have divided the total evolution time into $M=\beta/\delta$ steps, where
$\delta$ represents the infinitesimal imaginary-time step. Each term of the Hamiltonian  $H_i$ is a sum of commuting terms, so that, we can rewrite $e^{H_i \delta}$ in 
Eq. (\ref{eq:time_evolution_PEPS}) as a product of two-site operators
\beq 
e^{H_i \delta}= \prod e^{-h^{[\boldsymbol{x},\boldsymbol{x}^{\prime}]} \delta} = \prod g^{[\boldsymbol{x},\boldsymbol{x}^{\prime}]}.
\eeq 
We start by addressing only the update of the tensors $A$ and
$B$ defining a $r$ link, after applying $e^{-\delta H_r}$ to $|\psi \rangle$, which can then be generalised to all of the remaining links. We thus assume that the gate $g$ is applied on just one of the $r$ links.  
After applying the gate, we obtain a new iPEPS  $| \psi_{A^{\prime}, B^{\prime}} \rangle$ which is characterized by a tensor $A$ and $B$ everywhere except for the two tensors connected by the link where the gate acted. The effect of the gate
$g$ is to increase the bond dimension of the iPEPS. For this reason, the iPEPS $|\psi_{A^{\prime}, B^{\prime}} \rangle$ needs to be approximated by a new one defined by
two approximated tensors $\tilde{A}$ and $\tilde{B}$, where these two tensors again have the same bond dimension $D$ fixed at the beginning of the algorithm. In particular, the new PEPS  is calculated by minimising the distance to $|\psi_{A^{\prime}, B^{\prime}} \rangle$
\beq 
\min_{\tilde{A},\tilde{B}} \parallel \psi_{A^{\prime}, B^{\prime}} \rangle- |\psi_{\tilde{A}, \tilde{B}} \rangle \parallel^2 =\min_{\tilde{A},\tilde{B}} d\left( \tilde{A},\tilde{B} \right),
\eeq
where we have introduced
\begin{multline}  \label{eq:distance_tensors}
d\left( \tilde{A},\tilde{B} \right)= \langle \psi_{A^{\prime}, B^{\prime}} |\psi_{A^{\prime}, B^{\prime}} \rangle + \langle \psi_{\tilde{A}, \tilde{B}} |\psi_{\tilde{A}, \tilde{B}} \rangle \\
-\langle \psi_{\tilde{A}, \tilde{B}} | \psi_{A^{\prime}, B^{\prime}} \rangle -\langle \psi_{A^{\prime}, B^{\prime}} |\psi_{\tilde{A}, \tilde{B}} \rangle.
\end{multline}
Therefore, in order to optimally update the iPEPS,  minimizing the distance $d\left( \tilde{A},\tilde{B} \right)$ in Eq.~(\ref{eq:distance_tensors}), we need to {\it (i)} compute the environment for that specific $r$ link following the corner transfer matrix (CTM) sheme, which was originally derived by Baxter \cite{baxter1978variational,baxter2016exactly}. This  leads to an approximate representation of the environment in terms of corner matrices and transfer tensors~\cite{orus2009simulation}. After this step, we need to 
{\it (ii)} determine the optimal new tensors $\tilde{A}$ and $\tilde{B}$ for the link,
using the optimization techniques proposed in \cite{jordan2008classical,phien2015infinite}, which are referred to the full update scheme.
This represents a clean and accurate protocol for performing the tensor update during the imaginary time evolution in which the effect of the entire wave function on the bond tensors is considered including the environmental tensor network. Let us now discuss some further important details of the algorithm.

\begin{figure*}[t]
\centering
\includegraphics[width=1.0\textwidth]{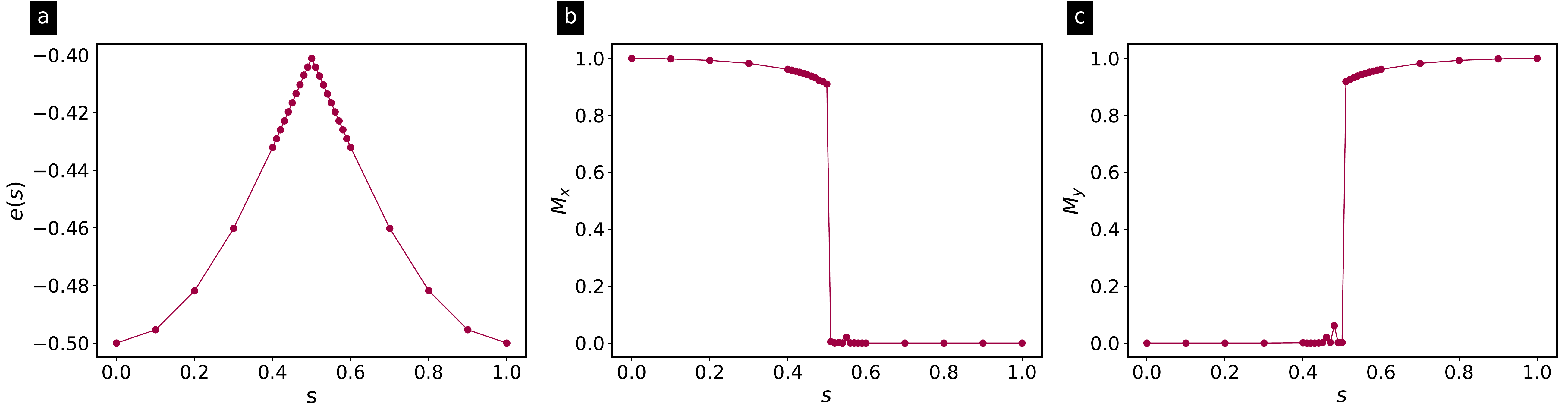} \\
\caption{{\bf First-order phase transition in the zero-field 90$^{\rm o}$ compass model:} {\bf (a)} Energy of the groundstate per link $e(s)$  on an infinite square lattice obtained by using the iPEPS algorithm
with $D=2$. The energy $e(s)$ has a sharp kink with a discontinuous first derivative, signalling the first-order nature of the
transition. {\bf (b)-(c)} Expectation values of the local order parameters  $M_x=\langle \tau^x\rangle$ and $M_y=\langle \tau^y\rangle $ for the ground state $| \psi_{GS}(s) \rangle$, which clearly display a discontinuity at the self-dual point $J_x=J_y$ for $s=1/2$, marking the  occurrence of a first-order phase transition.} 
\label{fig:first_order_pt_peps}
\end{figure*} 

\textit{(i) Environment approximation:} To compute the environment approximation, we contract the PEPS using the corner transfer matrix renormalization group (CTMRG), as suggested in \cite{nishino1996corner}. To explain this method, let $a$ and $b$ be the reduced tensors
\beq
a=\sum_{s=1}^d A_s \otimes A^{\star}_s,
\eeq
\beq
b=\sum_{s=1}^d B_s \otimes B^{\star}_s,
\eeq
with double bond indices such as $\bar{u}=(u,u^{\prime})$. 
Then, the scalar product $\langle \psi | \psi \rangle$ can be expressed as a two-dimensional network $\epsilon (\boldsymbol{x}_1,\boldsymbol{x}_2)$ made of infinitely many copies of $a$ and $b$ (see Fig. \ref{fig:peps_ctmrg_sketch} {\bf (a)}). 
The exact environment of sites $\boldsymbol{x}_1$ and $\boldsymbol{x}_2$ is obtained for $\epsilon \left(\boldsymbol{x}_1,\boldsymbol{x}_2\right)$ by removing the tensors $a$ and $b$, as shown in Fig. \ref{fig:peps_ctmrg_sketch} {\bf (b)}. The goal of the CTM method is to compute an approximation $\mathcal{G} \left(\boldsymbol{x}_1,\boldsymbol{x}_2\right)$ to $\epsilon \left(\boldsymbol{x}_1,\boldsymbol{x}_2\right)$ by finding the fixed point of the four CTMs.
This effective environment is given in terms of a set of four $\chi \times \chi$ } corner transfer matrices $\left \lbrace C_1, C_2, C_3, C_4 \right \rbrace$, eight half transfer row/column tensors
$\left \lbrace T_{a_1},T_{a_2},T_{a_3},T_{a_4},T_{b_1},T_{b_2},T_{b_3},T_{b_4} \right \rbrace$ and the two tensors $a$ and $b$ (see Fig.~\ref{fig:peps_ctmrg_sketch} {\bf (c)}). 
The twelve tensors of $\mathcal{G} \left(\boldsymbol{x}_1,\boldsymbol{x}_2\right)$ are updated according to four directional coarse-graining moves, namely left, right, up and down moves, which are iterated until the environment converges. Given an effective 
environment $\mathcal{G} \left(\boldsymbol{x}_1,\boldsymbol{x}_2\right)$ a move, e.g. to left, 
consists to the following three main steps:
\begin{itemize}
    \item \textbf{Insertion:} Insert a new column made of tensors $T_{b_1}$, $a$, $b$ and $T_{a_3}$ (see Fig. \ref{fig:peps_ctmrg_sketch} {\bf (d)}).
    \item \textbf{Absorption:} Contract tensor $C_1$ and $T_{b_1}$, tensor $C_4$ and $T_{a_3}$,  tensor $T_{b_4}$ and $a$, and also the tensor $T_{a_4}$ and $b$, resulting in two new CTMs $\tilde{C}_1$ and $\tilde{C}_2$ and two new
    transfer tensors $\tilde{T}_{a_4}$ and $\tilde{T}_{b_4}$, represented by shaded green areas in Fig. \ref{fig:peps_ctmrg_sketch} {\bf (d)}). 
    \item \textbf{Renormalization:} The renormalization step requires introducing two isometries $Z$ and $W$. This produces renormalized CTM's 
    $C^{\prime}_1=Z^{\dagger} \tilde{C}_1$, $C^{\prime}_4=Z^{\dagger} \tilde{C}_4$ and  half-raw transfer matrices $T^{\prime}_{a_4}$ and $T^{\prime}_{b_4}$ (see Figs. \ref{fig:peps_ctmrg_sketch} {\bf (g)-(i)}).   
\end{itemize}
A proper choice of the isometries $Z$ in the renormalization step is of great importance. Here, we consider instead the eigenvalue decomposition of 
$\tilde{C}_1 \tilde{C}^{\dagger}_1+\tilde{C}_1 \tilde{C}^{\dagger}_1=\tilde{Z} D_z \tilde{Z}^{\dagger}$ (see first equation of Fig. \ref{fig:peps_ctmrg_sketch} {\bf (e)}), and use the isometry $Z$ that results from keeping to entries of $\tilde{Z}$ corresponding to the $\chi$  largest eigenvalues of $D_z$.
Instead, for the isometry $W$, we decompose $\tilde{Q}_1+\tilde{Q}^{\dagger}_1+\tilde{Q}_4+\tilde{Q}^{\dagger}_4=\tilde{W} D_w \tilde{W}^{\dagger}$ (see second equation of Fig. \ref{fig:peps_ctmrg_sketch} {\bf (f)}). The numerical cost of implementing these steps scales with $D$ and 
$\chi$ as $O(D^6 \chi^3)$. The net result is a new effective environment $\mathcal{G}^{\prime}$ for sites $\boldsymbol{x}_1$ and $\boldsymbol{x}_2$. By composing the four moves of the directional CTMs, we can recover one iteration of the CTMRG.

\textit{(ii) Tensor optimization:} To find the optimal tensors $\tilde{A}$ and
$\tilde{B}$ after a single step of imaginary-time evolution, we have to minimize the distance in Eq. (\ref{eq:distance_tensors}). We proceed as follows:

\begin{itemize}
    \item We fix tensor $\tilde{B}$ to some initial tensor or, alternatively, to the tensor
obtained from the previous iteration. In order to find $\tilde{A}$, we rewrite Eq. (\ref{eq:distance_tensors}) as a quadratic scalar expression for the tensor, namely
    \beq
       d(\tilde{A},\tilde{A}^{\dagger}) = \tilde{A}^{\dagger} R \tilde{A} - \tilde{A}^{\dagger} S - S^{\dagger} \tilde{A} + T
    \eeq
where $R$, $S$, and $T$ can be obtained from the appropriate tensor contractions
including the effective environment around the $r$-link. 
    \item We then find the minimum of $d(\tilde{A},\tilde{A}^{\dagger})$ with respect to
    $\tilde{A}^{\dagger}$, which is given by $\tilde{A}=R^{-1} S$.
    \item Next, we fix the tensor $\tilde{A}$, and search for an optimal $\tilde{B}$ using the corresponding procedure described in the two previous steps.
\end{itemize}
The above steps are iterated until the cost function $d\left( \tilde{A},\tilde{B} \right)$ converges to a sufficiently small value, which can be set to a specific desired value in the algorithm.
Once the optimal tensors are found, these are replaced over the entire  lattice considering the four-site unit cell layout, which thus approximates simultaneously  the effect of all the gates $g$ acting over the infinitely-many links of the same type, here $r$ links. Such a procedure defines the updated infinite PEPS  $|\psi_{\tilde{A},\tilde{B}} \rangle$ in terms of the two new tensors.
Finally, the same procedure is repeated for the $l-$, $u-$ and
$d-$links to complete one full  step of imaginary-time evolution. These steps can then be concatenated until we reach the stationary state that approximates the groundstate within the PEPS family with a specific  accuracy.

\begin{figure}[t]
\centering
\includegraphics[width=0.5\textwidth]{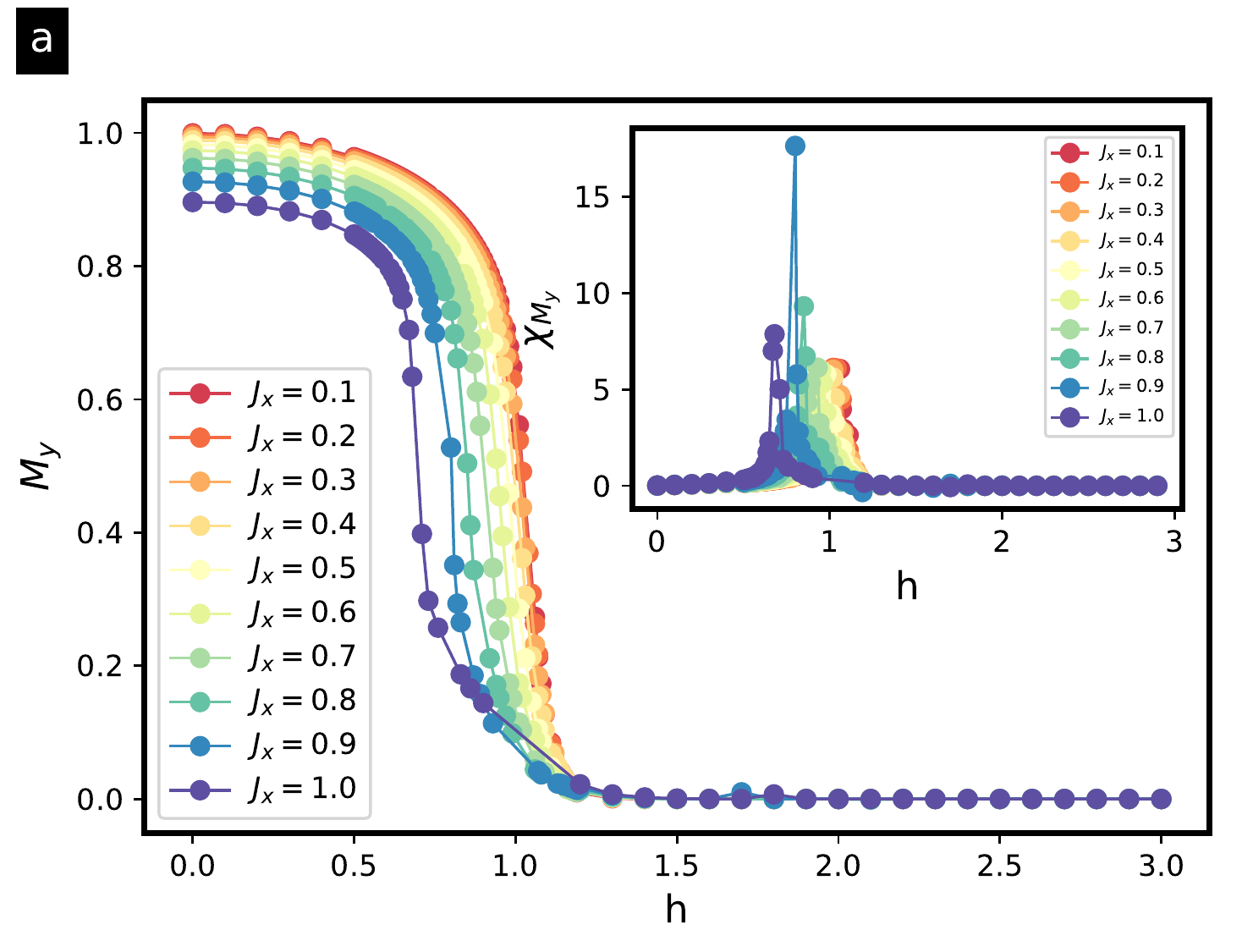} \\
\includegraphics[width=0.5\textwidth]{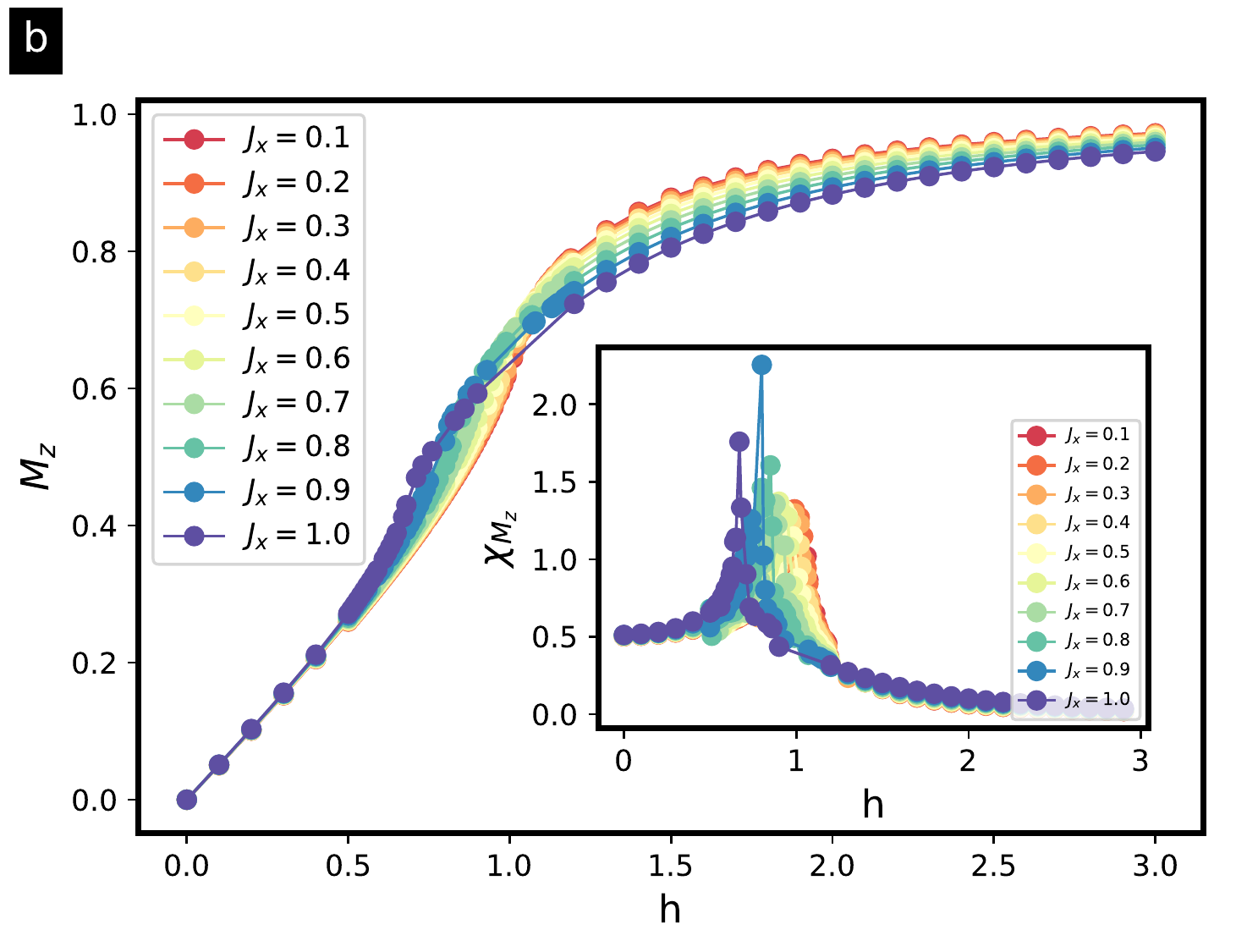}
\caption{{\bf Magnetizations and magnetic susceptibilities:} {\bf (a)} Ferromagnetic magnetization along the $y$ direction for fixed coupling strength $J_y=1$, and for different couplings $J_x$ and magnetic fields $h$. The system develops a non-zero expectation value for transverse fields below a critical value $h<h_{\rm c}$. In the inset, we show the magnetic susceptibility $\chi_{M_x}$ for the same parameters, which develops  peaks at those  critical points. {\bf (b)} Paramagnetic magnetization along the $z$ direction. In the inset, we show the magnetic susceptibility $\chi_{M_z}$, which shows peaks that coincide with those of {\bf (a)}. }
\label{fig:peps_mx_mz}
\end{figure} 

\subsubsection{Quantum compass phase transitions}
In this section, we show the results obtained by using the above  iPEPS algorithm  to compute the groundstate $\ket{\Psi_{GS}}$ of the 90$^{\rm o}$ compass model $H_{\rm eff}$~\eqref{eq:compass} in a non-vanishing transverse field, working directly in the infinite-lattice limit. 
In particular, we have computed the ground state wave function $|\psi_{GS} \rangle$ of the system by performing the imaginary-time evolution  for different values of the spin couplings $J_x$, $J_y$ and the transverse magnetic field $h$, and then  evaluated observable quantities on it, such as the groundstate energy and the local order parameters related to the aforementioned ferromagnetic phases.

In the following, we benchmark our PEPS numerical routine by fixing the bond dimension to $D=2$ and setting $h=0$. As shown in~\cite{orus2009first} for the zero-field limit $h =0$, iPEPS with $D = 2$ already yields better results than those obtained by combining fermionization with mean-field theory~\cite{chen2007quantum}.
As discussed in this reference, the values of the groundstate energy per bond agree with those obtained through a rough extrapolation to the thermodynamic limit of exact diagonalization and Green’s function Monte Carlo results for finite systems presented in \cite{dorier2005quantum}. In the limit $h=0$, as one
tunes the couplings across the symmetric  self-dual point $J_x=J_y$, a first-order phase transition between two gapped ferromagnetic orders occurs, i.e. $M_x=\langle \tau_x \rangle \neq 0$, when $J_x > J_y$ and $M_y=\langle \tau_y \rangle \neq 0$ when $J_y > J_x$. To study this  phase transition, we will consider adimensional couplings, and  restrict  them to the range $J_x, J_y \in [0,1]$ by the parametrization 
$J_x=\cos(s \pi/2)$ and $J_y=\sin(s \pi/2)$ with $s\in [0,1]$. In Fig. \ref{fig:first_order_pt_peps} {\bf (a)}, we show the ground state energy per 
lattice link $e(s)=\bra{\Psi_{GS}}H_{\rm eff}\ket{\Psi_{GS}}/2N_1N_2$. Our results show the presence of a sharp kink at $s=1/2$, which
is compatible with the existence of a first order phase transition at $J_x=J_y$, and agrees with the numerical results presented in~\cite{orus2009first}. 
Other indicators for this phase transition are the magnetization $M_x=\bra{\Psi_{GS}} \tau^x \ket{\Psi_{GS}}$ and $M_y=\bra{\Psi_{GS}} \tau^y\ket{\Psi_{GS}}$ displayed in  Fig. \ref{fig:first_order_pt_peps} {\bf (b)-(c)}, which show a clear  discontinuity at $s=1/2$ that separates the ${\rm FM}_x$ and ${\rm FM}_y$ phases that where identified with our mean-field ansatz. We interpret this fact as 
conclusive evidence of the existence  of a first order phase transition, which agree with the results presented in~\cite{orus2009first}, and thus gives compelling evidence for the validity of our iPEPS routine. Having benchmarked this limit, let us now switch on the transverse magnetic field to a non-zero value $h>0$, which favours a paramagnetic phase PM with all spins pointing along the $z$-axis.

In contrast to the $h=0$ limit, the transverse-field compass model has not been studied so thoroughly. In the previous section, we used a mean-field ansatz to predict a critical line~(\ref{criticalcompass}) separating  the symmetry-broken orbital ferromagnets from this paramagnet via second-order phase transitions. These  critical lines are represented by  green lines in Fig. \ref{fig:ferrophases}. In order to test the validity of these mean-field predictions, we use our iPEPS algorithm for $D=2$ with the hope that, as occurs for the $h=0$ limit, it will also give a quantitatively-better account of this second-order phase transition than the previous mean-field ansatz. By measuring the paramagnetic and ferromagnetic
magnetizations, we confirm that these quantities can be used to identify the critical points also for a non-zero magnetic field $h>0$. 
In Fig. \ref{fig:peps_mx_mz} {\bf (a)}, we present the magnetization $M_y=\langle \tau^y  \rangle$ as a  function of the transverse  magnetic field $h$, setting $J_y=1$ and exploring different values of 
$J_x<1$. This figure shows that, for weak transverse fields, the  magnetization attains a non-zero value that signals a symmetry-broken FM$_y$. In the inset, we show that the corresponding  magnetic susceptibility $\chi_{M_y}=\partial M_y / \partial h$ peaks at a specific value of the transverse field, which can be used to locate the corresponding critical points. This flow of the FM$_y$-PM critical point $h/J_y|_c$ as a function of $J_x$ is to be expected,    as the $J_x\to 0$ limit corresponds to a set of decoupled rows, each of which chain with a well-known critical point $h/J_y|_c=1$~\cite{PFEUTY197079}. We find that, as $J_x\to J_y=1$, the critical point shifts towards $h/J_y|_c=0.7$. In Fig. \ref{fig:peps_mx_mz} {\bf (b)}, we depict the transverse magnetization $M_z=\langle \tau^z  \rangle$, which is not an order parameter of the model, and can display a non-zero value for arbitrary value of the couplings, saturating at  $M_z=1$ when all spins are perfectly aligned with the $z$ axis for $h\gg 1$. In the inset, we represent the   susceptibility $\chi_{M_z}=\partial M_z / \partial h$, which  peaks at the transition points, in analogy to the transverse-field Ising chain.

Let us now discuss how these results can be be applied to recover the iPEPS analogue of the strong-coupling mean-field critical lines in Eq.~\eqref{criticalcompass}. Note that, according to Eq.~\eqref{eq:spin_couplings}, the ratio of the spin couplings is set by the lattice anisotropy  $J_x/J_y=(a_1/a_2)^2=\xi_2^2$. In Fig.~\ref{fig:chern}, we consider the planes in parameter space for $\xi_2=1$, and  $\xi_2=1/2$, such that $J_x/J_y=1$, and $J_x/J_y=1/4$ respectively. For this particular ratio, our iPEPS algorithm for $D=2$ estimates the critical points at  $h/J_y|c\approx 0.7$ and $h/J_y|c\approx 1.07$, respectively. Note that there is a large deviation in both cases with respect to the corresponding mean-field prediction $h/J_y|_c=2$.  Using the expressions for the transverse field and spin coupling strength in terms of the microscopic parameters of the four-Fermi-Wilson model~\eqref{eq:spin_couplings}, we obtain a pair of critical lines that  are represented by the red lines in  Fig.~\ref{phasediagram} {\bf (a)-(b)} below. As can be seen from these results, although the analytical mean-field predictions~(\ref{criticalcompass}) capture the correct parametric dependence of the critical lines,  the iPEPS predicts a smaller region for the symmetry-broken FM$_y$ phase. Given the fact that iPEPS treats correlations more accurately, we believe that these iPEPS result capture the correct trend and, although better estimates will be achieved by increasing the bond dimension beyond $D=2$, the extend of the FM$_y$ will in any case be smaller than that predicted by the mean-field ansatz. Moreover, we have checked that, for $h \neq 0$, iPEPS also provides significantly-lower variational energies than the one obtained by the separable-state mean-field ansatz, which typically under-estimates the effect of the transverse field. Future studies may take these results as a starting point, and study  the specific scaling relations, shedding light into the universality class of the 90$^{\rm o}$ compass model in a transverse field.  

\section{\bf Auxiliary-field gap equations at large $N$} \label{gapequations}

In the following two sections, we  use an alternative tool  to characterize   the phase diagram of  the four-Fermi-Wilson model~\eqref{eq:total_H}. This technique will allow us to connect our discussion of the $N$-flavoured Chern insulators and the trivial band insulators of Sec.~\ref{sec: chern} with the  ferromagnetic FM$_x$, FM$_y$ and paramagnetic PM phases  discussed in Secs.~\ref{sec:var_mf} and~\ref{sec:var_tn}.  As described below, the number of flavours $N$ will play a key role in this endeavour. We note that the inclusion of $N$  flavours  permits building QFTs with  internal symmetries, as for example the $U(N)$  global symmetry of our four-Fermi QFT~\eqref{continuum}. One may expect that including more flavours  would lead to  further complexity and modify  the properties of the QFT, as the number of coupled degrees of freedom increases with $N$. What can be surprising at first sight is that  the limit of a very large flavour number $N\to\infty$  can actually turn out to simplify the theory and allow to explore non-perturbative effects in a controlled and well-defined framework.   

A paradigm in this regard  is the scalar $O(N)$ model, which generalizes the $\lambda_0\phi^4$ field theory to $N$ flavours  interacting through a rotationally-invariant quartic term $\frac{\lambda_0}{N}\left(\sum_{\mathsf{f}}\phi_{\mathsf{f}}\phi_{\rm f}\right)^2$. Here, the increased number of flavours   $N>1$ changes the physics substantially,  as the  breakdown of the $O(N)$ symmetry   is accompanied by  the appearance of $N-1$ massless excitations  above the symmetry-broken groundstate, the Goldstone bosons~\cite{PhysRev.127.965}. In the symmetric phase, the  flavour number  also changes  crucial aspects of this QFT, such as the renormalization-group (RG) fixed points, which yield  scaling dimensions and critical exponents that generally depend on $N$~\cite{PhysRevD.7.2911,RevModPhys.45.589}. This dependence  is a consequence of the increased complexity: Feynman diagrams arising in  perturbative RG calculations, the so-called radiative corrections~\cite{PhysRevD.7.1888},  can have different contributions depending on how one decorates the graph  with different flavour indexes. In the simplest case, one finds loops where the internal flavour index is summed over yielding  a contribution that scales with $N$, but also other single-flavour loops with exactly the same graph structure  that do not scale with $N$. Although this increases the complexity {\it a priori\/}, since  different contributions cannot be  distinguished by the topological structure of the diagrams, it is also the key for the development of large-$N$ techniques where these QFTs actually become simpler~\cite{coleman_1985}. When this scaling is combined with that of the interaction vertex, which scales with $1/N$, one can group the Feynman diagrams by their order $\mathcal{O}(1/N^\alpha)$, and retain only the leading ones  for $\alpha=0$ in the limit $N\to\infty$, and calculate  corrections to this limit systematically  for $\alpha\in\{1,2,\cdots\}$. It is in this particular limit  where  the QFT becomes simpler than any finite-$N$ instance. Moreover, these large-$N$ methods allow one to go beyond ordinary perturbation theory in the coupling  $\lambda_0$, addressing strong-coupling phenomena such as the quantum-mechanical contributions to  spontaneous symmetry breaking~\cite{PhysRevD.10.2491,PhysRevD.9.3320}. 

For QFTs with four-Fermi terms, such as the Gross-Neveu model in (1+1) dimensions~\cite{PhysRevD.10.3235}, the situation gets  richer as, in addition to the  $U(N)$ flavour symmetry, one also has  chiral symmetry. Once again,  Feynman diagrams in general depend on $N$, which can modify a perturbative RG approach~\cite{PhysRevD.7.2911}. However, the fermionic case introduces 
further possibilities, as the breakdown of chiral symmetry can occur by the process of dynamical mass generation~\cite{PhysRev.122.345}. In 1+1 dimensions,  the vacuum of massless Dirac fermions is unstable towards a scalar condensate, which  forms at any non-zero coupling strength $g^2>0$. As a consequence,  the fermions acquire a mass that depends non-analytically on the coupling strength $g^2$, and cannot be thus captured to any finite order of perturbation theory~\cite{PhysRevD.10.3235}. In (2+1) dimensions, a  chiral-invariant~\eqref{eq:gammas_4} Gross-Neveu  model~\eqref{continuum} still displays such a chiral symmetry breaking which, in contrast, takes place at  a non-zero coupling strength $g^2\to g^2_{\rm c}$. It is in the vicinity of this strong-coupling point where  one  obtains a renormalizable large-$N$ QFT, allowing for estimates of the scaling dimensions and critical exponents that improve as one increases the order $\alpha$~\cite{HANDS199329,hep-lat/9706018}.

\begin{figure}[t]
	\centering
	\includegraphics[width=0.5\textwidth]{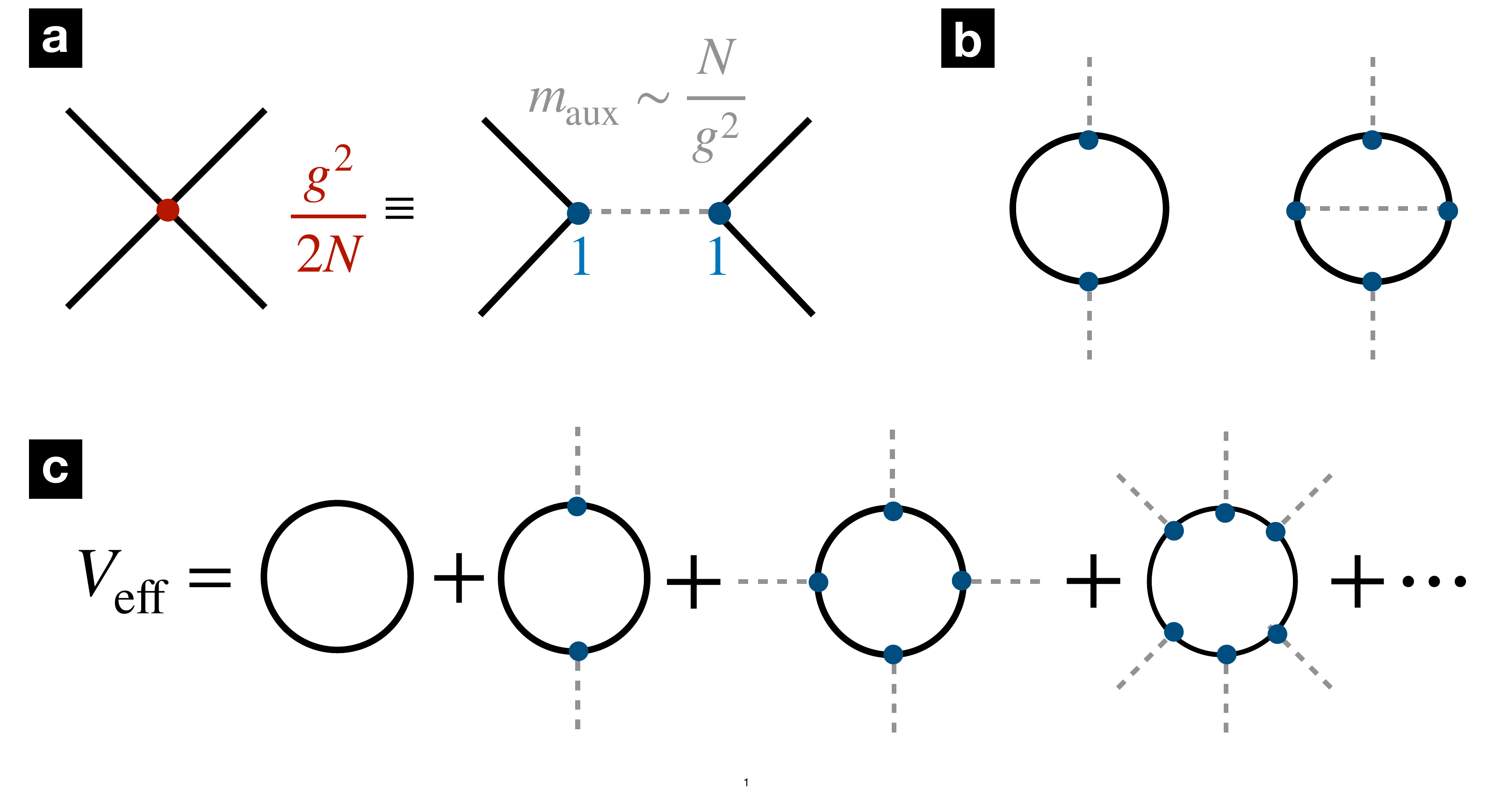}
	\caption{{\bf Auxiliary field and large-$N$ diagrams:} {\bf (a)} The four-Fermi vertex $g^2/2N$ (red dot) can be rewritten in terms of an auxiliary field of mass $m_{\rm aux}\sim N/g^2$ and no kinetic energy that mediates the interaction with new vertices of order $\mathcal{O}(N^0)$ (blue dots). {\bf (b)} 1-particle irreducible diagrams with more internal auxiliary lines, such as the second one, involve extra powers of the inverse auxiliary mass, and thus contribute with higher powers of $1/N$ that can be neglected for $N\to \infty$. {\bf (c)} The effective potential in the large-$N$ limit  can be obtained by resumming all the Feynman diagrams obtained by a single fermion loop, and an even number of external auxiliary lines.}
	\label{fig:large_N_FD}
\end{figure}

A convenient starting point for the large-$N$ analysis is the introduction of an auxiliary  field with zero kinetic energy and a mass that scales linearly with $N$. The  bare propagator, represented by a dotted line in Fig.~\ref{fig:large_N_FD}{\bf (a),} scales with $1/N$ and serves to mediate the  four-Fermi terms  via new interaction vertices that do not scale with $N$. Thanks to these auxiliary fields, the Feynman diagrams with a distinct $N$-scaling can be readily identified, as they are now endowed with a different topological structure. For instance,  the leading-order one-particle irreducible diagrams, which cannot be split in two disconnected pieces by cutting an internal line (see Fig.~\ref{fig:large_N_FD}{\bf (b)}), must minimize the number of internal  propagators of the auxiliary field, as each of these is suppressed by an additional power of  $1/N$, yielding sub-dominant terms in the large-$N$ limit. As discussed in more detail below, the leading-order radiative corrections are thus formed by a single fermion loop which,  due to  the algebraic properties of the gamma matrices,  can only be dressed by an even number of external auxiliary lines~\cite{coleman_1985} (see Fig.~\ref{fig:large_N_FD}{\bf (c)}).  Remarkably, the large-$N$ radiative corrections can be resummed   to all orders of  the coupling strength $g^2$, which allows  one  to address non-perturbative phenomena~\cite{coleman_1985}, such as dynamical mass generation, chiral symmetry breaking, and dimensional transmutation in $(1+1)$ dimensions~\cite{PhysRevD.10.3235}. 

We note that in the continuum QFT~\eqref{continuum}, it suffices to introduce a single auxiliary $\sigma$ field, which acquires a non-zero vacuum expectation value proportional to the scalar  condensate and contributes to the fermion mass. This auxiliary $\sigma$ field is an instance of the so-called Hubbard-Stratonovich fields~\cite{startonovich,PhysRevLett.3.77}, which allow for various mean-field approximations with applications in condensed matter~\cite{coleman_2015}. From this perspective, one typically needs  to explore a variety of  possible symmetry-breaking channels, which  requires introducing multiple  Hubbard-Stratonovich auxiliary fields. In the rest of this section, we use the results from the strong-coupling analysis to guide our choice of auxiliary fields, and show how a large-$N$ method can provide a detailed account of the phase diagram of  our four-Fermi-Wilson model~\eqref{eq:total_H} by functional-integral techniques. We obtain  a set of non-linear equations, the gap equations, the solution of which gives  access to the critical lines, as well as the symmetry-broken phases with non-zero order parameters. In the following Sec.~\ref{effectivepotential}, we shall exploit the above resummation of the leading Feynman diagrams to calculate the effective potential, which also gives us access to the symmetric regions, and in particular to a characterization of the correlated QAH phase, and the equivalence of the PM and TBI phases.

\subsection{Hamiltonian field theory and gap equations}
\label{sec:H_ft_gap_eqs}

Let us start by focusing on the canonical partition function $Z={\rm Tr}(\ee^{-\beta H})$  of the four-Fermi-Wilson Hamiltonian~\eqref{eq:total_H}, where $\beta=1/T$ is the inverse temperature  in natural units, and rewrite it as a functional integral  in Euclidean time $\tau \in [0,\beta)$~\cite{negele_orland_2019}. We use the over-complete  basis of fermionic coherent states  at each point of the Brillouin zone $\boldsymbol{k}\in \text{BZ}$,   and introduce a set of anti-commuting  Grassmann vectors $\Psi_{\boldsymbol{k}}(\tau)$, $\Psi^{\star}_{\boldsymbol{k}}(\tau)$, each of which contains  2 spinor and $N$ flavor components, and satisfies anti-periodic boundary conditions in the imaginary-time direction $\Psi_{\boldsymbol{k}}(\tau+\beta)=-\Psi_{\boldsymbol{k}}(\tau)$, $\Psi^{\star}_{\boldsymbol{k}}(\tau+\beta)=-\Psi^{\star}_{\boldsymbol{k}}(\tau)$. The partition function can be expressed as a functional integral over these Grassmann variables
\begin{equation}
Z=\int [{\rm d}\Psi^\star {\rm d}\Psi]\ee^{-S_E[\Psi^\star,\Psi]}, \label{partition}
\end{equation}
where we have introduced  the functional integral measure $[{\rm d}\Psi^\star {\rm d}\Psi]$, and the Euclidean action 
\begin{equation}
S_E\!=\!\! \int_{0}^{\beta}\!\!\!\!\!{\rm d}\tau\!\! \left(\!\sum_{\boldsymbol{k}\in \text{BZ}}\!\!\! \!\Psi^\star_{\boldsymbol{k}}(\tau)\!\big(\partial_{\tau}+h_{\boldsymbol{k}}(m)\big)\! \Psi_{\boldsymbol{k}}(\tau)+V_g[\Psi^\star,\Psi]\!\!\right)\!\!. \label{rawaction2}
\end{equation} 
Here, we have used the single-particle  Hamiltonian $h_{\boldsymbol{k}}(m)$ of Eq.~\eqref{freehamiltonian}, and introduced $V_g[\Psi^\star,\Psi]$ as the expectation value of the four-Fermi interaction~\eqref{interaction} in such a coherent-state basis. Let us now describe how to rewrite this interaction in terms of auxiliary fields, using the strong-coupling results as a guide.

First of all, we need to consider the auxiliary $\sigma$ field which, as discussed above, can attain a non-zero value proportional to the scalar  condensate in a continuum QFT
\beq
\label{eq:scalar_condensate}
\Sigma\propto\langle\overline{\Psi}(\boldsymbol{x})\Psi(\boldsymbol{x})\rangle.
\eeq
Considering the discretized model in the single-flavor $N=1$ limit,   and the bilayer perspective of Fig.~\ref{hoppings}, this scalar condensate is simply proportional to the magnetization of the orbital spins~\eqref{eq:orbital_spins} along the $z$-axis $\Sigma\propto\langle\tau^z_{\boldsymbol{n}}\rangle$. We know from the previous section that, for any non-zero transverse field $h\neq 0$, this magnetization is always non-vanishing, and there is thus no spontaneous symmetry breaking associated with its non-zero value. From the perspective of  relativistic LFTs, the fermion masses  introduced by the Wilson-type discretization~\eqref{wilson}-\eqref{eq:bare_mass} are responsible for the non-zero value of the $\sigma$ field, and the spontaneous condensation is only expected to be recovered in the vicinity of a critical point where a continuum QFT emerges. For the particular choice of gamma matrices~\eqref{eq:gammas_4},  this  QFT would be endowed with an emergent chiral symmetry, and the formation of the scalar condensate is related to chiral symmetry breaking via dynamical mass generation. On the other hand, for the current choice~\eqref{eq:gammas}, although chiral symmetry cannot be defined,  the emergent QFT will have important connections to the QAH effect and the nature of the topological phase transitions discussed in Sec.~\ref{sec: chern} as one enters the strong-coupling regime.

As discussed in the previous section, regardless of the absence of chiral symmetry, there are other possible symmetry-breaking channels that can be activated by increasing the interactions of the four-Fermi-Wilson model~\eqref{eq:total_H}. In the $N=1$ limit, this occurs due to the spontaneous breakdown of a $\mathbb{Z}_2$ symmetry, and the appearance of the ferromagnetic long-range orders, either along the $x$-axis $\langle\tau^x_{\boldsymbol{n}}\rangle$~\eqref{eq:x_mag}, or the  $y$-axis  $\langle\tau^y_{\boldsymbol{n}}\rangle$~\eqref{eq:y_mag}. In light of Eqs.~\eqref{eq:gammas} and~\eqref{eq:orbital_spins}, these order parameters can be readily generalized to the $N$-flavor case by introducing  two additional auxiliary $\pi$ fields $\Pi_1(\boldsymbol{x}),\Pi_2(\boldsymbol{x})$ and 
two possible  $\pi$-condensates 
\beq 
\label{eq:pi_condensates}
\Pi_1\propto \braket{\overline{\Psi}(\boldsymbol{x}) (\mathbb{I}_N\otimes\gamma^1) \Psi(\boldsymbol{x})},\hspace{1ex}  \Pi_2\propto \braket{\overline{\Psi}(\boldsymbol{x})(\mathbb{I}_N\otimes\gamma^2) \Psi(\boldsymbol{x})}.
\eeq
Once again, these condensates are not related to the spontaneous breakdown of the continuous internal  symmetry $U(N)$  but, instead, to  inversion symmetry on the discrete lattice 
\beq
\label{eq:inv_symm}
\Psi(\boldsymbol{x})\mapsto\big(\mathbb{I}_N\otimes\gamma^0\big)\Psi(-\boldsymbol{x}),\hspace{2ex}\forall\boldsymbol{x}\in\Lambda_s.
\eeq
Let us note that this transformation resembles the parity symmetry of Dirac fermions in even-dimensional spacetimes~\cite{Peskin:1995ev},  the breakdown of which may occur  via the  pseudo-scalar condensate $\Pi_5\propto \braket{\overline{\Psi}(\boldsymbol{x})(\ii \mathbb{I}_N\otimes\gamma^5) \Psi(\boldsymbol{x})}$~\cite{PhysRevD.30.2653,PhysRevD.58.074501}. In our case, however, Eq.~\eqref{eq:inv_symm}  corresponds to a planar rotation of angle $\theta=\pi$ and,  as discussed in Sec.~\ref{optical}, thus belongs to the Lorentz group of continuous transformations $SO(1,2)$. To define parity in odd-dimensional spacetimes,   one can search for a transformation that flips an odd number of spacetime axes, e.g.  $\Psi({x_1,x_2})\mapsto(\mathbb{I}_N\otimes\gamma^2)\Psi(x_1,-x_2)$ in our case, such that the $\Sigma$ and $\Pi_2$ condensates are parity-odd while $\Pi_1$ is parity-even,  which follows from the Clifford algebra fulfilled by the gamma matrices.  
 In contrast, using the inversion symmetry of Eq.~\eqref{eq:inv_symm},   the homogeneous scalar condensate~\eqref{eq:scalar_condensate} is invariant  $\Sigma\mapsto\Sigma$, whereas a non-zero value  of any of the $\pi$ condensates~\eqref{eq:pi_condensates} breaks it $\Pi_j\mapsto-\Pi_j$, such that the $\pi$ fields are treated on equal footing. 
 
 In comparison to the pseudo-scalar condensate that may arise in chiral-invariant lattice models~\cite{PhysRevD.30.2653,PhysRevD.58.074501},  we find that the number of possible symmetry-breaking channels is doubled when chiral symmetry is absent from the oustet~\eqref{eq:gammas}. Moreover, at exact isotropy, any combination of  the  condensates $\cos\phi\Pi_1+\sin\phi\Pi_2$ can be stabilised in the groundstate. Let us note that either of these $\pi$ condensates not only breaks inversion symmetry, but  also forbids the recovery of invariance   with respect to specific Lorentz boosts in the long-wavelength limit. We shall thus refer to them as Lorentz-breaking condensates. Accordingly, when approaching the critical point from the symmetry-broken phases to recover the continuum QFT,  the effective QFT would not be Lorentz-invariant unless we precisely hit the critical point. This would obviously change if we approach the critical point from the symmetry-preserving phase. 

Before proceeding with the large-$N$ approximation, let us discuss these auxiliary fields from the perspective of the Hubbard bilayer. Having $N$ flavours is equivalent to stacking $N$ Hubbard bilayers on top of each other, which only get coupled through the quartic Hubbard-type interactions. From the perspective of mean-field theory, the $\sigma$ field is related to the so-called Hartree decoupling of the interactions, which introduces terms that are proportional to the densities and is  thus  responsible for a shift of the bare mass in Eq.~\eqref{freehamiltonian_d_vector_aux}. On the other hand, the $\pi$ fields include the so-called Fock contributions, since they lead to terms that modify the inter-layer tunnelling, which is equivalent to the exchange spin-flip terms  induced by the $\gamma^1$ and $\gamma^2$ matrices in the language of the orbital spins. Therefore, our large-$N$ formalism is related to a self-consistent Hartree-Fock method in condensed-matter Hubbard-type models, and becomes exact in the limit of an infinite number of bilayers $N\to\infty$.

 Once we have identified the relevant auxiliary fields, we should apply a Hubbard-Stratonovich transformation to rewrite the action~\eqref{rawaction2} in terms of them. In Appendix~\ref{app:two_condensates} we show that, except for the isotropic limit $a_1=a_2$, the two $\pi$ fields cannot condense simultaneously. We can thus introduce these fields by two separate Hubbard-Stratonovich transformations, and compare the corresponding energies to determine which $\Pi_j$ condensate corresponds to the groundstate in the event of a spontaneous breakdown of inversion symmetry. We thus  consider the transformations of the four-Fermi term 
\begin{align}
(\overline{\Psi}(\boldsymbol{x})\Psi(\boldsymbol{x}))^2\to
\frac{1}{2}\big((\overline{\Psi}(\boldsymbol{x})\gamma^1\Psi(\boldsymbol{x}))^2+(\overline{\Psi}(\boldsymbol{x})\Psi(\boldsymbol{x}))^2\big),\\
(\overline{\Psi}(\boldsymbol{x})\Psi(\boldsymbol{x}))^2\to
\frac{1}{2}\big((\overline{\Psi}(\boldsymbol{x})\gamma^2\Psi(\boldsymbol{x}))^2+(\overline{\Psi}(\boldsymbol{x})\Psi(\boldsymbol{x}))^2\big),
\label{single2}
\end{align}
  both of which are  exact identities  in the single-flavour ${N=1}$ limit. In the following calculations, we assume that the lattice translational invariance is preserved in the event of condensation, and thus consider that the auxiliary  fields are homogeneous, i.e. ${\Sigma(\boldsymbol{x})=\Sigma}$, $\Pi_j(\boldsymbol{x})=\Pi_j$. 
  
  Tied to the condition that, after integrating over these auxiliary fields, the original action $(\ref{rawaction2})$ must be recovered with the corresponding four-Fermi term~\eqref{single2},  the transformed partition function  $Z=\int  [{\rm d}\Psi {\rm d}\Psi^\star {\rm d}\Sigma {\rm d}\Pi_j  \exp(-S_E[\Psi^\star,\Psi,\Sigma,{\Pi}_j])$ leads to an action that depends on the auxiliary fields
\beq
\label{eq:free_action_aux}
\begin{split}
S_E\!= \!\!\!\int_{0}^{\beta}\!\!\!\!{\rm d}\tau \bigg( &\frac{NA_{s}}{g^2}\left(\Sigma^2+{\Pi}_j^2 \right)\\+&\sum_{\boldsymbol{k}\in {\rm BZ}}\!\!\! \Psi^\star_{\boldsymbol{k}}(\tau)\Big(\partial_\tau+{h}_{\boldsymbol{k}}(m+\Sigma,{\Pi}_j)\Big)\Psi_{\boldsymbol{k}}(\tau) \!\bigg),
\end{split}
\eeq
where  the  single-particle Hamiltonian~\eqref{freehamiltonian} gets modified to \beq{h}_{\boldsymbol{k}}(m)\to{h}_{\boldsymbol{k}}(m+\Sigma,{\Pi}_j)=\mathbb{I}_N\otimes\big( {\boldsymbol{d}}_{\boldsymbol{k}}(m+\Sigma,{\Pi}_j)\cdot  \boldsymbol{\sigma}\big).
\eeq
Here,   the vector~\eqref{freehamiltonian_d_vector} whose winding is related to the Chern number  of the QAH phase~\eqref{eq:chern_winding}, also gets modified $\boldsymbol{d}_{\boldsymbol{k}}(m)\to\boldsymbol{d}_{\boldsymbol{k}}(m+\Sigma,{\Pi}_j)$ due to the presence of the auxiliary  field 
\beq
\boldsymbol{d}_{\boldsymbol{k}}(m+\Sigma,{\Pi}_j)=\boldsymbol{d}_{\boldsymbol{k}}(m)+\Pi_j\boldsymbol{e}_j+\Sigma\boldsymbol{e}_3,
\label{freehamiltonian_d_vector_aux}
\eeq
where we remark that the two cases $j=1,2$ are considered separately since, as shown in Appendix~\ref{app:two_condensates}, the $\pi$ fields do not condense simultaneously for generic anisotropies. 

Since we are interested in quantum phase transitions, we consider the zero-temperature limit in which $\tau\in[0,\infty)$. By using a Fourier transform in  imaginary time, we  introduce the so-called Matsubara frequencies $\omega_{n_0}=\frac{(2n_0+1)\pi}{\beta}$ with $n_0\in\mathbb{Z}$, which become continuous variables in this limit,  spanning  the range $\omega \in (-\infty,\infty)$. Performing a series of Gaussian integrals over the Grassmann variables~\cite{negele_orland_2019}, the Euclidean action can be expressed as $S_E= \beta A_sN\mathsf{s}_E$, where the action per unit 'volume' and fermion flavor is  
\begin{align}
\label{eq:eff_action_fields}
\begin{split}
\mathsf{s}_E\!=\! \frac{1}{g^2} \!\Big(\!\Sigma^2+{\Pi}_j^2\!\Big)\!-\!\!\int_{\omega,\boldsymbol{k}}\!\!\log\!\big({\omega^2+\epsilon_{\boldsymbol{k}}({m}+\Sigma,{\Pi}_j)^2}\big).
\end{split}
\end{align}
Here, we have introduced  $\int_{\omega,\boldsymbol{k}}=\int_{\omega}\int_{\boldsymbol{k}}= \int_{\mathbb{R}} \frac{{\rm d}\omega}{2\pi}\int_{\text{BZ}}\frac{{\rm d}^2k}{4\pi^2}$, and  the   dispersion relation modified by the auxiliary fields
\begin{align} 
\epsilon_{\boldsymbol{k}}({m}+\sigma,{\Pi}_j)= \parallel\boldsymbol{d}_{\boldsymbol{k}}(m+\Sigma,{\Pi}_j)\parallel. \label{singleparticleenergy}
\end{align}

We thus observe that the Euclidean action is proportional to the flavour number, which plays the role of an inverse Planck's constant $\hbar_{\rm eff}\propto1/N$. Accordingly,  the large-$N$ limit implies $\hbar_{\rm eff}\to 0$, such that the quantum fluctuations of the auxiliary fields are suppressed, and the partition function can be approximated by its saddle point, and determined by 
\beq
\frac{\partial \mathsf{s}_E}{\partial \Sigma}\bigg|_{\Sigma,{\Pi}_j}=\frac{\partial \mathsf{s}_E}{\partial \Pi_j}\bigg|_{\Sigma,{\Pi}_j}=0.
\eeq
These saddle-point equations lead to the so-called gap equations,  a system of  non-linear  equations which, upon using contour-integration over the Matsubara frequencies,  read
\begin{align}
\,\,\, \frac{\Sigma}{g^2}&=\frac{1}{2}\int_{\boldsymbol{k}} \, \frac{m+\Sigma+{m}_W(\boldsymbol{k})}{\epsilon_{\boldsymbol{k}}(m+\Sigma,{\Pi}_j)},\label{firstequation}\\
 \,\,\,\frac{\Pi_j}{g^2}&=\frac{1}{2}\int_{\boldsymbol{k}}  \, \frac{\frac{1}{a_j}\sin( k_ja_j)+\Pi_j}{\epsilon_{\boldsymbol{k}}({m}+\Sigma,{\Pi}_j)},\label{secondequation}
\end{align}
where  the contribution of the Wilson term~\eqref{wilson} to the mass is encoded in the following expression
\beq
\label{eq:wilson_cont_mass}
{m}_W(\boldsymbol{k})=\frac{1}{a_1}\Big(1-\cos(k_1 a_1)\Big)+\frac{1}{a_2}\Big(1-\cos(k_2 a_2)\Big).
\eeq

In this section and the following one,  the main goal  is to determine the phase boundary separating the various phases of matter that have been discussed so far, namely the QAH, TBI,  FM$_x$, FM$_y$ and PM phases. As will be discussed below, the TBI and PM phases are actually adiabatically connected, and can thus be described as different limits of the same phase.  To determine the critical lines delimiting the ferromagnetic phases, we would need to find  the points where the above equations are fulfilled, and  the $\pi$ fields go to zero  $\Pi_j=0$ starting from the corresponding symmetry-broken phases. Note that, in practice, one divides Eq.~\eqref{secondequation} by the corresponding value of the $\Pi_j$ condensate,   such that there are divergences associated to the  $\Pi_j=0$ point. However,  one can get as close to  this point as required by the  accuracy with which  the critical points are to be determined. In order to draw the ($ma_1,\xi_2,g^2/a_1$) phase diagram in analogy to Fig.~\ref{fig:ferrophases}, we perform the integrals that appear in the gap equations numerically, and vary the vacuum expectation values of the auxiliary fields $\Sigma,\Pi_j$  until  the  above  equations are fulfilled. 
Since we always get the combination $M=m+\Sigma$ in the equations, we start by assigning a value to $M$ for a given coupling strength $g^2$ and lattice anisotropy $\xi_2=a_1/a_2$, and then solve the self-consistent equations~\eqref{secondequation}  to find the values of  $\Pi_j$. 

\begin{figure}[t]
	\centering
		\centering
		\includegraphics[width=0.5\textwidth]{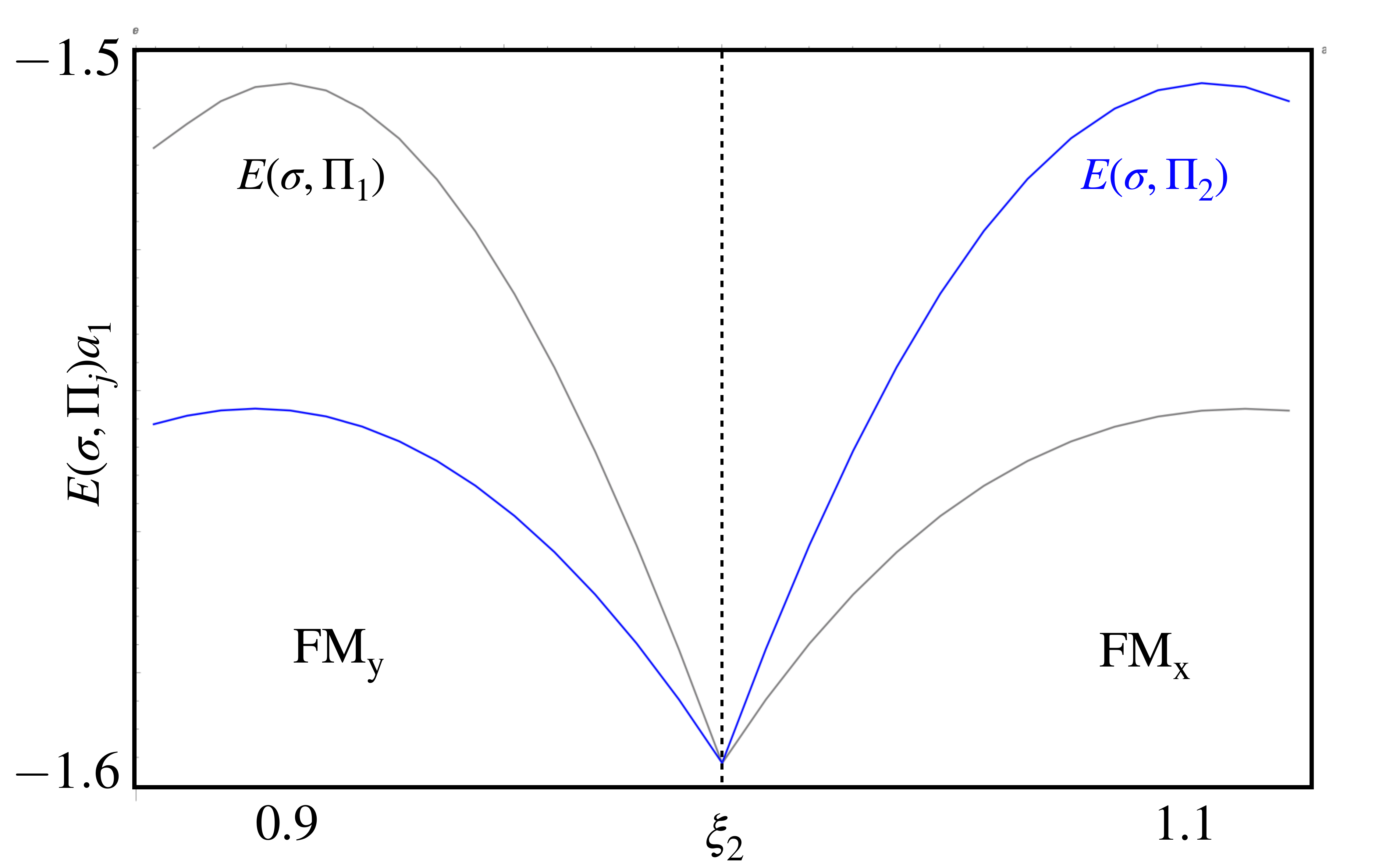}
		\caption{{\bf Preferred Lorentz-breaking channel:} The groundstate energies~\eqref{energycomparison2} for the two possible symmetry-breaking channels are represented as a function of the spatial anisotropy $\xi_2=a_1/a_2$ for $ma_1=-2$ and $g^2/a_1=5$. For $a_1>a_2$, the $\Pi_1$ condensate is preferred, whereas for $a_2>a_1$, the $\Pi_2$ condensate dominates.These correspond, respectively, to the FM$_x$ and FM$_y$ phases identified in the strong-coupling limit.}
		\label{fig:energy_comparison}
		\end{figure}

Once this is done, we can solve equation~\eqref{firstequation} to find the  the scalar  condensate $\Sigma$, and finally extract the corresponding  mass $m$.		Note however that we have two potential solutions $(\Pi_1,0)$ and $(0,\Pi_2)$,  and we must determine  which of the two possible $\pi$-condensates  occurs depending on the specific anisotropy, i.e. ${a_1>a_2}$ or ${a_2>a_1}$. The preferred symmetry-breaking channel can be determined by comparing the two groundstate energies $j\in\{1,2\}$ per unit area and flavour number at a certain point in the ($ma_1,\xi_2, g^2/a_1$) parameter space
\begin{equation}
E(\sigma,\Pi_{j}) = \frac{1}{g^2}\!\left(\Sigma^2+\Pi_{j}^2\right) -\!\!\int_{\boldsymbol{k}}\!\!{\rm d}^2k\epsilon_{\boldsymbol{k}}({m}+\Sigma,\Pi_{j}). \label{energycomparison2}
\end{equation}
This energy consists of the single-particle terms integrated over the Brillouin zone, and thus assumes a half-filled system with homogeneous  auxiliary-field terms. By finding the two solutions  for a chosen anisotropy, one can evaluate equation~(\ref{energycomparison2}) numerically in both cases, and figure out which channel has the lower groundstate energy. We note that this procedure is non-trivial since the $(ma_1,\xi_2,g^2/a_1)$ point of parameter space is not completely set from the start, but found recursively from the numerical routine just discussed. The numerical comparison of both channels requires an exhaustive numerical analysis to find sufficient solutions to densely cover the   parameter space, which can then be compared.
As shown in Fig.~\ref{fig:energy_comparison}, we have found that the dominant order parameter is $\Pi_1\propto \braket{\overline{\Psi}(\boldsymbol{x}) (\mathbb{I}_N\otimes\gamma^1) \Psi(\boldsymbol{x})}$  if ${a_2<a_1}$,  and $\Pi_2\propto \braket{\overline{\Psi}(\boldsymbol{x}) (\mathbb{I}_N\otimes\gamma^2) \Psi(\boldsymbol{x})}$ if ${a_2>a_1}$ , which is consistent with the results found for the variational mean-field and iPEPs methods applied to the quantum compass model  in the strong-coupling and single-flavour regimes.

\begin{figure}[t]
	\centering
		\centering
		\includegraphics[width=0.45\textwidth]{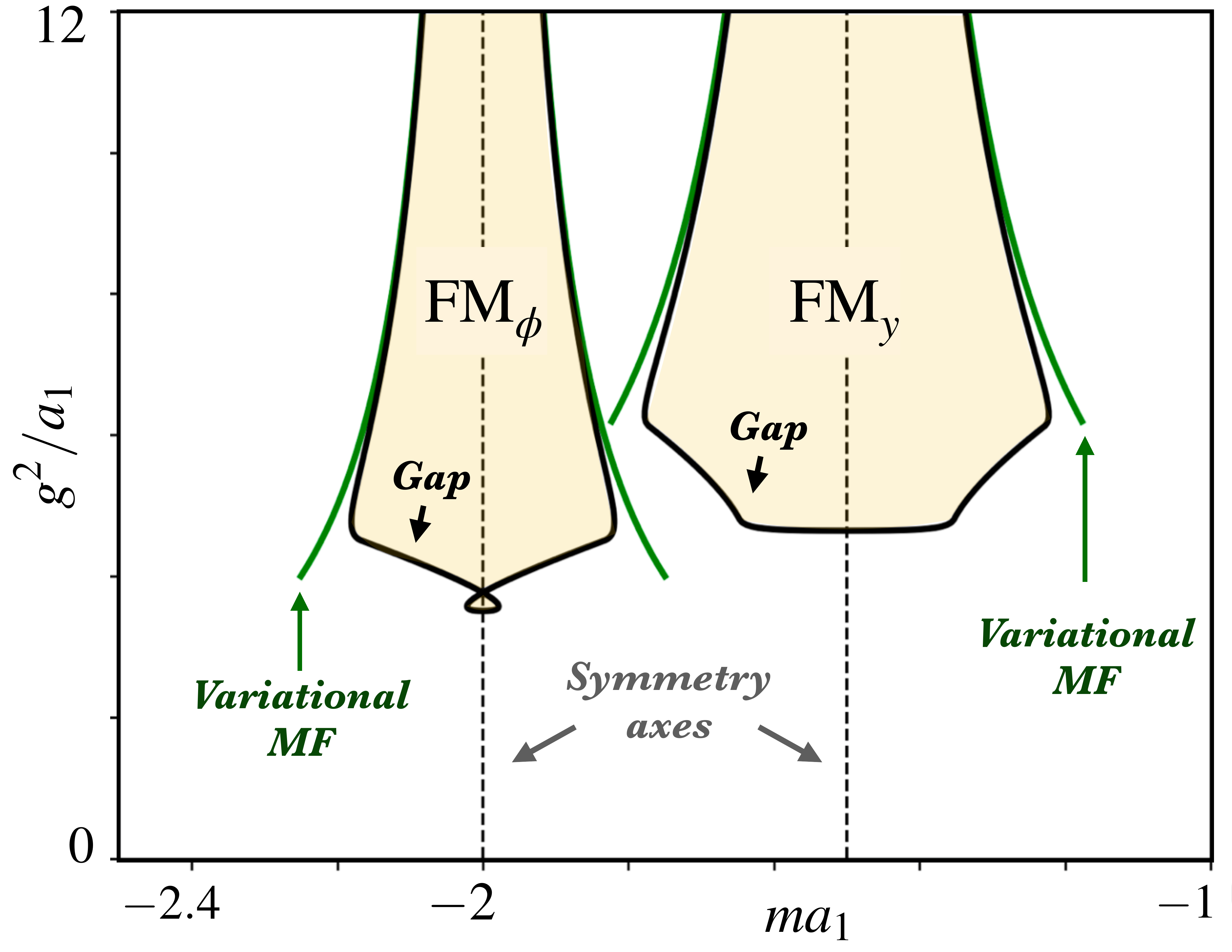}
		\caption{{\bf Phase diagram with Lorentz-breaking condensates:} The left figure represents the critical points  predicted by the large-$N$ calculation (black), and those obtained by the variational mean-field ansatz (green), both of which display a  vertical symmetry line at ${ma_1=-2}$ for the isotropic case $\xi_2=1$. The right figure represents the critical points  in the anisotropic case ${\xi_2=0.5}$ with the vertical symmetry line at ${ma_1=-1.5}$.}
		\label{continuumboth}
		\end{figure}

Let us now present a detailed discussion of our findings by analyzing the
two exemplary plots of the critical lines   presented  in figure~\ref{continuumboth}, which explore the role of interactions along the two vertical planes of Fig.~\ref{fig:chern}. In these plots, the black solid lines correspond to the critical points obtained from the solution of the gap equations~\eqref{firstequation}-\eqref{secondequation}.
The region inside this black line represents the inversion-broken phases. 
The left black solid line in  Fig.~\ref{continuumboth}  corresponds to the isotropic case $a_1=a_2$, where no comparison is required as both equations-\eqref{secondequation} for $j=1,2$ are equivalent. In close similarity to the variational compass-model result, this point is special in the sense that we cannot determine which linear combination $\cos\phi\Pi_1+\sin{\phi}\Pi_2$ is the correct order parameter, which is the situation found in Eqs.~\eqref{derivatives}-\eqref{derivatives2} by setting the exchange couplings~\eqref{eq:spin_couplings} to $J_x=J_y$ for equal $a_1=a_2$. This thus corresponds to the large-$N$ version of the FM$_\phi$ phase. In contrast,  the  right  black solid line of  Fig.~\ref{continuumboth} represents  the critical line  for lattice spacings  ${a_1=a_2/2}$, where we have found that  the enclosed region hosts a non-zero value of the $\Pi_2$ condensate, energetically preferred with respect to the $\Pi_1$ condensate (see Fig.~\ref{fig:energy_comparison}). As we keep on decreasing the ratio  $\xi_2=a_1/a_2$, this region moves progressively to the upper right, such that the appearance of the inversion-broken regions occurs for smaller absolute values of the  bare mass and larger interaction strengths. Although not shown in this figure,  for opposite anisotropies $a_2<a_1$, this behaviour is reversed, as we have found that the symmetry-broken $\Pi_1$ condensates move to the left, and thus seek larger absolute values of the  bare mass and smaller interaction strengths.

We note that both  ordered phases reflect mirror symmetry of the gap equations around the symmetry axis $m=-\frac{1}{a_1}-\frac{1}{a_2}$, which is depicted by a dashed line. In the language of the strong-coupling compass model~\eqref{eq:spin_couplings}, this symmetry  corresponds to the vanishing of the transverse field $h=0$, which is achieved within the symmetry plane of Fig.~\ref{fig:ferrophases}. We note that the $\sigma$ field and, with it, the scalar condensate vanish along this symmetry line $\langle\overline{\Psi}\Psi\rangle\propto\Sigma=0$. As discussed in more detail below,  this cancels the additive mass renormalisations arising
from the  fermion doublers. This feature is not a large-$N$ artefact, but actually connects to the so-called central-branch Wilson fermions,  which have interesting implications for Monte Carlo studies of  lattice gauge theories~\cite{10.1093/ptep/ptaa003,PhysRevD.102.034516}.
 
 In Fig.~\ref{continuumboth}, we  also represent in  green solid lines the  critical points~\eqref{criticalcompass} obtained by the variational mean-field ansatz of the 90$^{\rm o}$ compass model in a transverse field~\eqref{eq:compass} with microscopic parameters~\eqref{eq:spin_couplings}. It is quite remarkable to see that these two predictions match so well as one increases the interactions, since they emerge from completely different mean-field perspectives. In order to compare them, let us rewrite the strong-coupling 90$^{\rm o}$ compass model~\eqref{eq:compass}  as an effective action that depends solely on the auxiliary fields. Considering  the corresponding Euler-Lagrange equations
 \beq
\label{eq:aux_field_psi}
\sigma(\boldsymbol{x})=-\frac{g^2}{2}\overline{\Psi}(\boldsymbol{x})\Psi(\boldsymbol{x}),\hspace{1ex} \pi_j(\boldsymbol{x})=-\frac{g^2}{2}\overline{\Psi}(\boldsymbol{x})\gamma^j\Psi(\boldsymbol{x}),
\eeq
 the compass Hamiltonian leads to an action with nearest-neighbour couplings 
 \beq
 \label{eq:eff_action_compass}
 \begin{split}
 S_{\rm eff}=\!\int\!\!{\rm d}x^0a_1a_2\!\!\sum_{\boldsymbol{x}\in\Lambda_{s}}\!\!\Big(&\tilde{J}_1\pi_1(\boldsymbol{x})\pi_1(\boldsymbol{x}+a_2\boldsymbol{e}_2)\\
 +&\tilde{J}_2\pi_2(\boldsymbol{x})\pi_2(\boldsymbol{x}+a_1\boldsymbol{e}_1)+h\sigma(\boldsymbol{x})\Big),
 \end{split}
 \eeq
 where we have introduced the adimensional nearest-neighbour couplings for the $\pi$ fields  $\tilde{J}_1=(a_1/g^2)^2$ and $\tilde{J}_2=(a_2/g^2)^2$, whereas the coupling $h$ to the $\sigma$ field has the dimension of mass and corresponds directly to the transverse field in Eq.~\eqref{eq:spin_couplings}. This action clearly differs from the effective Euclidean action obtained in the large-$N$ approximation after integrating over the fermionic fields~\eqref{eq:eff_action_fields}. To compare both expressions on equal footing, we note that the real-space version of Eq.~\eqref{eq:eff_action_fields} would be expressed in terms of the logarithm of a  fluctuation determinant, which is highly non-local in contrast to Eq.~\eqref{eq:eff_action_compass}. On other hand, the Euler-Lagrange constraints~\eqref{eq:aux_field_psi} together with the Gutzwiller projector onto singly-occupied sites mentioned below Eq.~\eqref{eq:spin_couplings}, imply that the $\sigma$ and $\pi$ fields in Eq.~\eqref{eq:eff_action_compass} are not independent, but rather constrained to   
 \beq
 \sigma^2(\boldsymbol{x})+\pi_1^2(\boldsymbol{x})+\pi_2^2(\boldsymbol{x})=\left(\frac{g^2}{a_1a_2}\frac{1}{2}\boldsymbol{\tau}_n\right)^{\!\!\!2}=\frac{3}{4}\left(\frac{g^2}{a_1a_2}\right)^{\!\!\!2}\!.
 \eeq
 A simple rescaling of   these auxiliary fields shows that, for the   compass model, they  are constrained to lie on the unit sphere $\mathsf{S}_2$. Note that this type of strong-coupling  constraint  also arises in  the aforementioned $O(N)$ models~\cite{PhysRevD.7.2911,RevModPhys.45.589}, which  in the case of $N=3$ flavours leads  to the $O(3)$ non-linear sigma model   as one takes the strong-coupling limit $\lambda_0\to\infty$.   By contrast,  the compass action~\eqref{eq:eff_action_compass} can be understood as an anisotropic discretized version of a non-linear sigma model with a $\mathbb{Z}_2$ inversion symmetry $(\sigma(\boldsymbol{x}),\boldsymbol{\pi}(\boldsymbol{x}))\mapsto(\sigma(-\boldsymbol{x}),-\boldsymbol{\pi}(\boldsymbol{x}))$ instead of the continuous $O(3)$ symmetry. This is a generalisation of the rotationally-invariant Heisenberg spin model, which is known to be discretizations of the $O(3)$ non-linear sigma model~\cite{HALDANE1983464,PhysRevLett.50.1153}.

 The agreement of these two different mean-field methods  serves as a partial benchmark of both approaches, and hints to the validity of the conclusions drawn from   Fig.~\ref{continuumboth}: the non-interacting QAH and TBI phases of Sec.~\ref{sec: chern} will eventually disappear in favor of the symmetry-broken ferromagnetic phases or the paramagnetic phases  of  Secs.~\ref{sec:var_mf} and~\ref{sec:var_tn}. The solution of the gap equations tells us precisely for which bare parameters the spontaneous breakdown of inversion symmetry, and the formation of the condensates, will take place. As shown in the figure, we can thus extend the strong-coupling predictions to the regime of intermediate interactions, and find the whole extent of the critical line that surrounds the orbital FM$_x$ and FM$_y$ phases. Additionally, since there is no critical line that separates the orbital PM and the TBI, we can conclude that both states are limiting cases of the same phase.
 
 Let us note that,  in contrast to our previous results for the Gross-Neveu-Wilson model in (1+1) dimensions~\cite{BERMUDEZ2018149}, the critical lines obtained from the gap equations do not extend towards the region of low interactions   $g^2/a_1\approx0$, where the weakly-correlated QAH phase is expected to be. Therefore, we cannot use them to delimit the regions with different   topological phases characterized by opposite Chern numbers~\eqref{eq:critical_lines_chern}, nor the transitions to the disordered PM or the trivial band insulating TBI  phases. In order to overcome these limitations, we need to explore regions of the phase diagram where the $\pi$ fields are zero, and estimate the topological invariant for this large-$N$ approximation. We will show in Sec.~\ref{effectivepotential} that the effective potential~\cite{PhysRevD.10.2491,PhysRevD.9.3320} plays a key role in this regard. Moreover, it will allow us to explore the small  lobe that forms in the symmetry-broken region around the symmetry axis for $a_1=a_2$ which, although not shown in the figure, persists for small spatial anisotropies $a_1\approx a_2$. The effective potential will allow us to explore this region further,  and connect it to a  first-order phase transition rather than a second-order one. 
 
 However, the effective potential is most naturally formulated for a Euclidean lattice where imaginary time is also discretized. In the following subsection, in order to know how to recover the continuum-time limit from this Euclidean-lattice formulation, we will analyze the results of the gap equations for a Euclidean lattice, and show that there can be additional additive renormalizations. These contributions must be carefully accounted for if one aims to describe the physics of a correlated QAH effect, which is ultimately defined in the continuum-time limit. Equipped with these results, we will be able to formulate in Sec.~\ref{effectivepotential} a description for the effective potential that allows us to fully characterize the phase diagram.

\subsection{Large-$N$ gap equations on the Euclidean Lattice}

In the LFT community,  the   discretization is an artificial scaffolding for the fields that serves to regularize the QFT, but has  no physical reality. Accordingly, the imaginary time can also be  discretized, such that spacetime coordinates are treated on equal footing~\cite{PhysRevD.10.2445,gattringer_lang_2010}. In the present context, the rectangular spatial lattice~\eqref{eq:lattice_an} must be upgraded to include the discretised time with yet another lattice spacing. This leads to a simple orthorhombic Bravais lattice defined by
 \beq
 \label{eq:Euc_lattice}
 \Lambda_E= \{(n_0a_0,n_1a_1,n_2a_2): n_{\alpha}\in \mathbb{Z}_{N_\alpha},\,\forall \alpha\in\{0,1,2\} \}
 \eeq
where $N_\alpha$ is the number of sites along the $\alpha$-axis with lattice spacing $a_\alpha$, such that the Euclidean 'volume' is   $Q=\prod_\alpha N_\alpha a_\alpha$. We note that higher-dimensional versions of these anisotropic lattices have  been exploited in the context of lattice gauge theories. If one is interested in the time-continuum limit and the connection to Hamiltonian field theories, as in our work, the temporal anisotropy is mandatory~\cite{KARSCH1982285,BURGERS1988587}. Moreover, using a  smaller temporal lattice spacing  sometimes allows for a higher precision in Monte Carlo calculations~\cite{ALFORD1997377,KLASSEN1998557,PhysRevD.60.034509,PhysRevD.65.094508,PhysRevLett.111.172001,aarts2019spectral}. Interestingly, one can also get improvements by exploiting the anisotropy along the spatial directions~\cite{hadron_scatt_2007}. In the context of the domain-wall-fermion approach~\cite{KAPLAN1992342},  one of the spatial directions is considered as an auxiliary dimension, such that a pair of distant domain-wall profiles of the bare mass allow one to recover a QFT of chiral fermions for anisotropic volumes in which the wall separation is sufficiently large, which can be exploited for studies on chiral-invariant four-Fermi models~\cite{Hands2015,PhysRevD.99.034504,PhysRevD.102.094502}. If the anisotropy occurs at the level of the lattice spacings rather than the volume, one can also formulate domain-wall fermions by abruptly changing the lattice spacing values in a particular location of the auxiliary dimension~\cite{PhysRevD.102.094520}. 

 A possible  advantage of formulating the four-Fermi-Wilson model with a  discrete time  is that future studies may use the extensive LFT machinery based on Monte Carlo simulations to corroborate our predictions beyond the large-$N$ or strong-coupling limits. Despite the infamous sign problem for fermions, which plagues many Monte Carlo simulations, unless one is interested in   doping our system above/below half filling, the Monte Carlo update schemes should give reliable results. However, one should be careful with the interpretation of these results in the context of QAH phases and the phase diagram of topological materials. Whereas the spatial discretization is imposed by the underlying crystal, the time discretization is a  computational artifact, and the physics should be extracted by taking a continuum-time limit. This limit is not recovered by simply sending the corresponding  lattice spacing $a_0 \rightarrow 0$, as the time discretization introduces spurious (time)doublers~\cite{gattringer_lang_2010}. These will affect the theory, as they  carry their own Wilson masses, such that one would  expect the topological invariant~\eqref{eq:Chern_free} to be modified if this continuum-time limit is not considered carefully.  As will be shown in this section, in order to recover the results of the continuum-time gap equations (\ref{firstequation}) and (\ref{secondequation}), there are additive renormalizations that must be carefully considered.
  
 Let us start by formulating the problem on the Euclidean cubic lattice~\eqref{eq:Euc_lattice}. Here, we  use fermionic coherent states and   Grassmann vectors   $\Psi_{\boldsymbol{x}},\overline{\Psi}_{\boldsymbol{x}}$ for all Euclidean spacetime points $\boldsymbol{x}\in\Lambda_{E}$,  each containing both flavour and spinor components, and respecting periodic (antiperiodic) boundary conditions along the spatial (time) directions. The partition function can be expressed by a functional integral  $Z=\int [{\rm d}\Psi^\star {\rm d}\Psi]e^{-S_E[\Psi^\star,\Psi]}$
with the following Euclidean action 
\begin{equation}
S_E[\Psi^\star,\Psi]= a_0 a_1 a_2\left(S_E^0[\Psi^\star,\Psi]+V_g[\Psi^\star,\Psi]\right). \label{rawaction3}
\end{equation} 
 The free Wilsonian action in the Euclidean lattice reads
\begin{align}
\label{eq:free_action_euclidean}
\begin{split}
S_E^0[\Psi^\star,\Psi]= \sum_{\boldsymbol{x} \in \Lambda_E}&\overline{\Psi}_{\boldsymbol{x}}\Big(m+\sum_\alpha \frac{1}{a_\alpha}\Big) \Psi_{\boldsymbol{x}}\\
+\!\!\!\sum_{\boldsymbol{x} \in \Lambda_E}\! \sum_{\alpha\phantom{\Lambda}\!\!\!}\! \sum_{s=\pm 1} &\overline{\Psi}_{\boldsymbol{x}}\!\left(\frac{s(\mathbb{I}_N\otimes\tilde{\gamma}_\alpha)}{2a_\alpha}-\frac{1}{2 a_\alpha}\right)\Psi_{\boldsymbol{x}+sa_\alpha \boldsymbol{e}_\alpha},\\
\end{split}
\end{align}
where the Euclidean  gamma matrices are 
\beq
\label{eq:gamma_euc}
{\tilde{\gamma}_0= \gamma^0}=\sigma^z, \hspace{1ex} {\tilde{\gamma}_1=-\ii \gamma^1=\sigma^y},\hspace{1ex}{\tilde{\gamma}_2=-\ii\gamma^2=- \sigma^x}, 
\eeq
and satisfy   Clifford's algebra ${\{\tilde{\gamma}_\alpha,\tilde{\gamma}_\beta\}=2 \delta_{\alpha \beta}}$ for the  Euclidean metric ${\delta={\rm diag}(1,1,1)}$. In order to introduce the auxiliary fields, we consider the  results of the previous section, and only consider a single $\pi$-channel. In the following, all calculations are presented for  the $\tilde{\gamma}_2$ interaction corresponding to the $\Pi_2$ condensate, as this is the symmetry-breaking channel for $\xi_2<1$. Note, however, that there is no conceptual difference for the calculations for the $\tilde{\gamma}_1$ channel.
The four-Fermi term on the Euclidean  lattice reads
\begin{align}
\begin{split}
	V_g[\Psi^\star,\Psi]=\frac{-g^2}{4 N}\sum_{\boldsymbol{x}\in \Lambda_E}\left((\overline{\Psi}_{\boldsymbol{x}}\Psi_{\boldsymbol{x}})^2+(\overline{\Psi}_{\boldsymbol{x}}\ii\tilde{\gamma}_2\Psi_{\boldsymbol{x}})^2\right).
\end{split}\label{Euclideancubeinteraction}
\end{align}

 We now formulate this problem using dimensionless fields $\tilde{\Psi}_{\boldsymbol{x}}$, which can be defined as 
 \beq
 \label{eq:diemsnionless_fields}
 \Psi_{\boldsymbol{x}}= ({a_0 a_1+a_1 a_2+a_0 a_2})^{-\half}\,\tilde{\Psi}_{\boldsymbol{x}},
 \eeq
  and likewise for the adjoints $\tilde{\overline{\Psi}}_{\boldsymbol{x}}$. This  makes direct contact with  the standard formulations based on Wilson fermions in  LFTs~\cite{montvay_munster_1994,gattringer_lang_2010}. Similar to the continuous-time field theory~\eqref{rawaction2}, the quadratic terms can be diagonalized by going to $\boldsymbol{k}$-space,  $S^0_{E}=\sum_{\boldsymbol{k}}\tilde{\overline{\Psi}}_{\boldsymbol{k}}S^0_{\boldsymbol{k}}(\tilde{m})\tilde{\Psi}_{\boldsymbol{k}}$,  where 
 \begin{equation}
 \label{eq:free_action_momnetum}
S^0_{\boldsymbol{k}}(\tilde{m})= \mathbb{I}_N \otimes\left(\tilde{m}+1-\sum_\alpha 2\kappa_\alpha \ee^{-\ii k_\alpha a_\alpha\tilde{\gamma}_\alpha}\right).
\end{equation}
Here,  we  introduced the dimensionless mass and tunnelings
\beq
\label{eq:mass_rescaling}
\tilde{m}=\frac{ma_1}{1+\xi_1+\xi_2},\hspace{1ex}\kappa_\alpha=\frac{a_1}{2a_\alpha}\frac{1}{(1+\xi_1+\xi_2)},
\eeq
which, in addition to the spatial anisotropy $\xi_2$ defined in Eq.~\eqref{eq:spatial_anisotropy}, also depend on the temporal anisotropy
\beq
\label{eq:temporal_anisotropy}
\xi_1=\frac{a_0}{a_1}.
\eeq
 In addition,  due to the anti-periodicity in the time direction, the  momenta $\boldsymbol{k}=(k_0,k_1,k_2)^{\rm t}$ in the Brillouin zone are
\beq
\label{eq:Euc_rec_lattice}
k_0=-\frac{\pi}{a_0}+\frac{2\pi(n_0+\half)}{N_0a_0},\hspace{2ex} k_{j}=-\frac{\pi}{a_j}+\frac{2\pi n_{j}}{N_ja_j},
\eeq 
where we recall that $j\in\{1,2\}$, and  $n_\alpha\in\mathbb{Z}_{N_\alpha}$.

 Following the previous section, we can introduce  dimensionless versions of the auxiliary Hubbard-Stratonovich  fields $\tilde{\Sigma}$, $\tilde{\Pi}_2$. The action corresponding to the partition function $Z=\int {\rm d}\tilde{\overline{\Psi}}{\rm d}\tilde{\Psi} {\rm d}\tilde{\Sigma} {\rm d}\tilde{\Pi}_2 \exp(-S_E[\tilde{\overline{\Psi}},\tilde{\Psi},\tilde{\Sigma},\tilde{\Pi}_2])$ now reads
\begin{align}
\label{eq:adimensional_HS_transf}
\begin{split}
S_E= N \frac{\tilde{Q}}{\tilde{g}^2}\Big(\tilde{\Sigma}^2+\tilde{\Pi}_2^2\Big)+\sum_{\boldsymbol{k}}\tilde{\overline{\Psi}}_{\boldsymbol{k}}S^0_{\boldsymbol{k}}(\tilde{m}+\tilde{\Sigma},\tilde{\Pi}_2)\tilde{\Psi}_{\boldsymbol{k}},
\end{split}
\end{align}
where we have  assumed homogeneous auxiliary fields, and  introduced the dimensionless volume and coupling strength
\beq
\label{eq:coupling_rescaling}
\tilde{Q}= N_0N_1N_2,\hspace{1ex}\tilde{g}^2=\frac{g^2}{a_1}\frac{\xi_1\xi_2}{(1+\xi_1+\xi_2)^2}.
\eeq
Additionally, in analogy to the continuum-time case in Eqs.~\eqref{eq:free_action_aux}
 and~\eqref{freehamiltonian_d_vector_aux}, the free action in momentum space~\eqref{eq:free_action_momnetum} gets modified by the presence of the auxiliary fields to
\beq
S^0_{\boldsymbol{k}}(\tilde{m}+\tilde{\Sigma},\tilde{\Pi}_2)=S^0_{\boldsymbol{k}}(\tilde{m})+\mathbb{I}_N\otimes(\tilde{\Sigma}+\tilde{\Pi}_2\tilde{\gamma}_2).
\eeq

One now proceeds by integrating  out the fermions to find an effective action for the auxiliary fields  $Z=\int d\tilde{\Sigma} d\tilde{\Pi}_2\,\, e^{-N\tilde{Q} \mathsf{s}_E[\tilde{\Sigma},\tilde{\Pi}_2]}$, where the action per unit volume and flavour number is 
\begin{align}
\label{eq:adimensional_action_aux_fields}
\begin{split}
\mathsf{s}_E[\tilde{\Sigma},\tilde{\Pi}_2]= \frac{1}{\tilde{g}^2}(\tilde{\Sigma}^2+\tilde{\Pi}_2^2)-\frac{1}{\tilde{Q}}\sum_{\boldsymbol{k}}\log\left(\tilde{\mathsf{s}}_{\boldsymbol{k}}^2(\tilde{m}+\tilde{\Sigma},\tilde{\Pi}_2)\right).
\end{split}
\end{align}
In this expression, we have introduced the function
\begin{align}
\label{eq:action_momentum_aux}
\begin{split}
\tilde{\mathsf{s}}_{\boldsymbol{k}}^2(\tilde{m}+\tilde{\Sigma},\tilde{\Pi}_2)&=\big(\tilde{m}+\tilde{\Sigma}+\tilde{m}_E(\boldsymbol{k})\big)^2\\
&+\!\sum_\alpha\big(2\kappa_\alpha \sin(k_\alpha a_\alpha)+\tilde{\Pi}_\alpha\delta_{\alpha,2}\big)^2,
\end{split}
\end{align}
and the dimensionless Euclidean version of the Wilson-term contribution to the mass in Eq.~\eqref{eq:wilson_cont_mass}, namely
\beq
\label{Euclidean_wilson_contribution}
\tilde{m}_{E}(\boldsymbol{k})= 1-\sum_\alpha 2\kappa_\alpha \cos(k_\alpha a_\alpha).
\eeq
Let us also note that for the $\tilde{\gamma}_1$ channel, one simply changes $\delta_{\alpha,2}\to\delta_{\alpha,1}$ in Eq.~\eqref{eq:action_momentum_aux}.

As occurred for  continuous time, the flavour number $N$ plays the role of an inverse Planck's constant $h_{\rm eff}\propto1/N$, and the large-$N$ limit is controlled by the semi-classical limit where the auxiliary fields do not fluctuate around the saddle-point configurations $h_{\rm eff}\to 0$. The gap equations corresponding to this Euclidean saddle point  are
\begin{align}\label{discretegap1}
\frac{\tilde{\Sigma} }{\tilde{g}^2}&= \frac{1}{\tilde{Q}}\sum_{\boldsymbol{k}} \frac{\tilde{m}+\tilde{\Sigma}+\tilde{m}_E(\boldsymbol{k})}{\tilde{\mathsf{s}}^2_{\boldsymbol{k}}(\tilde{m}+\tilde{\Sigma},\tilde{\Pi}_2)},\\
\frac{\tilde{\Pi}_2}{\tilde{g}^2}&=\frac{1}{\tilde{Q}}\sum_{\boldsymbol{k}} \frac{2\kappa_2 \sin(k_2 a_2)+\tilde{\Pi}_2}{\tilde{\mathsf{s}}^2_{\boldsymbol{k}}(\tilde{m}+\tilde{\Sigma},\tilde{\Pi}_2)}.\label{discretegap2}
\end{align}
Note that the structure resembles that of equations (\ref{firstequation}) and (\ref{secondequation}), allowing us to use the same  algorithm for their numerical solution, albeit having an extra  mode sum for the time-like direction stemming from the Euclidean discretization. In the continuum-time case, this mode sum would correspond to the integration over the Matsubara frequencies, which was performed analytically to arrive at Eqs.~(\ref{firstequation}) and~(\ref{secondequation}).

\subsection{Time doublers and  mass renormalisations}

As advanced at the beginning of the previous subsection, the spurious time doublers  must be carefully accounted for in order to recover the correct phase diagram in the continuum-time limit. This situation was first noted for the (1+1) Gross-Neveu model with a Wilson-type discretization~\cite{BERMUDEZ2018149}. In the present context,  by inspection of Eq.~\eqref{eq:free_action_momnetum}, one can readily see that the expansions around ${k}_0\in\{0,\pi/a_0\}$ yield long-wavelength actions that resemble those of a continuum massive Dirac fermion. Paralleling the discussion around of the long-wavelength Hamiltonian QFT~\eqref{eq:continuum_theory}, 
there are  points in $\boldsymbol{k}$-space, $\boldsymbol{K}_{\boldsymbol{n}_d}=(\pi n_{d,0}/a_0, \pi n_{d,1}/a_1,\pi n_{d,2}/a_2)$ for $\boldsymbol{n}_d=(n_{d,0},n_{d,1},n_{d,2}) \in \{0,1\}\times \{0,1\}\times\{0,1\}$,  around which we can define a set of   Dirac  spinors $\{\Psi_{\boldsymbol{n}_d}(\boldsymbol{k})\}_{\boldsymbol{n}_d}$ with $N$ flavours governed by the long-wavelength Euclidean action
 \beq
 \label{eq:continuum_theory_euclidean}
 {S}_F\!=\!\!\!\int\!\!\frac{{\rm d}^3k}{(2\pi)^3}\!\!\sum_{\boldsymbol{n}_d} 
\overline{\Psi}_{\boldsymbol{n_d}}\!(\boldsymbol{k})\! \left(\ii\big(\mathbb{I}_N\otimes\tilde{\gamma}_{\boldsymbol{n_d}}^\mu\big) k_\mu+m_{\boldsymbol{n_d}} \right)\!\Psi_{\boldsymbol{n_d}}\!(\boldsymbol{k}).
 \eeq
 Here, the repeated-index sum is performed with respect to the Euclidean metric, and  the Euclidean gamma matrices for each Dirac fermion are labelled by  $\boldsymbol{n}_{d}\in\mathbb{Z}_2^3$ and read
 \beq
\label{eq:gammas_doubler_euc}
\tilde{\gamma}^0_{\boldsymbol{n}_d}=(-1)^{n_{d,0}}\tilde{\gamma}_0,\hspace{1ex} \tilde{\gamma}^1_{\boldsymbol{n}_d}=(-1)^{n_{d,1}}\tilde{\gamma}_1,\hspace{1ex}\tilde{\gamma}^2_{\boldsymbol{n}_d}=(-1)^{n_{d,2}}\tilde{\gamma}_2,
\eeq
whereas the corresponding Euclidean Wilson masses are
\begin{equation}
\label{eq:WIlson_masses_euc}
m_{\boldsymbol{n}_d}=m+\frac{2n_{d,0}}{a_0}+\frac{2n_{d,1}}{a_1}+\frac{2n_{d,2}}{a_2}.
\end{equation}
As we can see, in addition to the fermion doublers for $n_{d,0}=0$, which correspond precisely to the physical spatial doublers obtained in  the long-wavelength Hamiltonian QFT from the model defined on the physical  lattice~\eqref{eq:continuum_theory}, we get extra doublers at $n_{d,0}=1$, which are an artifact of the   discretization of time, and we refer to them as spurious time doublers.

Let us now discuss how these time doublers affect the phase diagram, and how one can recover the correct continuous-time limit. Although one may expect that sending $a_0\to 0$ makes these spurious doublers very massive~\eqref{eq:WIlson_masses_euc}, such that they have no effect on the low-energy physics,  it turns out that they can induce additive renormalizations that do not vanish as we approach such a continuous-time limit. As detailed in Appendix~\ref{app:add_renorm}, this renormalization can be obtained by comparing the gap equations of the discrete-time Euclidean formulation, in the limit $a_0\to0$, with those of the Hamiltonian approach that are directly derived for continuous times.

\begin{figure*}[t]
		\includegraphics[width=0.8\textwidth]{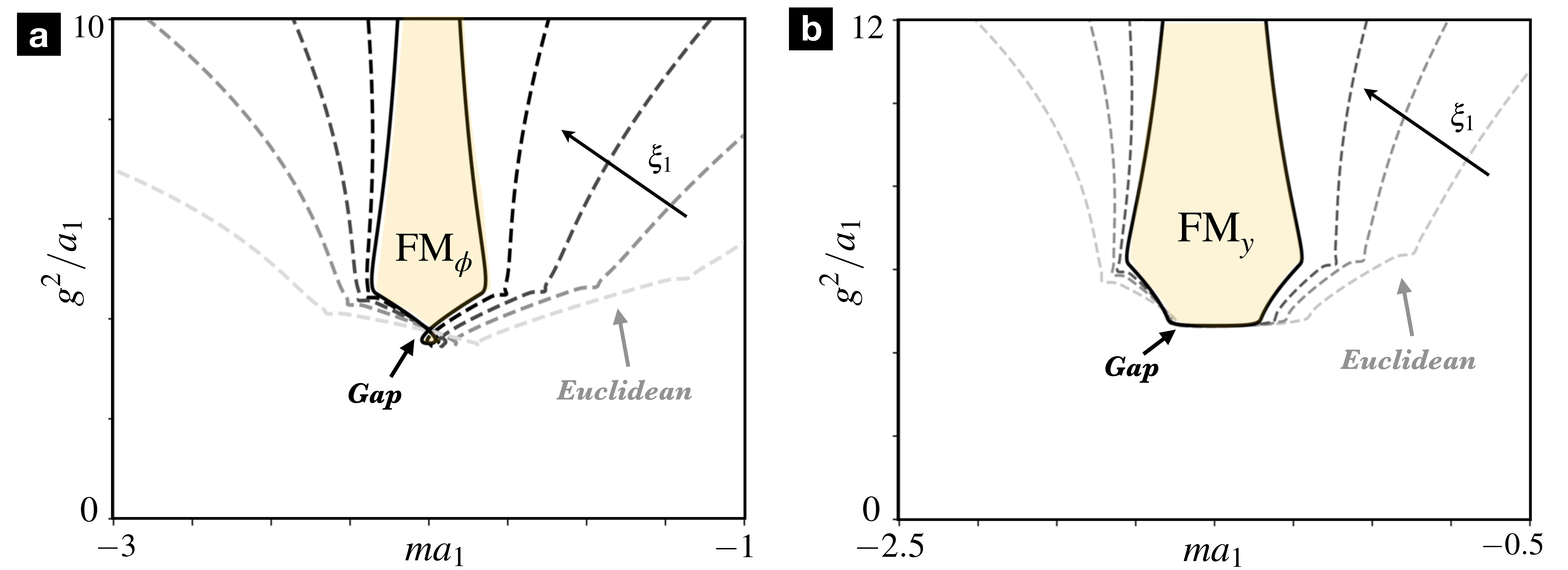}
	\caption{{\bf Phase diagram with Lorentz-breaking condensates on the Euclidean lattice:} The critical lines obtained by the solution of  the  the large-$N$ gap equations~\eqref{firstequation} and~\eqref{secondequation} in continuous time (black solid line) and those in Eqs.~\eqref{discretegap1}-\eqref{discretegap2} for discrete time (dashed lines with gray scale). The later require using the rescalings and renormalisations of Eq.~\eqref{eq:rescalings_parameters}, and correspond to temporal anisotropies $\xi_1=\{10,20,40,64\}$. In {\bf (a)} we present the results for $\xi_2=1$, whereas {\bf (b)} corresponds to   $\xi_2=0.5$. In both cases, as the temporal anisotropy increases,  the discrete-time critical lines tend towards the continuum ones, and we expect a perfect agreement for   $\xi_1\to\infty$. }
	\label{discrete}
\end{figure*}

In particular, we find that the gap equation for the Lorentz-breaking $\Pi_j$ condensates become equal in this limit, but those of the scalar $\Sigma$ condensate differ by an additive term that scales with the coupling strength $g^2$ and does not vanish when $a_0\to 0$. As discussed in the Appendix, such a term can be identified with the contribution of  the spurious time doublers at $n_{d,0}=1$, and leads to an additional shift of the bare mass. Hence, if we want to recover the correct phase diagram of the model by solving the dimensionless version of the Euclidean-lattice gap equations in Eqs.~\eqref{discretegap1}-\eqref{discretegap2}, we need to combine this renormalization with the rescalings of Eqs.~\eqref{eq:mass_rescaling} and~\eqref{eq:coupling_rescaling}. This can be summarised in the redefinition of the phase-diagram axes 
 \beq
\label{eq:rescalings_parameters}
 \begin{split}
 \frac{g^2}{a_1}&=\frac{(1+\xi_1+\xi_2)^2}{\xi_1\xi_2}\tilde{g}^2,\\
 \hspace{1ex}ma_1&=(1+\xi_1+\xi_2)\tilde{m}+\frac{(1+\xi_1+\xi_2)^2}{2\xi_1}\tilde{g}^2.
 \end{split}
 \eeq
 Likewise, the condensates from the Hamiltonian and Euclidean-lattice approached are connected as follows  $\Sigma a_1=(1+\xi_1+\xi_2)\tilde{\Sigma}$ and $\Pi_2a_2=(1+\xi_1+\xi_2)\tilde{\Pi}_2$.

In summary, we  see that discretization of the time-like direction introduces spurious doublers that would lead to a modified phase diagram if left unnoticed. This is not so important for LFTs, as one typically concentrates on the properties of the continuum QFT that arises around the critical line, but its specific location in terms of the bare parameters is not of relevance. However, for the application of this method to understand the phase diagram of an interacting QAH phase, e.g. understanding the robustness of the Chern insulator to interactions, it is crucial to keep track of these renormalization effects. In Fig.~\ref{discrete}, we represent the phase diagram obtained by  solving the discrete-time gap equations~\eqref{discretegap1}-\eqref{discretegap2} using the re-scalings and renormalizations of Eq.~\eqref{eq:rescalings_parameters}. We explore different temporal and spatial anisotropies ${\xi_1=\{10,20,40,64\}}$, $\xi_2=1$ (left panel), and ${\xi_1=\{20,40,64\}}$, $\xi_2=0.5$ (right panel). Note that the time-continuum limit corresponds to $\xi_1=a_1/a_0\to\infty$, while the spatial anisotropies translate into $a_1=a_2$ (Fig.~\ref{discrete} {\bf (a)}) and $a_1<a_2$ (Fig.~\ref{discrete} {\bf (b)}). The corresponding critical lines are depicted by dashed lines, where the gray scale becomes darker for increasing temporal anisotropy. The black solid lines correspond to the results depicted in Fig.~\ref{continuumboth}, and thus to the numerical solution of the Hamiltonian-field theory gap equations~\eqref{firstequation} and~\eqref{secondequation}, which have no contribution from spurious time doublers.  Both figures show how the results obtained from the Euclidean-lattice formulation with a discrete time, and both dimensionless fields and couplings,  converges to the  time-continuum phase diagram as one increases the   temporal anisotropy $\xi_1\to\infty$. 

The advantage of the Euclidean-lattice formulation is now two-fold. On the one hand, Monte Carlo techniques developed in the LFT community that employ Wilson fermions in the fermionic sector of QCD could be readily applied to go beyond the current gap equations, provided that one takes the lesson learned from the   parameter rescaling and renormalizations~\eqref{eq:rescalings_parameters}. On the other hand, we can use the Euclidean-lattice formalism to calculate the effective potential, as discussed in the following section and, by virtue of the relations~\eqref{eq:rescalings_parameters}, also explore the regions of the phase diagram where the correlated QAH phase is expected to be found.

\section{\bf Effective potential and  large-$N$ Chern insulators } \label{effectivepotential}

As briefly discussed in the introduction of Sec.~\ref{gapequations}, the large-$N$ radiative corrections of  four-Fermi QFTs   can be represented by a collection of one-particle irreducible (1PI) amputated   diagrams  composed of a single fermion loop and an even number of external auxiliary lines (see Fig.~\ref{fig:large_N_FD}), all of which have zero external momenta and no external fermion lines, i.e. amputated. This type of   Feynman diagrams appears naturally in the context of   the so-called effective action $S_{\rm eff}$, which acts as the generating functional of  proper vertex functions, defined   as the inverse of the dressed $n$-point propagators~\cite{Peskin:1995ev}. For instance, the 2-point vertex function  contains all the non-perturbative information encoded in  the self-energy $\Sigma_{s}(\boldsymbol{k})$, which will play a key role in the description  of correlation effects in the QAH phases. Let us now discuss why these are the relevant diagrams in the large-$N$ limit, and how they can be resummed to obtain the effective potential. 

In general, the effective action $S_{\rm eff}[\varphi_{\rm c}(x)]$ is  a functional of the symmetry-breaking order parameter, which can be seen as a classical field  $\varphi_{\rm c}(x)$. In the context of  chiral-invariant four-Fermi QFTs,  this order parameter is the aforementioned scalar condensate~\eqref{eq:scalar_condensate}, which we recall is proportional to the vacuum expectation value of the $\sigma(x)$ field playing the role of $\varphi_{\rm c}(x)$. In a translationally-invariant situation $\partial_\mu\varphi_{\rm c}(x)=0$, analogous to the case $\sigma(x)=\Sigma,\,\,\,\forall x$ explored in the previous section, this effective action can be expressed as  the spacetime integral of an effective potential $S_{\rm eff}[\Sigma]=\int{\rm d}^DxV_{\rm eff}(\Sigma)$. We note that, by going to momentum space, this effective action can only generate vertex functions with zero external momentum. Accordingly, the effective action for a translationally-invariant classical field is built from  all the amputated 1PI Feynman diagrams  evaluated at zero external momentum. These diagrams  allow one to understand how the  radiative corrections of a purely quantum-mechanical origin affect the process of spontaneous symmetry breaking in scalar QFTs~\cite{PhysRevD.7.1888,PhysRevD.9.1686}. Moreover, the effective potential also allows to go beyond perturbation theory, providing a neat instance of  large-$N$ methods~\cite{PhysRevD.10.2491,coleman_1985}.

The standard discussion of the effective potential for four-Fermi QFTs~\cite{coleman_1985} must be reconsidered in  our case~\eqref{eq:total_H}, as we have already argued that, for our choice of gamma matrices~\eqref{eq:gammas}, chiral symmetry is absent from the outset. In fact, we have also noted that, except for the specific couplings corresponding to the symmetry line of Fig.~\ref{continuumboth},  the $\sigma$ field displays a non-zero vacuum expectation value that is not connected to any spontaneous symmetry breaking. Therefore, the $\sigma(x)$ field cannot play the role of the classical field $\varphi_{\rm c}(x)$ in our case. On the other hand, as discussed in the previous section, there are two alternative symmetry-breaking channels in which inversion~\eqref{eq:inv_symm}, instead of  chirality~\eqref{eq:chiral_symmetry}, is the symmetry that is actually broken at the phase transition. We recall that these channels  were activated by  non-zero values of the auxiliary $\pi$ fields~\eqref{eq:pi_condensates} rather than the $\sigma$ field~\eqref{eq:scalar_condensate}. From this perspective, we need to revisit the discussion of the resummation of the leading-order Feynman diagrams to obtain an adequate effective potential. In this subsection, we present the details corresponding to the $\tilde{\Pi}_2$ channel, giving a detailed account of the effective potential $V_{\rm eff}(\tilde{\Pi}_2)$ in the large-$N$ limit,  although we also note that the calculations for the $\tilde{\Pi}_1$ channel are completely analogous.
Armed with this effective potential, we can now explore the regions where inversion symmetry remains intact, and use the effective potential to characterize the extent of the QAH phase as interactions are increased. 

\subsection{Fermion condensates and  the effective potential}

Let us consider the anisotropic regime $a_1<a_2$, which allows us to focus on  a single Lorentz-breaking   field $\tilde{\Pi}_2$. We recall that, by introducing the auxiliary field via a Hubbard-Stratonovich transformation, one sees that the field does not have kinetic energy and is subjected to a classical potential 
\beq
\label{eq:class_pot}
V_{\rm cl}(\tilde{\Pi}_2)=\frac{N }{2\tilde{g}^2}\tilde{\Pi}_2^2.
\eeq
This expression accounts for the $\tilde{g}^2/N$  scaling of the auxiliary-field lines/propagators mentioned  in the introduction of Sec.~\ref{gapequations}. In the present subsection, we  use the adimensional formulation of the model~\eqref{eq:adimensional_HS_transf}, so that all adimensional quantities appear with a tilde, and we should apply the required rescalings and renormalizations in Eq.~\eqref{eq:rescalings_parameters} at the end.

At the classical level, this potential~\eqref{eq:class_pot} finds its minimum at $\tilde{\Pi}_{2,\rm c}=0$, such that there is no spontaneous breakdown of inversion symmetry. However, this is not the full picture, as  one should also introduce quantum-mechanical corrections, leading to the full effective potential
\beq
\label{eq:full_eff_V}
V_{\rm eff}(\tilde{\Pi}_2)= V_{\rm cl}(\tilde{\Pi}_2)+\delta V_{\rm q}(\tilde{\Pi}_2)
\eeq
 In the $N\rightarrow\infty$ limit, these quantum corrections $\delta V_{\rm q}(\tilde{\Pi}_2)$ correspond to  all of the amputated 1PI diagrams at zero external momentum which, after introducing the auxiliary field (see Fig.~\ref{fig:large_N_FD}), can be constructed by combining fermion loops and auxiliary $\tilde{\Pi}_2$ lines. Note  we can decorate these diagrams with any number of external auxiliary lines and still obtain contributions to the vertex functions with zero external momentum. These external lines must be connected to a fermion loop via  the coupling term  $\tilde{\Pi}_2(\tilde{\overline{{\Psi}}}\tilde{\gamma}_2\tilde{\Psi})$, which does not scale with $1/N$ in comparison to the original four-Fermi term~\eqref{interaction}. We can thus introduce an arbitrary number of external auxiliary fields without altering the scaling $\mathcal{O}(N^\alpha)$ of the specific order $\alpha$ in a large-$N$ expansion. On the contrary,  inserting  internal auxiliary lines, as in Fig.~\ref{fig:large_N_FD} {\bf (b)}, is penalized by an extra $1/N$ scaling for  each auxiliary-field propagator, as the mass of this auxiliary field scales with $N$ (see Fig.~\ref{fig:large_N_FD} {\bf (a)}). Likewise, since extra fermion loops can only be introduced in 1PI diagrams  by means of additional internal  lines of the auxiliary field, they will also give sub-leading contributions. The leading-order contribution in the $N\to\infty$ limit can thus be obtained by resumming the series of Feynman diagrams with a single fermion loop and increasing numbers of external auxiliary lines (see Fig.~\ref{fig:large_N_FD} {\bf (c)}). 

 \begin{figure*}[t]
		\includegraphics[width=0.85\textwidth]{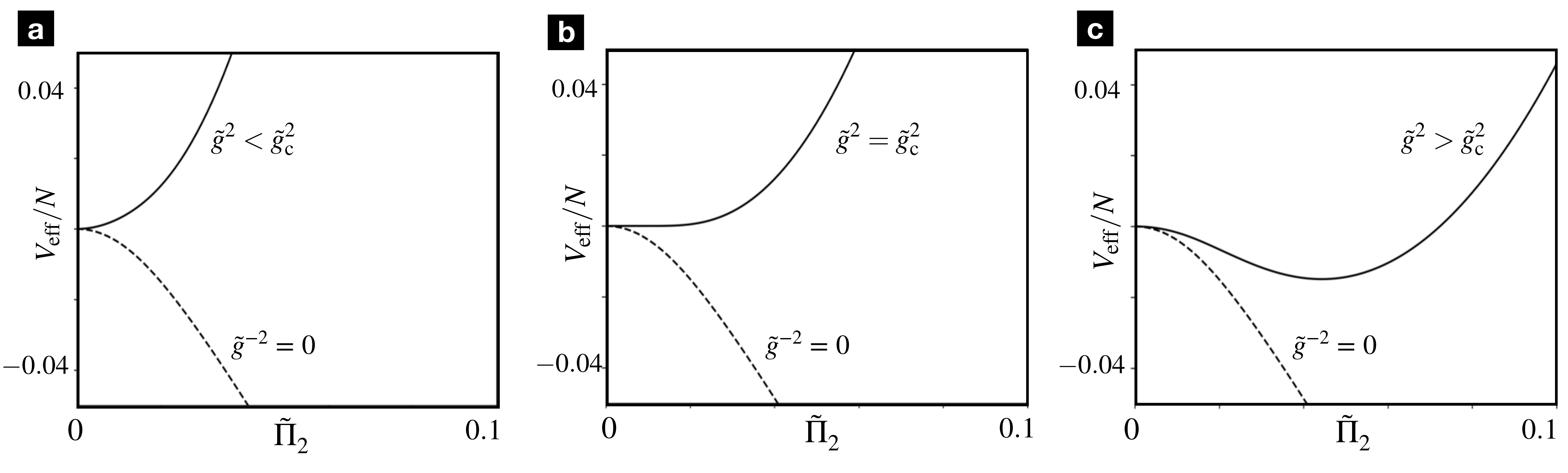}
	\caption{{\bf Effective potential and the spontaneous breakdown of inversion symmetry:} We represent the numerical values of $V_{\rm eff}(\tilde{\Pi}_2)$ obtained by evaluating Eq.~\eqref{eq:eff_potential_discrete} for different couplings $\tilde{g}^2$ (solid line) and  the isolated quantum corrections labelled by $\tilde{g}^{-2}=0$ (dashed line). {\bf (a)} For $\tilde{g}^2<\tilde{g}^2_c$, the  minimum occurs for $\tilde{\Pi}_2=0$. {\bf (b)} For $\tilde{g}=\tilde{g}_c$, the effective potential changes from a quadratic to a quartic dependence for small values of  the symmetry-breaking $\tilde{\Pi}_2$ field. {\bf (c)} For $\tilde{g}^2>\tilde{g}^2_c$, a minimum with non-zero value of  $\tilde{\Pi}_2$ appears, signalling the breakdown of inversion symmetry.}
	\label{fig:potential}
\end{figure*}

In Appendix~\ref{effective} we show how, in analogy to the chiral-invariant case, one only needs to decorate this single fermion loop  with even numbers of external auxiliary lines. These large-$N$ diagrams contribute to the effective potential  as 
\begin{align}
\begin{split}
\delta V_{\rm q}(\tilde{\Pi}_2)=N \sum_{n=1}^{\infty} \frac{1}{2n} \int_{\boldsymbol{p}}{\rm Tr}\left(-\ii \tilde{\gamma}_2 \frac{\tilde{\Pi}_2}{(\ii \slashed{p}+\tilde{m}(\boldsymbol{k}))}\right)^{\!\!\!2n}, 
\label{veffbasic}
\end{split}
\end{align}
where we have introduced $\slashed{p}=\tilde{\gamma}^\alpha p_{\alpha}$, and  
\beq
\label{eq:eff_potential_parameters}
p_{\alpha}=2\kappa_\alpha\sin k_\alpha,  \hspace{1ex}\tilde{m}(\boldsymbol{k})=\tilde{m}+\tilde{\Sigma}+\tilde{m}_E(\boldsymbol{k}),
\eeq
 where  $\tilde{m}_E(\boldsymbol{k})$ is the Euclidean Wilson-type contribution to the mass  that already appeared in Eq.~\eqref{Euclidean_wilson_contribution}. For a continuum QFT, $p_{\alpha}$ would be the Euclidean momentum with an ultra-violet  cutoff, e.g. $p\leq\Lambda_{\rm c}$, and $\tilde{m}$ the adimensional bare mass. We also note that the repeated-index summation is performed  using the Euclidean metric, and we use a mode-sum discretization of the integral $\int_{\!\boldsymbol{p}}=\frac{1}{\tilde{Q}}\sum_{\boldsymbol{k}^{\phantom{k^o}\!\!\!\!\!\!\!\!\!\!}}$ over the  reciprocal BZ~\eqref{eq:Euc_rec_lattice}.

In addition to these straightforward differences due to the lattice regularization, the above expression~\eqref{veffbasic} differs from the standard effective potential of continuum   QFTs like Eq.~\eqref{continuum} by the appearance of the $\tilde{\gamma}_2$ matrix. As discussed in Appendix~\ref{effective}, this  complicates  considerably the resummation method with respect to the standard calculations~\cite{coleman_1985}, and leads to new quantum-mechanical sources of radiative corrections  $\delta V_{\rm q}(\tilde{\Pi}_2)=\delta V_{\rm q,1}(\tilde{\Pi}_2)+\delta V_{\rm q,2}(\tilde{\Pi}_2)$, which will play a key role below. In addition to the standard contribution
\beq
\label{eq:q_corrections_1}
\delta V_{\rm q,1}(\tilde{\Pi}_2)=-N\int_{\boldsymbol{p}} \log\left(1+\frac{\tilde{\Pi}_2^2}{p^2+\tilde{m}^2(\boldsymbol{k})}\right),
\eeq
which is also found for chiral invariant QFTs in the continuum by simply letting $\tilde{\Pi}_2\to\tilde{\Sigma}$ ~\cite{coleman_1985},
we find that the new Lorentz-breaking channel has new radiative corrections given by  
\begin{align}
\label{eq:q_corrections_2}
\begin{split}
\delta V_{\rm q,2}(\tilde{\Pi}_2)\!=-\frac{N}{2}\!\int_{\boldsymbol{p}} \!\log\!\!\left(\!1-\frac{4p^2_2 \tilde{\Pi}_2^2}{\left(p^2+\tilde{m}^2(\boldsymbol{k})+\tilde{\Pi}_2^2\right)^2}\!\right)\!\!.
\end{split}
\end{align}

Let us now connect to our  discussion  of the gap equations in Sec.~\ref{gapequations}. Note that both  the classical and quantum contributions to the effective potential scale with $N$ at this leading order, such that an effective   Planck's constant $\hbar_{\rm eff}\propto1/N$ shall vanish in the large-$N$ limit. Accordingly,  the leading-order solution is found by searching for the minima of $V_{\rm eff}(\tilde{\Pi}_2)$ at different points in the ($\tilde{m}$,$\tilde{g}^2$)-plane. With respect to the classical case~\eqref{eq:class_pot}, which only allows for a zero Lorentz-breaking condensate, the quantum corrections can lead to new minima in which a non-zero value of $\tilde{\Pi}_2$ develops.  Although not apparent at first sight, these minima correspond to the saddle points of the action per unit volume and number of flavors $\mathsf{s}_E[\tilde{\Sigma},\tilde{\Pi}_2]$ introduced in Eqs.~\eqref{eq:adimensional_action_aux_fields}-\eqref{eq:action_momentum_aux}, which was obtained by integrating over the fermionic Grassmann variables after the Hubbard-Stratonovich transformation, and led to the previous gap equations~\eqref{discretegap1}-\eqref{discretegap2}. Although not directly apparent in Eqs.~\eqref{eq:adimensional_action_aux_fields}-\eqref{eq:action_momentum_aux}, the action $\mathsf{s}_E[\tilde{\Sigma},\tilde{\Pi}_2]$  is invariant under the transformation $\tilde{\Pi}_2\to-\tilde{\Pi}_2$, which can be seen by making $\boldsymbol{k}_2\to-\boldsymbol{k}_2$ in the momentum integrals, and thus amounts to inversion symmetry~\eqref{eq:inv_symm}. Using the  notation of the present section~\eqref{eq:eff_potential_parameters}, we could rewrite this action as 
\begin{equation}
\begin{split}
\mathsf{s}_E[\tilde{\Sigma},\tilde{\Pi}_2]\!&=\!\half\!\!\left(\mathsf{s}_E[\tilde{\Sigma},\tilde{\Pi}_2]+\mathsf{s}_E[\tilde{\Sigma},-\tilde{\Pi}_2]\right)\!\!=\!\frac{1}{\tilde{g}^2}\!\!\left(\tilde{\Sigma}^2+\tilde{\Pi}_2^2\right)\\
&+\half\int_{\boldsymbol{p}}\!\log\big[\!\!\left(p_+^2+\tilde{m}^2(\boldsymbol{k})\right)\!\!\left(p_-^2+\tilde{m}^2(\boldsymbol{k})\right)\!\!\big]
\end{split}
\end{equation}
where we have introduced the displaced momenta $p_{\pm}=(p_0,p_1,p_2\pm\tilde{\Pi}_2)$. Using the properties of the logarithm, one can then express the  effective potential with the radiative corrections of Eqs.~\eqref{eq:q_corrections_1} and~\eqref{eq:q_corrections_2} as
\beq
V_{\rm eff}(\tilde{\Pi}_2)=N(\mathsf{s}_E[\tilde{\Sigma},\tilde{\Pi}_2]-\mathsf{s}_E[\tilde{\Sigma},0])
\eeq
where we subtract $\mathsf{s}_E[\tilde{\Sigma},0]=\tilde{\Sigma}^2/\tilde{g}^2-\int_{\boldsymbol{p}}\!\log\left(p^2+\tilde{m}^2(\boldsymbol{k})\right)$. Since this last term does not depend on the symmetry-breaking order parameter $\tilde{\Pi}_2$, this subtraction  does not modify the position of the minima, and  the current diagrammatic derivation~\eqref{eq:q_corrections_1}-\eqref{eq:q_corrections_2} and  the effective action obtained by integrating out the fermions~\eqref{eq:adimensional_action_aux_fields}-\eqref{eq:action_momentum_aux} are thus consistent. The diagrammatic derivation, however, separates novel quantum corrections~\eqref{eq:q_corrections_2} from those appearing in   chiral-invariant QFTs~\eqref{eq:q_corrections_1}, and thus allows to identify new effects that can be brought up by these radiative corrections. As explored below in Sec.~\ref{sec:order}, an important consequence of these additional radiative corrections is both to displace critical lines,  and moreover to change the nature of the inversion-breaking phase transition in the neighbourhood of the line of symmetry $ma_1=-2$ from second to first order.

It is important to stress that the effective potential~\eqref{eq:full_eff_V} provides similar information to the  gap equations~\eqref{discretegap1}-\eqref{discretegap2}, while not being restricted to a non-zero value of $\tilde{\Pi}_2$. This will be crucial to explore the full phase diagram of the model. Let us now describe our numerical method to obtain the minima of the effective potentials, which can be explicitly written as
\begin{align}
\label{eq:eff_potential_discrete}
\begin{split}
\frac{V_{\rm eff}(\tilde{\Pi}_2)}{N}=\! \frac{\tilde{\Pi}_2^2}{2\tilde{g}^2} -\frac{1}{\tilde{Q}}\! \sum_{\boldsymbol{k} } \!\log\!\!\left[1+\frac{\tilde{\Pi}_2^2}{\tilde{m}^2(\boldsymbol{k})+ \sum_\alpha\!\!4\kappa_\alpha^2 \sin^2(k_\alpha a_\alpha)\phantom{\left.\right)^2\hspace{-2ex}}}\right]\, \\
-\frac{1}{2\tilde{Q}} \sum_{\boldsymbol{k}} \log\!\!\left[1-\frac{16 \kappa_2^2 \sin^2(k_2a_2)\tilde{\Pi}_2^2}{\left(\tilde{m}^2(\boldsymbol{k})+ \sum_\alpha\!\! 4\kappa_\alpha^2 \sin^2(k_\alpha a_\alpha)+\tilde{\Pi}_2^2\right)^2}\right]\!\!.
\end{split}
\end{align}
As occurred for the   gap equation~\eqref{discretegap2}, the effective potential only depends on the combination of parameters $M=\tilde{m}+\tilde{\Sigma}$, as one readily finds by inspecting Eq.~\eqref{eq:eff_potential_parameters}. We can thus fix the value of $M$, evaluate the mode sums of Eq.~\eqref{eq:eff_potential_discrete} for a given lattice volume $\tilde{Q}$, and use a numerical minimization algorithm to find the value of the Lorentz-breaking condensate $\tilde{\Pi}_2$. 
In figure~\ref{fig:potential}, we represent   the effective potential as a function of the Lorentz-breaking condensate  for  three different values of the adimensional coupling strength  $\tilde{g}^2$, after fixing  ${M=-1.8}$. Since the potential is symmetric under $\tilde{\Pi}_2\to-\tilde{\Pi}_2$, we only represent it for positive-valued condensates. In all three figures, we represent  with a dashed line the quantum corrections in Eqs.~\eqref{eq:q_corrections_1}-\eqref{eq:q_corrections_2}, {which are labelled as $\tilde{g}^{-2}=0$.} When adding these radiative corrections to the classical potential, we observe three different regimes. In Fig.~\ref{fig:potential} {\bf (a)}, we observe that the effective potential has single minimum at $\tilde{\Pi}_2=0$, such that the discrete inversion symmetry is preserved. As we keep on increasing the coupling $\tilde{g}^2$, the curvature of the potential changes in  Fig.~\ref{fig:potential} {\bf (b)},  until the potential develops a double-well structure with a minimum at a non-zero value of the condensate,  corresponding the spontaneous  breakdown of inversion symmetry (see Fig.~\ref{fig:potential} {\bf (c)}). We can thus identify a critical interaction $\tilde{g}^2_{\rm c}$, which defines a critical point and indicates the location of the phase boundary. 

\begin{figure*}[t]
		\includegraphics[width=0.8\textwidth]{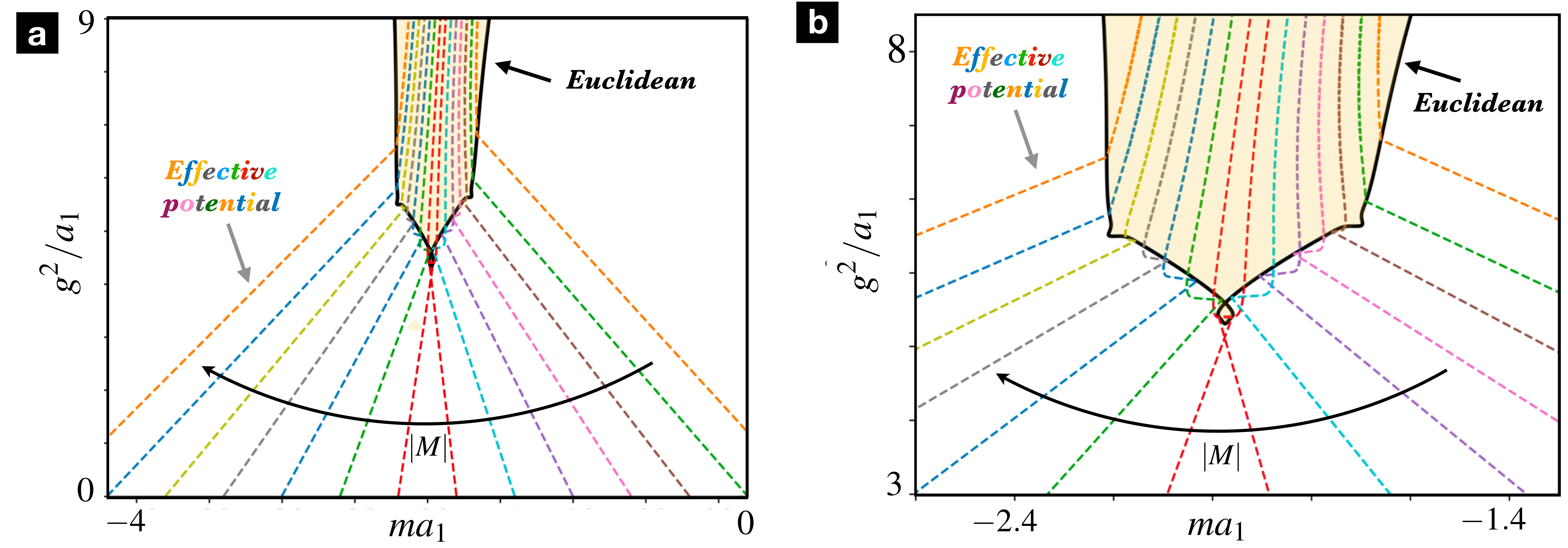}
	\caption{{\bf Phase diagram predicted by the large-$N$ effective potential}: {\bf (a)} We represent in dashed colored lines the constant-$M$ trajectories describing the bare parameters $(ma_1,g^2/a_1)$ where one obtains a minimum of the potential~\eqref{eq:eff_potential_discrete} for a fixed value of $M=\tilde{m}+\tilde{\Sigma}$, but varying bare parameters. In a black solid line, the results based on the  discrete-time gap equations~\eqref{discretegap1}-\eqref{discretegap2} for the same temporal anisotropy $\xi_1=64$ are presented. {\bf (b)} Zoom into the trajectories shown in {\bf (a)},  which shows the kinks of the dashed colored lines precisely  at the critical line where the inversion-breaking condensate forms. For this spatial isotropic case $\xi_2=1$, this condensate corresponds to the orbital FM$_\phi$ phase. }
	\label{fig:trajectories_M}
\end{figure*}

According to our previous discussion, this critical point must coincide with the corresponding point on the phase boundary predicted by our numerical solution of the gap equations~\eqref{discretegap1}-\eqref{discretegap2}. In order to compare both methods, we need to extract the  bare mass $\tilde{m}$, which  can be determined by plugging the value of the minimum $ \tilde{\Pi}_{2}$ into Eq.~\eqref{discretegap1}, and then solving it numerically to find the value of the scalar condensate $\tilde{\Sigma}$, after which one simply subtracts   $\tilde{m}={M}-\tilde{\Sigma}$. 
  By repeating this algorithm for different couplings $\tilde{g}^2$ one can draw trajectories of constant $M$ in the $(ma_1, g^2/a_1)$-plane, remembering that the dimensionless couplings have to be rescaled and renormalized according to Eq.~\eqref{eq:rescalings_parameters}. Each of the trajectories can be labelled by the input parameter $M$, which corresponds physically to the bare mass $\tilde{m}$ at zero coupling $\tilde{g}^2$=0, since the scalar condensate can only become non-zero as one switches on the interactions. For $a_1>a_2$, the numerical routine is similar, but we need to  exchange $\tilde{\Pi}_2,p_2\tilde{\Pi}_2\to\tilde{\Pi}_1,p_1\tilde{\Pi}_1$ in Eqs.~\eqref{eq:q_corrections_1}-\eqref{eq:q_corrections_2}. For $a_1=a_2$, both corrections give the same result, signalling that the symmetry-breaking can occur for any linear combination $\cos\phi\tilde{\Pi}_1+\sin\phi\tilde{\Pi}_2$.

  In Fig.~\ref{fig:trajectories_M} {\bf (a)}, some of these trajectories are plotted for the isotropic case $a_1=a_2$, using dashed lines with different colors for the different values of $M$. In this figure, we also plot with a solid black line the critical points predicted by the solution of the gap equations~\eqref{discretegap1}-\eqref{discretegap2} for the same volume $\tilde{Q}$. The structure of the trajectories is clear, they are straight lines until hitting the critical line, where they bend backwards before entering the symmetry-broken phase. By zooming into the critical region, as presented in figure~\ref{fig:trajectories_M} {\bf (b)}, this bending becomes clearer, and one sees that  it is actually preceded by a kink that coincides exactly with the parameters where $V_{\rm eff}(\tilde{\Pi}_2)$ develops a non-zero condensate $\Pi_2>0$. By inspecting Eq.~\eqref{discretegap1}  in light of the function~\eqref{eq:action_momentum_aux}, it is clear that the spontaneous formation of a non-zero condensate   $\tilde{\Pi}_2$ results  in a decreased value of the $|\tilde{\Sigma}|$, such that $\tilde{m}$ changes abruptly from increasing(decreasing) with $\tilde{g}^2$ into decreasing (increasing), depending on the side of the symmetry axis at $\tilde{m}=-2$ in which the trajectory resides. Therefore, the kink can be directly associated with the spontaneous  breakdown of inversion symmetry, which allows us to identify the critical line delimiting the $\tilde{\Pi}_2$ condensate. As shown in this figure,  this critical line agrees perfectly with the solutions predicted by the gap equations, which serves as a benchmark of our numerical method for the effective potential. Once the validity has been demonstrated, we can now exploit the effective potential to get insights in the symmetry-preserved region, going beyond the information that can be extracted from the gap equations.
  
\begin{figure*}[t]
	\centering
	\includegraphics[width=0.8\textwidth]{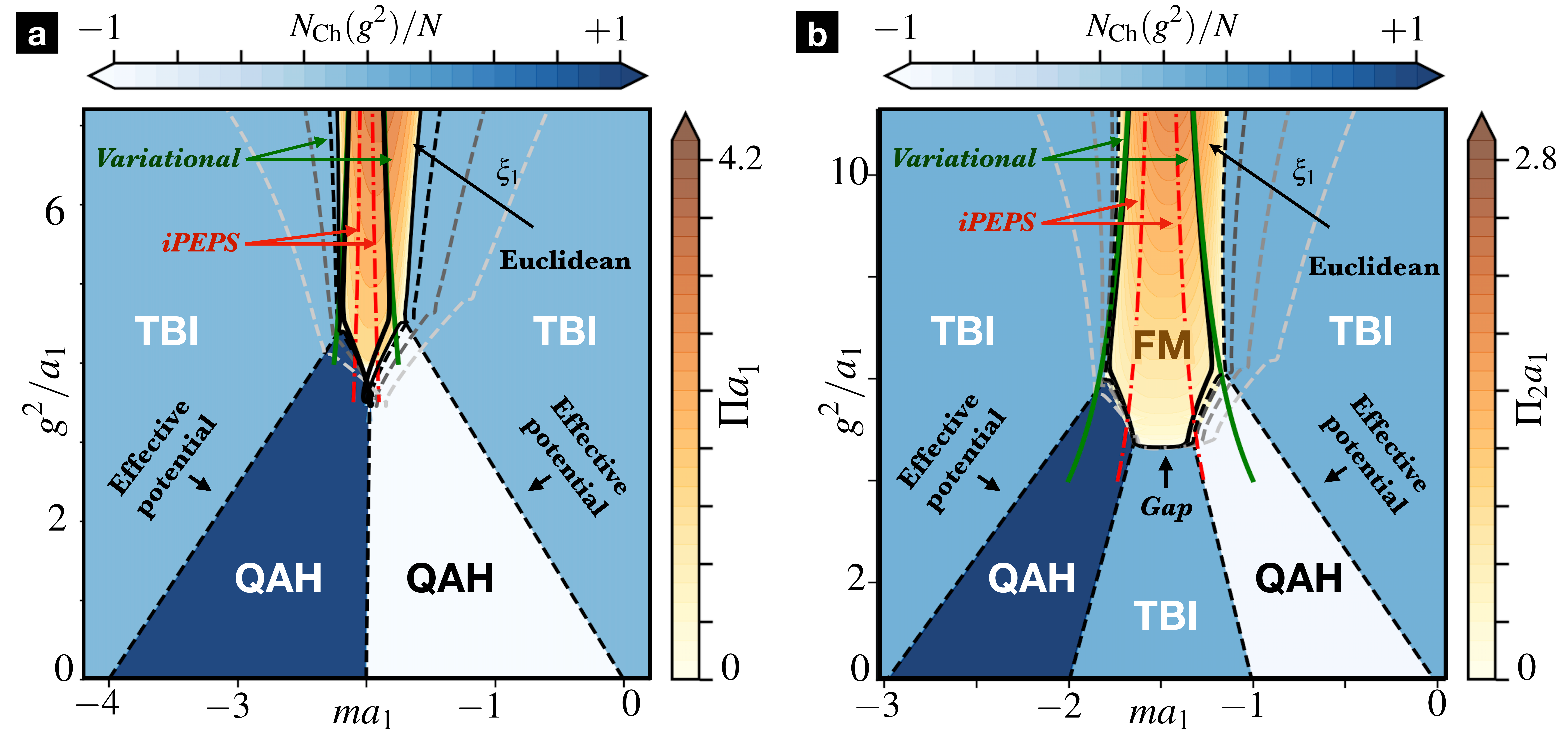}
	\caption{{\bf Full phase diagram with correlated large-$N$ QAH phases:} We represent, on a blue scale, a contour plot of the  topological invariant in Eq~\eqref{topoconst} that allows to locate the correlated QAH phase, and the phase transition to topologically-trivial phases. We also include for comparison,  all of the results of the previous sections. The large-$N$ calculations of the continuum-time gap equations~\eqref{firstequation} and~\eqref{secondequation}  obtained via  a Hamiltonian field theory for a single channel, are represented by a  solid black line. The discrete-time gap equations~\eqref{discretegap1}-\eqref{discretegap2} obtained via a Euclidean field theory are represented by dashed lines in a grey scale of increasing temporal anisotropy. The orange-scale contour plot of the symmetry-broken region is obtained by varying the non-zero value of the condensate in those gap equations.  The green line shows the critical points~\eqref{criticalcompass} predicted by the variational mean-field ansatz  in the compass model that emerges at strong couplings. The red dashed-dotted lines correspond to the critical points obtained by the iPEPS variational ansatz of the same compass model. }
	\label{phasediagram}
\end{figure*}

\subsection{Scalar condensate and  topological invariants}
\label{sec:self_energy_th}

In the   subsection above, we have seen how the effective potential can be used to determine the Lorentz-breaking condensate  which, as argued previously, is the large-$N$ version of the orbital ferromagnet FM$_\phi$ in the isotropic case $a_1=a_2$. In contrast to the gap equations, however, we can also explore regions of parameter space where this condensate is zero. We have also noted that the effective action, obtained by the spacetime integral of the effective potential, serves as the generating functional of any $n$-point proper vertex functions~\cite{Peskin:1995ev}. In this section, we focus on the $2$-point function in momentum space $\Gamma^{(2)}(\boldsymbol{p})$, which contain information about all the intermediate scattering processes in which particle-antiparticle pairs are virtually created, and can be expressed by a sum  of  1PI diagrams  leading to the self energy $\tilde{\Sigma}_s({p})$. From the perspective of  Euclidean QFTs of Dirac fermions in the continuum, the  $2$-point proper vertex is the inverse of the dressed Euclidean propagator $\Gamma^{(2)}\!(p)=\ii\slashed{p}+{m}+\tilde{\Sigma}_{ s}(p)$. Within the realm of the large-$N$ approximation,  the self energy is easily expressed in terms of the condensates as
\beq
\label{eq:self_energy_Dirac field}
{\Sigma}_s(p)=\big({\Sigma}+{{\gamma}}^1{\Pi}_1+{\gamma}^2{\Pi}^2\big)\delta^3(p),
\eeq
where the homogeneity of the condensates is responsible for the momentum independence of the self energy. 

In order to make a connection of these concepts with the topological characterization of the QAH phase, we need to rephrase this discussion in the context of condensed matter, where one defines the Euclidean-time single-particle Green's function in terms of the creation-annihilation operators  $G({x_1}-{x_2}) = \braket{\mathcal{T} \{\Psi^\dagger({x_1})\Psi({x_2})\}}$ for two spacetime points $x_1=(\ii t_1,\boldsymbol{x}_1)$, and $x_2=(\ii t_2,\boldsymbol{x}_2)$. In a translationally-invariant setting, one can perform  Fourier and Matsubara transforms to spatial-momentum $\boldsymbol{k}=(k_1,k_2)$ and frequency $k_0=\omega$ representations~\cite{negele_orland_2019,coleman_2015}, such that  the inverse  Green’s function can be expressed as 
\beq
G^{-1} (\ii k_0 ,\boldsymbol{k}) = \ii k_0 -h_{\boldsymbol{k}} + \Sigma_s(\ii k_0,\boldsymbol{k}),
\eeq
 where $h_{\boldsymbol{k}}$ is the single-particle Hamiltonian, like Eq.~\eqref{freehamiltonian} in our case, and $\Sigma_s(\ii k_0,\boldsymbol{k})$ is also called the self energy. Within the large-$N$ approximation, it can be readily connected to Eq.~\eqref{eq:self_energy_Dirac field} by simple algebra
\beq
\label{eq:self_energy_cond_mat}
{\Sigma}_s(\ii k_0,\boldsymbol{k})=\mathbb{I}_N\otimes\Big({\gamma}^0{\Sigma}+{\gamma}^0{\gamma}^1{\Pi}_1+{\gamma}^0{\gamma}^2{\Pi}_2\Big)\delta^3(k),
\eeq
which results from the different definition of the propagator (Green's function) in terms of the adjoint (creation) operator.

As advanced in the introduction,  topological invariants such as the Chern numbers~\eqref{eq:chern_winding} can be  generalized to the many-body case by means of these Green's functions~\cite{doi:10.1143/JPSJ.75.123601,PhysRevLett.122.146601}. As discussed in~\cite{PhysRevLett.105.256803,PhysRevB.83.085426,PhysRevX.2.031008,PhysRevB.86.165116,Wang_2013}, the static part of the self-energy $\Sigma_s(0,\boldsymbol{k})$ plays a key role in this topological characterization, as it offers a  practical route for the calculation of topological invariants beyond the non-interacting limit. Focusing on the inversion-symmetric phase, where we only have a non-zero value of the scalar condensate,  we can define the so-called topological Hamiltonian that contains these static contributions 
\beq 
h_t=h_{\boldsymbol{k}}(m)+\Sigma_s(0,\boldsymbol{k})=h_{\boldsymbol{k}}(m)+ {\gamma}^0\Sigma. 
\eeq
It is then a simple matter to realize that  the calculation of the Chern number in Eq.~\eqref{eq:chern_winding}  can be repeated with the bare mass being renormalized by the static self energy, i.e. $m+\Sigma$, which changes the mapping from the torus onto the unit sphere in Eq.~\eqref{freehamiltonian_d_vector} by $
\boldsymbol{d}_{\boldsymbol{k}}(m)\to\boldsymbol{d}_{\boldsymbol{k}}(m)+\Sigma\boldsymbol{e}_3$. Following the same calculation, the topological invariant in the presence of interactions  is given by
\begin{equation}
N_{\rm Ch}(g^2)=\frac{N}{2}\sum_{\boldsymbol{n}_d}(-1)^{(n_{d,1}+n_{d,2})}{\rm sign}(M_{n_d}) \label{topoconst},
\end{equation}
where we have introduced the masses of the spatial doublers renormalized by the scalar condensate
\begin{equation}
M_{\boldsymbol{n}_d}=m+\Sigma+\frac{2n_{d,1}}{a_1}+\frac{2n_{d,2}}{a_2},
\end{equation}
and we recall that 
$\boldsymbol{n}_d=(n_{d,1},n_{d,2}) \in \{0,1\}\times \{0,1\}$.

Note that on a Euclidean lattice,  the discretization of the time axis introduces the spurious time doublers previously discussed. However,  if we take the limit   $a_0\rightarrow 0$ with the appropriate rescaling and renormalization of the bare couplings~\eqref{eq:rescalings_parameters}, the model reduces to a two-dimensional large-$N$ Chern insulator, and these spurious time doublers have no influence on the long wavelength physics. In particular, given the discussion of the previous section, these time doublers have a mass with a contribution on the order of $1/a_0$, which becomes very large in the time-continuum limit and is responsible for the fact that the Wilson masses of these spurious time doublers  always carry the same sign. As a consequence, not only do they lie at very high energies and thus not appear in the long wavelength limit, but also their contribution to the topological invariant vanishes exactly since we have an even number of them, i.e. 4, giving cancelling contributions to the Chern number.

With these formulas at hand, the previous numerical algorithm that calculates the trajectories displayed in Fig.~\ref{fig:trajectories_M}
 is very useful, as the input parameter is the renormalized mass $M=\tilde{m}+\tilde{\Sigma}$. This means that one can assign a constant topological invariant to each of these trajectories using Eq.~(\ref{topoconst}), and delimit the phase boundaries separating correlated QAH phases from  trivial band insulators within the symmetry-preserved region. In Fig.~\ref{phasediagram}, all of the predictions for the phase diagram are presented together for temporal anisotropy $\xi_1=64$, and two spatial anisotropy ratios, { namely $\xi_2=1$ for Fig.~\ref{phasediagram} {\bf (a)} and $\xi_2=0.5$ for Fig.~\ref{phasediagram} {\bf (b)}.} These two anisotropy values correspond to the shaded planes of parameter space depicted in Fig.~\eqref{fig:chern}. The blue-scale contour plot was created by generating a dense set of trajectories like those  presented in Fig.~\ref{fig:trajectories_M} for different values of $M$ along the $x$-axis, and then calculating their topological invariant with Eq.~(\ref{topoconst}).  
 
 Let us start by focusing on  Fig.~\ref{phasediagram} {\bf (a)},  recalling  that the topological invariant of the non-interacting QAH effect~\eqref{eq:critical_lines_chern} changes when the  mass of an odd number of  spatial doublers is inverted, which occurs   at ${m\,a_1}\in\{0,-2,-4\}$ in the isotropic case $a_1=a_2$. For the large temporal anisotropies used,  the spurious  time doublers with masses proportional to $1/a_0$ would lead to additional phase transitions for very large negative values of the bare mass, eventually disappearing completely from the phase diagram when $\xi_1\to\infty$. This reflects Wilson's idea of turning the doublers into very heavy fermions that do not contribute significantly to the relevant physics. Let us note that, since $M$ collapses to $m$ in the non-interacting limit $g^2=0$, as the scalar condensate vanishes $\Sigma=0$, the trajectories for input values $Ma_1\in\{0,-2,-4\}$ (black dashed lines in the figure) turn out  to be the ones that separate the correlated topological phases with $N_{\rm Ch}(g^2)=\pm N$ from the trivial band insulators with a vanishing Chern number as one increases the interactions $g^2$. Remarkably, we find that these critical lines  touch exactly the corners of the  solid black line, which represents our numerical solution of the gap equations obtained via the   Hamiltonian formalism. For completeness, we also represent  with dashed grey-scale lines the results obtained by solving the gap equations in an Euclidean lattice, as one increases the temporal anisotropy towards $\xi_1=64$. According to these results, the lines of topological phase transitions meet the line of the symmetry-breaking phase transition, above which the $\pi$ condensate forms and inversion symmetry is spontaneously broken.   Regarding the latter, an orange-scale  contour plot was added, which was created by solving the gap equations for different non-zero values of $\Pi_2$, bearing in mind that the $\pi$ condensate can actually be any linear combination of the $\Pi_1$ and $\Pi_2$ fields in this isotropic limit. The resulting lines of constant values of this condensate retreat to the interior of the phase boundary with increasing $\Pi_2$, which could be found out by analyzing the scaling of the condensation of the $\Pi_2$-field. Let us finally note that the green solid line depict the critical lines obtained by the variational mean-field calculation of the compass model~\eqref{criticalcompass}, whereas the  red dashed-dotted line is  obtained by solving the compass model using the iPEPS variational algorithm for $J_x=J_y$. 
 
  In Fig.~\ref{phasediagram} {\bf (b)}, which corresponds to the spatial anisotropy $\xi_2=0.5$, a trivial phase arises separating the two correlated QAH phases that have an underlying large-$N$ Chern insulator. The trajectories that correspond to these topological phase transitions, represented again using dashed black lines, are obtained by setting ${M}a_1\in\{0,-1,-2,-3\}$, which connect  to critical points for the  the non-interacting Chern number~\eqref{eq:critical_lines_chern}   for ${m}a_1\in\{0,-1,-2,-3\}$ as $g^2\to0$, as displayed in Fig.~\ref{fig:chern}.  The behaviour is very similar to that found in the isotropic case with the novelty that a trivial band insulator now separates the two correlated QAH phases all the way up to the region where the Lorentz-breaking condensate appears.   Once again, the critical lines that mark these topological phase transitions extend as straight lines as the interactions $g^2$ are increased, until they meet the symmetry-breaking critical line precisely at four symmetric corners. In this figure, we also depicted in a red dashed-dotted line the results obtained by solving the compass model using the iPEPS variational algorithm  for $J_x=J_y/4$. We see that the region of a non-zero Lorentz-breaking condensate shrinks with respect to the mean-field-type methods, which is a general feature of the latter since the role of super-exchange interactions is typically overestimated, leading to larger regions with magnetic long-range orderings than those predicted by other methods that can better with  correlations.
  The unification of the results exhibits a coherent picture showing that the  methods used have been consistent, and can be applied to other similar models.

\subsection{First- and second-order phase transitions}
\label{sec:order}

 \begin{figure}[t]
	\centering
		\centering
		\includegraphics[width=0.45\textwidth]{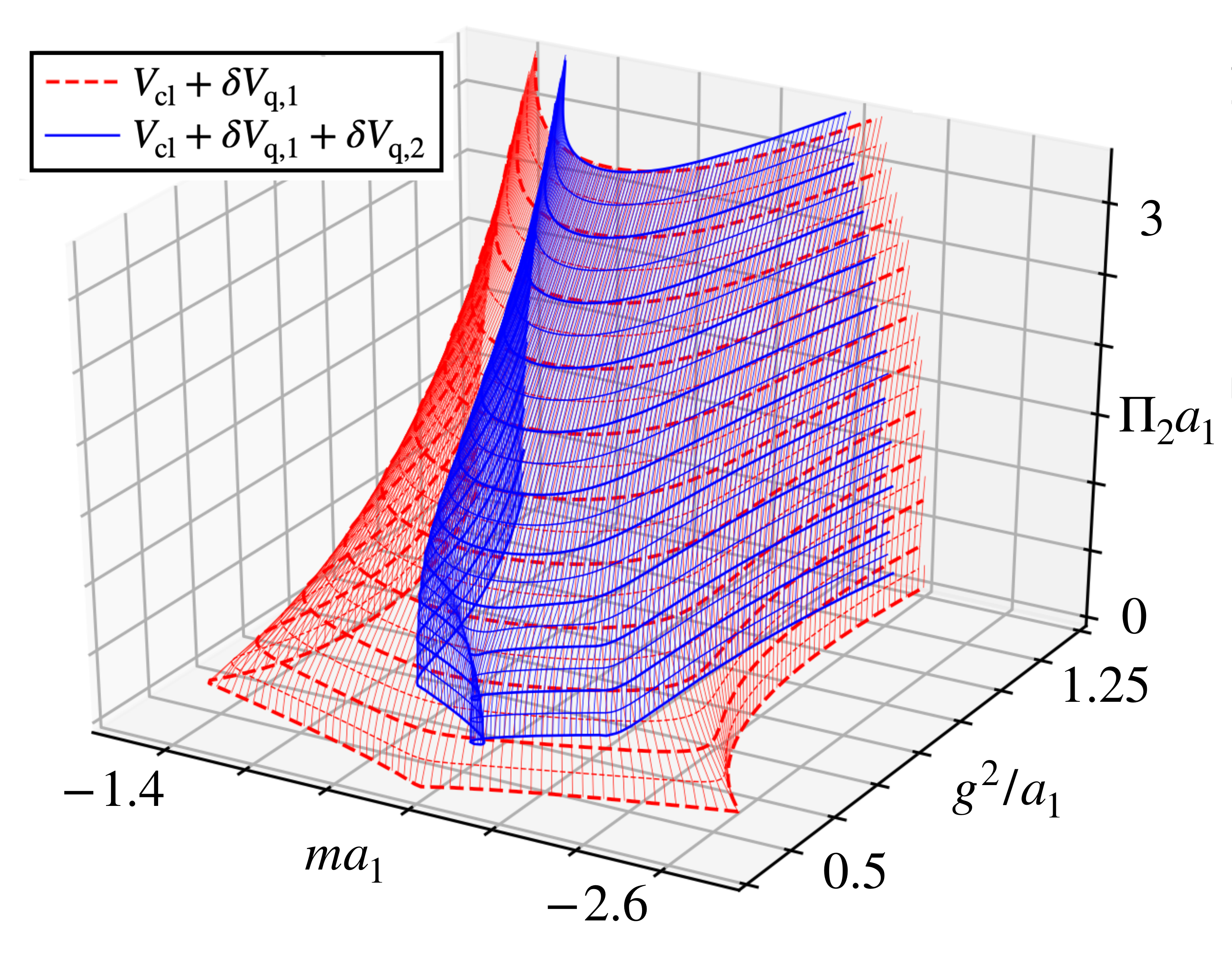}
		\caption{{\bf Comparison of Lorentz-breaking condensates:} The red (blue) surfaces represent the  Lorentz-breaking condensates as a function of $(ma_1,g^2/a_1)$ for $\xi_2=1$, which are obtained by solving the gap equations corresponding to the effective potential with radiative corrections in Eq.~\eqref{eq:q_corrections_1} (Eqs.~\eqref{eq:q_corrections_1}-\eqref{eq:q_corrections_2}).}
		\label{fig:comaprison_radiative_corrections}
		\end{figure}

As advanced at the end of Sec.~\ref{sec:H_ft_gap_eqs}, the small lobe containing  a Lorentz-breaking condensate in the isotropic limit $a_1=a_2$  (see Figs.~\ref{continuumboth} {\bf (a)} and~\ref{discrete} {\bf (a)}) actually  persists for weak spatial anisotropies $\xi_2\approx 1$.  As argued in this section, thanks to the formulation based on the effective potential, we can identify  the  additional radiative corrections~\eqref{eq:q_corrections_2} as the underlying source of this lobe structure. In Fig.~\ref{fig:comaprison_radiative_corrections}, we represent various non-zero values of the $\Pi_2$ condensate as a function of the bare parameters $ma_1,g^2/a_1$. The Lorentz-breaking condensate in red is obtained by solving the gap equations that correspond to an effective potential that only considers the radiative corrections common to chiral-invariant theories~\eqref{eq:q_corrections_1}. The blue surface represents the $\Pi_2$ condensate when  the new radiative corrections~\eqref{eq:q_corrections_2} are also considered. By comparing both plots, one readily sees that the loop-structure disappears if  these novel  radiative corrections~\eqref{eq:q_corrections_2} are not accounted for. Moreover, as one increases the value of the Lorentz-breaking  $\Pi_2$ condensate, the symmetry-broken phase in blue extends to a larger area in the ($ma_1,g^2/a_1)$ sections until the lobe eventually disappears. In the following, we present a more in-depth study of this re-entrant region via the effective potential, and show that the order of the phase transition can change from second to first order.

\begin{figure}[t]
	\centering
		\centering
		\includegraphics[width=0.45\textwidth]{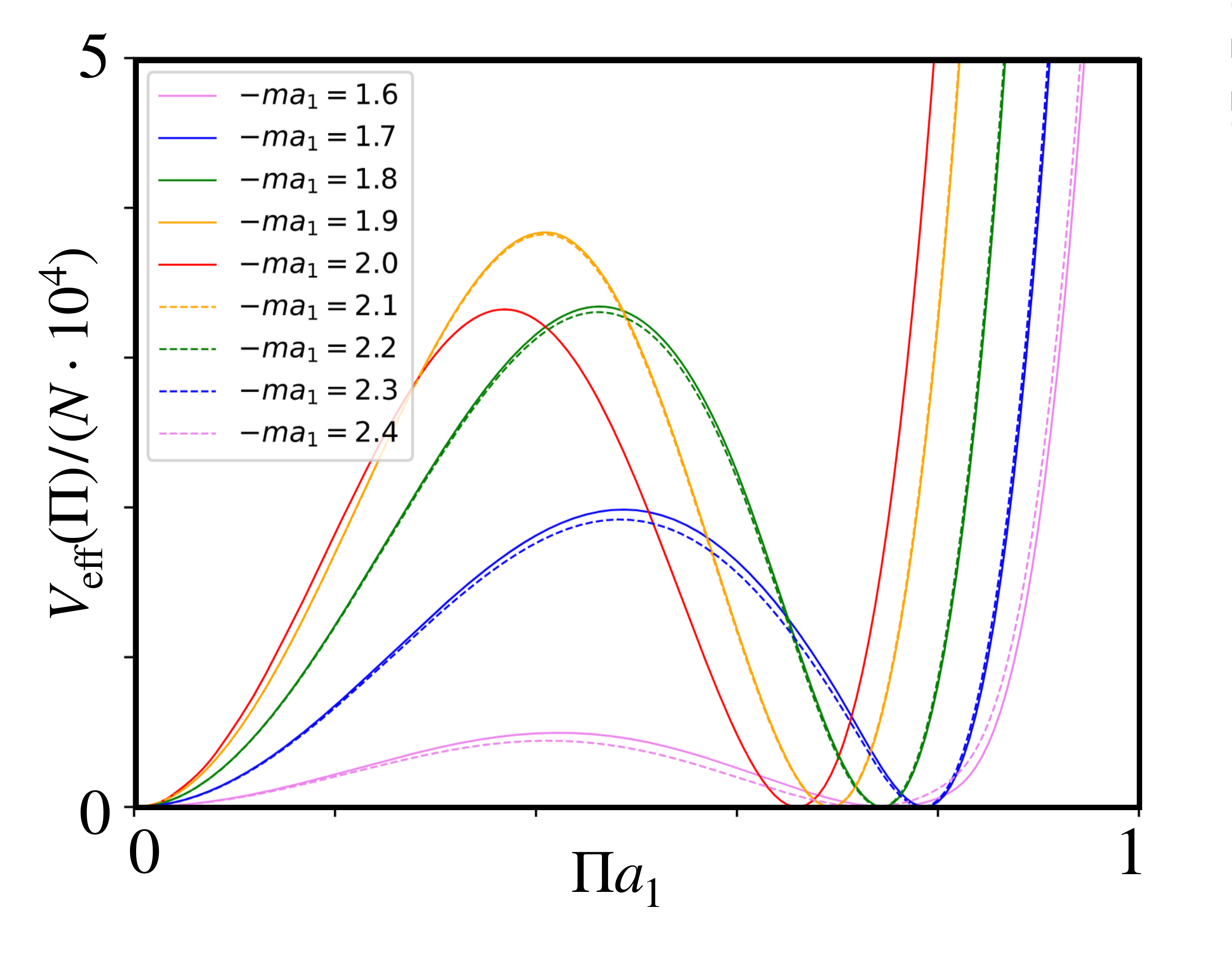}
		\caption{{\bf Critical effective potential near the line of symmetry:} $V_{\rm eff}(\Pi)$ is plotted for $\xi_1=64$, $\xi_2=1$ for selected values $1.5<-ma_1<2.5$ corresponding to constant-$M$ trajectories intersecting the lobe shown in Fig.~\ref{fig:trajectories_M}. Colours have been chosen to emphasise the symmetry under $ma_1\mapsto-ma_1-4$ expected to become exact as $\xi_1\to\infty$}
		\label{fig:Vcrits}
		\end{figure}

Let us now discuss how the effective potential can yield information on  the nature of the inversion-breaking phase transition, which we will illustrate for the  isotropic case $\xi_2=1$ so that the broken phase is labelled FM$_\phi$, and the condensate for any particular $\phi$ will be labelled as $\Pi$. Fig.~\ref{fig:Vcrits} plots $V_{\rm eff}(\Pi)$ for mass values $1.5<-ma_1<2.5$, corresponding to  trajectories similar to those shown in Fig.~\ref{fig:trajectories_M}, but focusing only on those that intersect the lobe of the gap-equation solution. In each case, the coupling  is tuned to the critical value $g_c^2(m)$ yielding two degenerate minima of the effective potential. As shown in Fig.~\ref{fig:Vcrits}, at this  point, groundstates with two distinct condensates $\Pi=0$ and $\Pi\not=0$ co-exist. This contrasts the behaviour presented in Fig.~\ref{fig:potential} where, recalling the symmetry $\Pi\to-\Pi$, the effective potential changes from a single- to a double-well structure. This is the standard scenario for a second-order phase transition, whereby condensates with a different value of $|\Pi|$ never co-exist. On the contrary, within the current range of bare masses $1.5<-ma_1<2.5$, the inversion-breaking transition is first-order. In this figure, we use solid and dashed lines with the same colours  to emphasise that, to very good approximation, the effective potentials calculated for $ma_1$ and $-ma_1-4$ are equal; this symmetry should become exact in the time-continuum limit $\xi_1\to\infty$. It is also apparent that the strength of the first-order transition defined by the barrier height separating the two minima at $\Pi=0$ and $\Pi\not=0$, which corresponds physically to the interface tension, initially grows as $m$ approaches the line of symmetry $ma_1=-2$, where $\Sigma$ vanishes from either direction, but then dips so that the barrier height at the symmetric point actually lies in a local minimum. 

As discussed above, the constant-$M$ trajectories shown in Fig.~\ref{fig:trajectories_M} exhibit a sharp kink precisely at the critical $g_c^2(m)$. However, a closer inspection of the figure reveals that this kink actually occurs within the symmetric phase, and thus {\em before\/} the phase boundary predicted by the gap equations \eqref{discretegap1}-\eqref{discretegap2} is reached. In the neighbourhood of the lobe, therefore, the gap equation is not finding the true transition, but rather tracing the locus of a local minimum of $V_{\rm eff}$. By contrast, in the regions $0<-ma_1<1.5$, $2.5<-ma_1<4$, $V_{\rm eff}(\Pi)$
has a unique minimum for all $g^2$. Accordingly,   there is no discontinuity in the value of the condensate  $\Pi(g^2)\vert_{V_{\rm eff}=V_{min}}$, which is consistent with a continuous second-order phase transition. In this case, the constant-$M$ trajectory kinks of Fig.~\ref{fig:trajectories_M} lie precisely on top of the phase boundary predicted by  the gap equations. Following the kink, the trajectory apparently remains for a while in the symmetric phase before curving upwards; however it can be shown that in this region the surface generated by contours of constant $\Pi$ in the broken phase actually curves back to overhang the symmetric phase (this can just be discerned in Fig.~\ref{fig:comaprison_radiative_corrections}). Every point in the phase diagram lying beneath the overhang is therefore intersected by two constant-$M$ trajectories, one corresponding to $\Pi=0$ and the other to $\Pi\not=0$, in apparent contradiction with the predicted second-order nature of the transition, It will require a more refined calculation of the full effective potential $V_{\rm eff}(\Sigma,\Pi)$, i.e. including loops with scalar auxiliary legs, to resolve this ambiguity; indeed, such a calculation will inevitably be needed to examine the nature of the topological phase transitions between the QAH and TBI phases shown in Fig.~\ref{phasediagram}.

\section{\bf Spin-orbit-coupled Fermi gases}\label{optical}

In the previous sections, we have presented a thorough study of
the phase diagram of the four-Fermi-Wilson model~\eqref{eq:total_H}, combining various tools developed by the condensed-matter, high-energy physics and quantum-information communities. We have seen that  large-$N$ methods yield a powerful tool to identify how the QAH phases are modified by interactions, and to understand the nature of the topological and symmetry-breaking phase transitions. By focusing on the strong-coupling limit, we have also shown that the effective super-exchange interactions  leading to a quantum compass model~\eqref{eq:compass} can yield more accurate estimates of the position of the critical lines and, eventually, the corresponding scaling of the underlying strongly-coupled fixed point. This follows from our comparison of  the large-$N$ predictions with two different variational methods for the compass model,  which shows a clear deviation of the critical lines using the more-accurate iPEPS algorithm. Thus, it would  be  interesting if future work could apply this method to the full fermionic model, rather than the effective compass model, in order to explore  arbitrary couplings. Another promising approach in this direction would be to use the discrete-time formulation based on Euclidean LFT~\eqref{eq:free_action_euclidean}-\eqref{Euclideancubeinteraction} with dimensionless fields~\eqref{eq:diemsnionless_fields}, in combination with Monte Carlo sampling techniques. In this way, one may evaluate thermodynamic observables based on the partition function of the model~\eqref{partition} beyond the large-$N$ limit. In this section, we discuss yet another alternative, that of quantum simulations (QSs)~\cite{Feynman_1982,Cirac2012}, where one exploits quantum-mechanical hardware to simulate a specific quantum many-body problem. We emphasise that these QSs have the potential of overcoming some of the limitations of the above alternatives to  large-$N$ methods, as they could probe real-time dynamics regardless of entanglement growth, and would not be limited by any sign problem as one explores finite fermion densities.

In this section, we focus on QSs based on ultracold atoms in optical lattices~\cite{Bloch_2008}. We present a detailed scheme for the QS of the four-Fermi-Wilson model using the so-called Raman optical lattices~\cite{PhysRevLett.112.086401,PhysRevLett.113.059901,Wu83,PhysRevLett.121.150401,Songeaao4748,liang2021realization}. These quantum simulators can be considered as  Fermi gases with a specific synthetic spin-orbit coupling~\cite{Galitski2013,zhai_2015,book_soc}, mimicking the  coupling of the intrinsic angular momentum of
the electron with its own motion~\cite{THOMAS1926} in the solid state~\cite{Bychkov_1984,PhysRev.100.580}. Spin-orbit coupling has turned out to be a source of important recent developments in condensed matter, as it underlies the experimental discovery~\cite{Konig766,Hsieh2008} of a new mechanism for the ordering of matter~\cite{PhysRevLett.95.146802,PhysRevLett.95.226801,Bernevig1757} in topological insulators and superconductors~\cite{RevModPhys.82.3045,RevModPhys.83.1057,classification_spt}. Given the special role of Chern insulators and the QAH effect within these topological phases, it does not come as a surprise that spin-orbit coupling is somehow disguised in our four-Fermi-Wilson model~\eqref{eq:total_H}. Additionally, given that the spin-orbit coupling is directly accounted for by the Dirac equation~\cite{10.2307/94981,10.2307/95359} and, ultimately,  by quantum electrodynamics~\cite{10.2307/j.ctv10crg18}, it is natural that our discretization of a QFT of self-interacting Dirac fermions is also connected to spin-orbit coupling.  Let us now discuss  this connection in detail for our representation of the Clifford algebra~\eqref{eq:gammas}.

Note that in $(2+1)$
spacetime dimensions, one can define rotations $R$  of angle $\theta$ around the normal  vector of the spatial plane, which are generated by
\beq
\boldsymbol{x}\to R\boldsymbol{x}=\ee^{\theta\mathsf{M}}\boldsymbol{x},\hspace{2ex} \mathsf{M}=\begin{pmatrix}
0 & 0 & 0\\
0 & 0 & -1\\
0 & 1 & 0
\end{pmatrix}.
\eeq
For our particular choice of gamma matrices~\eqref{eq:gammas}, the spinor representation of this rotation, which belongs to the Lorentz group $R\in SO(1,2)$, is generated by $\mathsf{S}=\frac{1}{4}[\gamma^1,\gamma^2]=-\frac{\ii}{2}\gamma^0$, such that the fields transform as
\beq
\label{eq:rotation}
\Psi(\boldsymbol{x})\to S(R)\Psi(R\boldsymbol{x})=\mathbb{I}_N\otimes\ee^{\ii\frac{\theta}{2}\gamma^0}\Psi(R\boldsymbol{x}).
\eeq
As noted in Sec.~\ref{sec:H_ft_gap_eqs}, the $\theta=\pi$ rotation   leads, up to an irrelevant phase, to the inversion symmetry defined in Eq~\eqref{eq:inv_symm}. Since this transformation can be generated infinitesimally, it does not correspond to parity symmetry.

In light of Eq.~\eqref{eq:rotation}, the two spinor components for each flavour $\psi_{\mathsf{f},1}(\boldsymbol{x}),\psi_{\mathsf{f},2}(\boldsymbol{x})$ in the original QFT~\eqref{continuum} can be identified with the spin up/down states of the fermions, respectively. From this perspective,   the  tunnellings of the naive discretization of the Hamiltonian field theory in
Eq.~\eqref{eq:naive}, namely ${\Psi}^\dagger(\boldsymbol{x})\frac{\ii\sigma^x}{2a_2}\Psi(\boldsymbol{x}+a_2\boldsymbol{e}_2)-{\Psi}^\dagger(\boldsymbol{x})\frac{\ii\sigma^y}{2a_1}\Psi(\boldsymbol{x}+a_1\boldsymbol{e}_1)$, are  understood as the    finite-difference discretization  of the so-called Rashba spin-orbit coupling $\boldsymbol{ e}_3\cdot(\boldsymbol{p}\wedge\boldsymbol{\sigma})=\ii\sigma^x\partial_y-\ii\sigma^y\partial_x$~\cite{Bychkov_1984}, when written in terms of fermionic creation/annihilation operators in second quantization. From this perspective, the  complete Wilson-type discretization in Eq.~\eqref{eq:total_H}  can be considered as  a Dirac-type spin-orbit coupling that generalises the aforementioned Rashba terms~\cite{book_soc}. 

Once the connection to spin-orbit coupling has been clarified, we can exploit the ideas underlying the cold-atom QSs of synthetic spin-orbit coupling in optical lattices and, in particular, we discuss how  the schemes in~\cite{PhysRevLett.112.086401,PhysRevLett.113.059901} can be adapted with minor modifications to realise our four-Fermi-Wilson model~\eqref{eq:total_H}, as briefly discussed in~\cite{ziegler2020correlated}. We also note that the recent experimental realization of the Qi-Wu-Zhang model  using Raman optical lattices~\cite{liang2021realization} is related to the non-interacting limit of our four-Fermi-Wilson model~\eqref{eq:total_H}.

\subsection{Raman optical lattices and spin-flip tunnellings}

\begin{figure*}[t]
	\centering
	\includegraphics[width=1\textwidth]{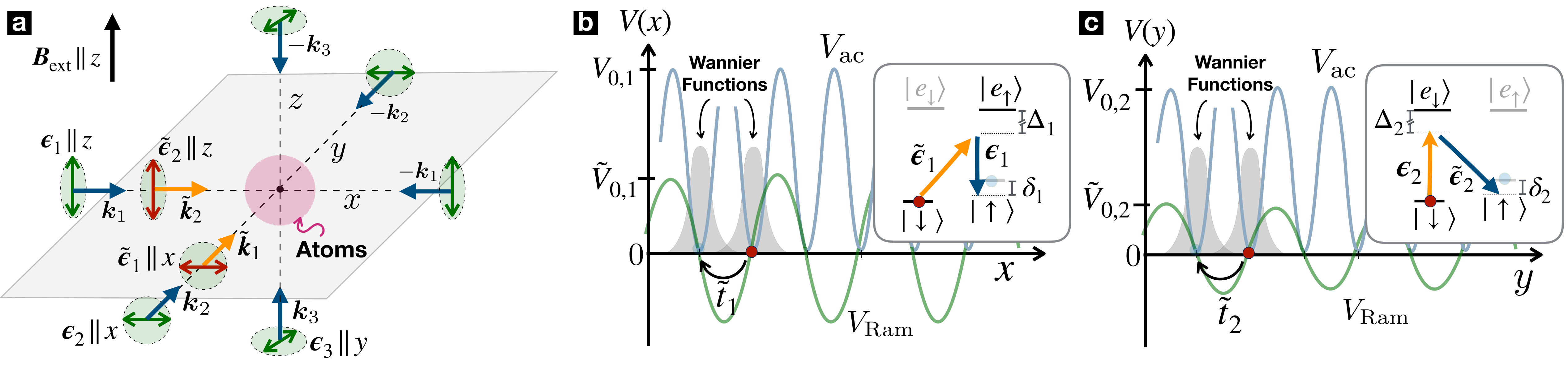}
	\caption{{\bf Spin-flip tunnelling in a Raman optical lattice:} {\bf (a)} Atom cloud subjected to a cubic optical lattice stemming from the three pairs of counter-propagating laser beams depicted by blue arrows, all of which have mutually orthogonal polarizations depicted by green arrows. In the $xy$ plane, we apply additional laser beams in a travelling-wave configuration (orange arrows) which can induce Raman transitions between two hyperfine groundstates $\ket{\uparrow}\leftrightarrow\ket{\downarrow}$  by absorbing a photon from this beam, and then subsequently emitting it in the standing wave, as depicted in the insets of {\bf (b)} and {\bf (c)}. In these two panels {\bf (b)} and {\bf (c)}, we depict the corresponding optical-lattice and Raman-lattice potentials, which lead to spin-flip tunnellings of strength $\tilde{t}_j$ due to the overlap of the neighbouring Wannier functions, mediated by the recoil of the laser beams.}
	\label{fig:SOC_scheme}
\end{figure*}

Let us consider a gas of fermionic atoms, such as the alkali-earth $^{87}$Sr gas, which are subjected to interfering laser beams that generate a cubic optical potential~\cite{GRIMM200095}, namely
\beq
\label{eq:Stark}
V_{\rm ac}(\boldsymbol{r})=\sum_{j}V_{0,j}\cos^2(k_{j}r_j).
\eeq
Here, $j\in\{1,2,3\}$ now labels all three spatial axes, $\boldsymbol{k}_j=k_{j}\boldsymbol{e}_j$ is the wave-vector of the laser beams with mutually-orthogonal polarizations $\boldsymbol{\epsilon}_j$ (see the blue and green  arrows in Fig.~\ref{fig:SOC_scheme}{\bf (a)}), which   interfere along the $j$-axis and lead to a standing-wave  pattern. We have also introduced   $V_{0,j}$ as the amplitude of the corresponding 
 ac-Stark shift experienced by the atoms in the groundstate manifold $\ket{^{1}{\rm S}_0,F,M}$, where $F=9/2$ is the total angular momentum, and $M\in\{-9/2,-7/2,\cdots,9/2\}$ are  the $10$ possible Zeeman sub-levels split by an additional external magnetic field $\boldsymbol{B}_{\rm ex}$ (we chose $z$ as the quantization axis in Fig.~\ref{fig:SOC_scheme}{\bf (a)}). In this work, it suffices to focus on two such hyperfine levels in order to define the spinor components $\ket{\sigma}\in\{\ket{\uparrow}=\ket{^{1}{\rm S}_0,F,M_\uparrow},\ket{\uparrow}=\ket{^{1}{\rm S}_0,F,M_{\downarrow}}\}$, choosing $M_\uparrow,M_{\downarrow}$ in a way that the electric-dipole selection rules allow  one to connect these levels via two-photon Raman transitions. These two states can be isolated from the remaining hyperfine levels in the groundstate manifold by exploiting a  Zeeman shift, or  an ac-Stark shift as in the case of  the  $^{87}$Sr gas~\cite{liang2021realization},   which must depend non-linearly on the magnetic  number $M_\sigma$.
 
 In order to induce the Raman transitions, one needs to drive  off-resonant couplings to  states within the excited-state manifold $\ket{e_\sigma}\in\ket{^{3}{\rm P}_1,F_\sigma',M_\sigma'}$, which requires using large detunings $\Delta_j$ to minimise spontaneous photon emission and the associated heating mechanisms, as depicted in the insets of Figs.~\ref{fig:SOC_scheme} {\bf (b)} and~{\bf(c)}. These drivings can be obtained by  two additional laser beams in a travelling-wave configuration,  selecting their wave-vectors $\tilde{\boldsymbol{k}}_j$ and polarizations $\tilde{\boldsymbol{\epsilon}}_j$ (see the orange  and red  arrows in Fig.~\ref{fig:SOC_scheme} {\bf (a)}), following the general ideas  of the schemes of synthetic spin-orbit coupling  to create a Raman potential~\cite{book_soc,PhysRevLett.112.086401,PhysRevLett.113.059901}. We note that  a standing-wave pattern for these  Raman beams can also be used which, in analogy to the optical-lattice potential~\eqref{eq:Stark},  can be obtained by exploiting retro-reflecting mirrors~\cite{PhysRevA.97.011605}, and underlies the recent implementation of synthetic spin-orbit coupling with the $^{87}$Sr gas~\cite{liang2021realization}. We would also like to remark that there have been other realizations of synthetic spin-orbit coupling that do not rely on  Raman lattice potentials, using alkaline-earth~\cite{Kolkowitz2017,Bromley2018} and    lanthanide~\cite{PhysRevX.6.031022,PhysRevA.94.061604,PhysRevLett.117.220401} atoms. In comparison to the alkalis, these atomic systems have  ultra-narrow optical transitions and a larger fine-structure splitting of the excited states, which can be exploited for the QS of synthetic spin-orbit coupling  minimizing the heating caused by the residual spontaneous emission from the excited-state manifold~\cite{PhysRevA.88.011601,PhysRevLett.116.035301}, as originally discussed in the context of synthetic gauge fields~\cite{gerbier_2010,gauge_qs_2014}. As discussed in the following paragraphs, these nice features can also be exploited for the generation of Raman lattice potentials.

 Let us start by focusing on 
the Raman transition along the $x$ axis (see  Fig.~\ref{fig:SOC_scheme} {\bf (b)}). This transition is implemented by the  standing-wave beams along the $x$ direction, and an additional Raman beam with  wave-vector $\tilde{\boldsymbol{k}}_1$, polarization $\tilde{\boldsymbol{\epsilon}}_1$ and  relative phase $\phi_1$, which  propagates  in a travelling-wave configuration along the $y$ axis (see  Fig.~\ref{fig:SOC_scheme} {\bf (a)}). As depicted in the inset of Fig.~\ref{fig:SOC_scheme} {\bf (b)}, since the standing wave is linearly polarised along the quantization direction, it can couple one of the groundstate spinors $\ket{\sigma}$ to an excited state $\ket{e_\sigma}$ with the same magnetic number. On the other hand, since the Raman beam is linearly polarised along a different direction, which corresponds to  a linear combination of  the two circular polarizations, it can impart the required angular momentum onto the atoms  to connect this excited state to a groundstate spinor component of a different magnetic number through a dipole-allowed transition, i.e. $M_\downarrow=M_{\uparrow}\pm1$.
For the Raman transition along the $y$ axis, the description  is analogous (see  Fig.~\ref{fig:SOC_scheme} {\bf (c)}), but it is now the standing wave which can impart momentum into the atoms, whereas the Raman beam  propagating along $\tilde{\boldsymbol{k}}_2$ has linear polarization $\tilde{\boldsymbol{\epsilon}}_2$ along the quantization axis. As discussed in more detail below, it is important to control the  phase $\phi_2$ of this second Raman process relative to $\phi_1$. 

As depicted in the insets of  Figs.~\ref{fig:SOC_scheme} {\bf (b)} and~{\bf (c)}, when the detunings   with respect to the excited states  $\Delta_j$ are very large, the two-photon processes only involve the spinor levels. Due to the participation of the standing wave, the Raman transition has a periodic intensity depicted with a green solid line in Figs.~\ref{fig:SOC_scheme} {\bf (b)}-{\bf (c)}, which has a doubled period with respect to the optical-lattice potentials depicted with  blue solid lines~\eqref{eq:Stark}. Altogether,  the groundstate spinors $\ket{\sigma}$    are subjected to the Raman  potential
 \beq
 \label{eq:Raman}
 \begin{split}
V_{\rm Ram}(\boldsymbol{r})=&\frac{\tilde{V}_{0,1}}{2}\cos(k_1x)\ee^{\ii (\tilde{k}_1 y-\delta_1 t+\phi_1)}\sigma^++{\rm H.c.}\\
+&\frac{\tilde{V}_{0,2}}{2}\cos(k_2y)\ee^{\ii (\tilde{k}_2 x-\delta_2 t+\phi_2)}\sigma^++{\rm H.c.},
\end{split}
\eeq
where we have introduced $\sigma^+=\ket{\uparrow}\bra{\downarrow}$,  $\tilde{V}_{0,j}$ ($\delta_j$) is the Rabi frequency (detuning) of the Raman transition driven by the $j$-th travelling wave, and we work in the interaction picture with respect to the atomic transition, such that $\delta_j$ are the detunings of the laser beatnotes with respect to the transition frequency  (see the insets of Figs.~\ref{fig:SOC_scheme} {\bf (b)}-{\bf (c)}). Let us note that, in principle, there can also be two-photon contributions from the standing- and travelling-wave beams that propagate along the same axis, but these will be highly off-resonant and contribute with higher-order shifts of the energy levels, which can be taken into account by adjusting  the laser frequencies.  
 
 The gas of neutral atoms of mass $m$ subjected to the total potential $V(\boldsymbol{r})=V_{\rm ac}(\boldsymbol{r})+V_{\rm Ram}(\boldsymbol{r})$ is described in second quantization, as customarily~\cite{doi:10.1080/00018730701223200,Bloch_2008}, and leads to the following non-relativistic Hamiltonian field theory
\beq
\begin{split}
\label{eq:atom_ft}
H&=\!\int\!\!{\rm d}^3r\!\!\sum_{\sigma,\sigma'}{\Phi}^{\dagger}_{\sigma}(\boldsymbol{r})\left(-\frac{\boldsymbol{\nabla}_{\!\!\boldsymbol{r}}^2}{2m}+\bra{\sigma}V(\boldsymbol{r})\ket{\sigma'}\right){\Phi}^{\phantom{\dagger}}_{\sigma'}(\boldsymbol{r})\\
&+\!\!\int\!\!\!{\rm d}^3r\!\!\!\int\!\!\!{\rm d}^3r'\!\!\sum_{\sigma,\sigma'}\!\!{\Phi}^{{\dagger}}_{\sigma}(\boldsymbol{r}){\Phi}^{{\dagger}}_{\sigma'}(\boldsymbol{r}')\frac{2\pi a_{s}}{m}\delta(\boldsymbol{r}-\boldsymbol{r}'){\Phi}^{\phantom{\dagger}}_{\sigma'}(\boldsymbol{r}'){\Phi}^{\phantom{\dagger}}_{\sigma}(\boldsymbol{r}),
\end{split}
\eeq 
where ${\Phi}^{\dagger}_{\sigma}(\boldsymbol{r}),{\Phi}^{\phantom{\dagger}}_{\sigma}(\boldsymbol{r})$ are the   creation/annihilation fields of fermionic atoms at position $\boldsymbol{r}$ in the internal state $\sigma$, and we have introduced the so-called $s$-wave scattering length $a_s$, which determines the strength of the contact two-body collisions for these dilute and ultra-cold atomic gases.

We consider that all  the laser beams have the same wavelength $k_j=\tilde{k}_j=:k=2\pi/\lambda,\,\,\forall j$, such that they can actually be generated from a single laser source, using acusto-optical modulators to control the detunings $\delta_1=\delta_2=:\delta$. To obtain a lattice field theory, we make use of the  Wannier basis~\cite{PhysRev.115.809} in
\beq
\label{eq:wannier_expansion}
{\Phi}^{\phantom{\dagger}}_{\sigma}(\boldsymbol{r})=\sum_{\boldsymbol{ n}}w(\boldsymbol{r}-\boldsymbol{r}^0_{\boldsymbol{n}})f_{\boldsymbol{n},\sigma}^{\phantom{\dagger}}.
\eeq
As depicted in Figs.~\ref{fig:SOC_scheme} {\bf (b)}-{\bf (c)},  the Wannier functions  $w(\boldsymbol{r}-\boldsymbol{r}^0_{\boldsymbol{n}})$ are  localised around the minima of the blue-detuned ac-Stark shift~\eqref{eq:Stark}, namely
\beq
\label{eq:minima}
\boldsymbol{r}^0_{\boldsymbol{n}}=\sum_{j}\frac{\lambda}{2}(n_j+\half)\boldsymbol{e}_j,
\eeq
where $n_j\in\mathbb{Z}_{N_j}$. In Eq.~\eqref{eq:wannier_expansion}, $f_{\boldsymbol{n},\sigma}^{{\dagger}}, f_{\boldsymbol{n},\sigma}^{\phantom{\dagger}}$ are dimensionless creation-annihilation operators of fermions in the lowest band of the optical lattice. As realised in the seminal works~\cite{PhysRevLett.81.3108,Greiner2002}, in the regime of deep optical lattices, where the potential barriers are much larger than the recoil energy $|V_{0,j}|\gg E_{\rm R}=k^2/2m$, the atoms are tightly confined within the minima, and the Hamiltonian field theory~\eqref{eq:atom_ft} can be expressed in terms of a lattice model with nearest-neighbor couplings. For instance, the kinetic, ac-Stark shift and interaction potentials lead to a spin-conserving tunnelling and density-density interactions
\beq
\label{eq:HUbbard}
H_{\rm sc}\!=\!\sum_{\boldsymbol{n}}\!\left(\!\sum_{\sigma,j}\! \left(\!-t_j f_{\boldsymbol{n},\sigma}^{{\dagger}}f_{\boldsymbol{n}+\boldsymbol{e}_j,\sigma}^{\phantom{\dagger}}+{\rm H.c.}\!\right)\!\!+\!\!\!\sum_{\sigma\neq\sigma'}\!\!\!\frac{U_{\uparrow\downarrow}}{2}n_{\boldsymbol{n},\sigma\phantom{'}\!\!}^{\phantom{\dagger}}n_{\boldsymbol{n},\sigma'}^{\phantom{\dagger}}\!\!\right)\!\!,
\eeq 
where the fermion number operators are $n_{\boldsymbol{n},\sigma}^{\phantom{\dagger}}=f_{\boldsymbol{n},\sigma}^{{\dagger}}f_{\boldsymbol{n},\sigma}^{\phantom{\dagger}}$. Here,   the tunnelling amplitudes 
\beq
t_j=\frac{4}{\sqrt{\pi}}E_{\rm R}\left(\frac{V_{0,j}}{E_{\rm R}}\right)^{\frac{3}{4}}\ee^{-2\sqrt{\frac{V_{0,j}}{E_{\rm R}}}},
\eeq
and interaction strengths
\beq
\label{eq:int_strength}
U_{\uparrow\downarrow}=\sqrt{\frac{8}{\pi}}ka_sE_{\rm R}\left(\frac{V_{0,1}V_{0,2}V_{0,3}}{E_{\rm R}^3}\right)^{\frac{1}{4}},
\eeq
are obtained by overlap integrals of the Wannier functions, weighted by the kinetic and ac-Stark potentials or the interaction potential, respectively~\cite{Bloch_2008}. 
Note that these overlaps can in principle couple neighbouring sites that lie further apart, but the corresponding strengths decay exponentially fast with distance~\cite{PhysRevLett.81.3108}, and are thus routinely neglected. By increasing the standing-wave intensity along the $z$-axis, such that $V_{0,3}\gg V_{0,1},V_{0,2}$, the dynamics of the atoms along the $z$ direction is effectively frozen, and Eq.~\eqref{eq:HUbbard} corresponds to the 2D Fermi-Hubbard model~\cite{PRSLSA_276_238,PhysRevLett.89.220407} which, as discussed in the introduction, is a paradigm in the physics of strongly-correlated materials~\cite{RevModPhys.70.1039,RevModPhys.78.17}. This Hamiltonian will be supplemented by additional tunnelling terms stemming from the Raman potential~\eqref{eq:Raman}, which we now discuss in detail.

A crucial ingredient of the QSs of synthetic spin-orbit coupling using Raman potentials~\cite{book_soc} is that, due to the specific form of the light interference in Eq.~\eqref{eq:Raman}, the corresponding overlaps of Wannier functions cannot contribute with local Raman transitions whereby an atom remains tightly trapped in a minimum of the optical potential, while its spin gets flipped $\ket{\uparrow}\leftrightarrow\ket{\downarrow}$. In the scheme of Figs.~\ref{fig:SOC_scheme} {\bf (b)}-{\bf (c)}, the vanishing of these on-site spin flips could be easily  understood in the limit of weak Raman potentials $|\tilde{V}_{0,j}|\ll |V_{0,j}|$, as a consequence of the zero value of this potential at the minima of the optical lattice~\eqref{eq:minima}. Remarkably, due to our choice of equal laser wave-vectors,  symmetry arguments allow to prove that this is not limited to weak potentials. Mathematically, whereas the Wannier functions localised with respect to a  single site are even with respect to lattice inversion about the site center~\eqref{eq:minima} (see the schematic drawing in Figs.~\ref{fig:SOC_scheme} {\bf (b)}-{\bf (c)}), the Raman potential~\eqref{eq:Raman} is odd, such that the corresponding overlap integrals vanish. The situation changes for the overlap between nearest-neighbour Wannier functions, as the above symmetry argument no longer applies in the  direction of tunnelling.  These overlaps lead  to a non-zero spin-flip tunneling with a complex-valued amplitude
\beq
\label{eq:spin_flipping}
H_{\rm sf}=-\sum_{\boldsymbol{n},j}\left(\ii \tilde{t}_{\! j}^{\phantom{\dagger}}\ee^{\ii(\delta t-\phi_{j,\boldsymbol{n}})}\!\!\left({f}^\dagger_{\boldsymbol{n},\uparrow}{f}^{\phantom{\dagger}}_{\boldsymbol{n}+{\bf e}_j,\downarrow}-{f}^\dagger_{\boldsymbol{n},\uparrow}{f}^{\phantom{\dagger}}_{\boldsymbol{n}-{\bf e}_j,\downarrow}\right)+{\rm H.c.}\right).\\
\eeq
 The modulus of the tunnelling, $\tilde{t}_j$, can be estimated analytically by approximating  the optical-lattice potential around the minima~\eqref{eq:minima} by a harmonic oscillator of frequencies $\nu_j=2\sqrt{V_{0,j}E_{\rm R}}$ along each axis, such that the Wannier functions become a separable product of Gaussians. When tunnelling along the $x(y)$ axis, the above symmetry argument can still be applied to neglect the ${\rm Re}\{\ee^{\ii\tilde{k}_1y}\}$ (${\rm Re}\{\ee^{\ii\tilde{k}_2x}\}$) contribution of the Raman potential~\eqref{eq:Raman} to the overlap integral along the $y$($x$) axis. On the other hand, the ${\rm Im}\{\ee^{\ii\tilde{k}_1y}\}$ (${\rm Im}\{\ee^{\ii\tilde{k}_2x}\}$) part gives a non-zero contribution multiplying the overlaps along the $x$($y$) axis, and leads to
\beq
\label{eq:spin_flipping_strength}
\tilde{t}_j=\frac{\tilde{V}_{0,j}}{2}\!\!\left(\ee^{-\frac{\pi^2}{4}\sqrt{\frac{V_{0,j}}{E_{\rm R}}}-\frac{1}{2}\sqrt{\frac{E_{\rm R}}{V_{0,j}}}}\right)\!\!.
\eeq
In addition, one obtains a site-dependent phase
\beq
\phi_{j,\boldsymbol{n}}=\phi_{j}-\pi(n_1+n_2).
\eeq
In the schemes of  Figs.~\ref{fig:SOC_scheme} {\bf (b)}-{\bf (c)}), the alternation of the signs of the spin-flip tunnellings $\ee^{\ii\phi_{j,\boldsymbol{n}}}=(-1)^{n_1+n_2}\ee^{\ii\phi_j}$ can be understood as a consequence of the doubled period of the Raman potential with respect to the optical lattice, which leads to alternating signs of the linear slopes of the Raman potential as one moves along the tunnelling direction. As discussed in the following subsection, this is another crucial property for the QS of our four-Fermi-Wilson model.

\subsection{Four-Fermi-Wilson quantum simulator}
 
Once we have obtained the microscopic Hamiltonian governing the dynamics of the two-component atoms in a deep optical Raman potential, we can discuss how this can be mapped via a $U(2)$ gauge transformation, followed by a rescaling,  to the four-Fermi-Wilson model of Eq.~\eqref{eq:total_H} in the single-flavour limit  $\mathsf{f}=1=N$. This transformation is
\beq
\label{eq:gauge_transformation}
\begin{split}
\psi^{\phantom{\dagger}}_{\mathsf{f},1}\!(\boldsymbol{x})&=\frac{1}{\sqrt{a_1 a_2}}\ee^{-\ii\frac{\delta}{2}t}{f}^{\phantom{\dagger}}_{\boldsymbol{n},\uparrow},\\ \psi^{\phantom{\dagger}}_{\mathsf{f},2}\!(\boldsymbol{x})&=\frac{1}{\sqrt{a_1 a_2}}\ee^{\ii \frac{\delta}{2}t+\ii\pi(n_1+n_2)}{f}^{\phantom{\dagger}}_{\boldsymbol{n},\downarrow}.
\end{split}
\eeq
The time-dependence can be understood as a change from the aforementioned interaction picture to the so-called rotating frame, such that the detunings correspond to an energy imbalance that will contribute to the bare mass. In addition, the site-dependent phase transformation allows us to rewrite the model in a translationally-invariant manner, making direct connection with the starting point for the lattice discretization of our four-Fermi-Wilson QFT discussed in Sec.~\ref{sec:discret}.

As discussed in~\cite{ziegler2020correlated}, the dependence on the relative phases $\phi_j$  generalises  previous schemes for synthetic spin-orbit coupling~\cite{PhysRevLett.112.086401,PhysRevLett.113.059901}, which can be exploited to connect precisely to our four-Fermi-Wilson models.  Setting the Raman-beam  phases  to $\phi_1=0,\phi_2=\pi/2$, we find that  the sum of the spin-conserving~\eqref{eq:HUbbard} and spin-flipping~\eqref{eq:spin_flipping} Hamiltonians maps directly to the  lattice field theory~\eqref{eq:total_H} with the following correspondence of the microscopic 
parameters 
\beq
\label{eq:parameters}
 a_j=\frac{1}{2\tilde{t}_j},\hspace{2ex} r_j=\frac{t_j}{\tilde{t}_j},\hspace{2ex} m=\frac{\delta}{2}-2(t_1+t_2),\hspace{2ex}g^2 =\frac{U_{\uparrow\downarrow}}{4\tilde{t}_1\tilde{t}_2}.
\eeq
Accordingly, the bare mass  $m$ can be controlled by the detuning of the Raman beams, whereas the coupling strength $g^2$ is proportional to the $s$-wave scattering length~\eqref{eq:int_strength}, and can thus be modified independently via a Feshbach resonance~\cite{RevModPhys.82.1225}. Alternatively, since it is inversely proportional to the spin-flip tunnellings~\eqref{eq:spin_flipping_strength}, one can tune this parameter by modifying the corresponding  potential depths $V_{0,j}$ or $\tilde{V}_{0,j}$. At this point, it is worth mentioning that the recent experimental realization of the Qi-Wu-Zhang model  with the $^{87}$Sr gas~\cite{liang2021realization} is related to the non-interacting limit of our four-Fermi-Wilson model~\eqref{eq:total_H} by a simple  $SU(2)$ rotation of gamma matrices~\eqref{eq:gammas}  that makes $\sigma^x\leftrightarrow \sigma^y$. In this case, the experiment uses standing-wave Raman beams, and their relative phase enters differently in the microscopic Hamiltonian. The only important point is that this phase difference is $\pi/2$. In the presence of the Hubbard interactions, this would only lead to an interchange of the two symmetry-breaking channels.  We thus believe that these experiments are a very promising route to explore the physics discussed in our work.

In this regard, it is important to note that the effective lattice spacings $a_j$ are not set by   the optical-lattice wavelength $\lambda/2$, but rather by the inverse of the spin-flip tunnelling strengths, which are proportional to the Rabi frequency of the Raman beams~\eqref{eq:spin_flipping_strength}, and can thus be tuned by changing the corresponding laser power. Since this laser power can be different along the $x,y$ axes, the cold-atom QS can also explore different anisotropies for the spatial  $\xi_2=a_1/a_2$, and Wilson parameters $r_j$. In this manuscript, we have explored in detail the limit $r_j=1$, which would require equal spin-conserving and spin-flip tunnellings. We note, however, that non-unity Wilson parameters $r_j\neq 1$ will simply rescale the axes but maintain the same shape of the non-interacting phase diagram. As the interactions are switched on, we expect that $r_j\approx1$ will not introduce additional strong-coupling phases. Exploring larger $r_j\gg 1$ or smaller values  $r_j\ll1$ is left for future studies, which could also be targeted by the proposed QS.

Let us now comment on  the observable consequences of the $U(2)$ gauge transformation~\eqref{eq:gauge_transformation}. The time-dependence due to the rotating frame is customary in quantum optics, and can be accounted for during the measurement process. The local site-dependent phase can change the interpretation of the symmetry-breaking order parameters related to the $\pi$-field condensates, while they do not change the $\sigma$-field scalar condensate. One can easily check that 
\beq
\label{eq:atoms_condensates}
\begin{split}
\Sigma\propto&\langle f^\dagger_{\boldsymbol{n},\uparrow}f^{\phantom{\dagger}}_{\boldsymbol{n},\uparrow}- f^\dagger_{\boldsymbol{n},\downarrow}f^{\phantom{\dagger}}_{\boldsymbol{n},\downarrow}\rangle,\\
\Pi_1\propto&(-1)^{n_1+n_2}\langle f^\dagger_{\boldsymbol{n},\uparrow}f^{\phantom{\dagger}}_{\boldsymbol{n},\downarrow}+ f^\dagger_{\boldsymbol{n},\downarrow}f^{\phantom{\dagger}}_{\boldsymbol{n},\uparrow}\rangle,\\
\Pi_2\propto&(-1)^{n_1+n_2}\langle \ii f^\dagger_{\boldsymbol{n},\downarrow}f^{\phantom{\dagger}}_{\boldsymbol{n},\uparrow}-\ii f^\dagger_{\boldsymbol{n},\uparrow}f^{\phantom{\dagger}}_{\boldsymbol{n},\downarrow}\rangle.\\
\end{split}
\eeq
Accordingly, in the language of the cold-atom QS, a non-zero value of the scalar condensate corresponds to an atomic density imbalance between the two hyperfine groundstates. For the Lorentz-breaking $\Pi_j$ condensates the situation changes, as a non-zero value marks an alternating pattern that turns the orbital ferromagnets FM$_x$ (FM$_y$) into N\'eel-ordered anti-ferromagnets   AFM$_x$ (AFM$_y$). In order to prepare these phases, as well as the correlated QAH or trivial band insulators discussed in this manuscript, one would start by preparing an ultra-cold spin-polarised Fermi gas by optical pumping, and then adiabatically ramping up the ac-Stark and Raman potentials to create the non-interacting groundsate for specific values of $ma_1,\xi_2$. One could then adiabatically change the coupling strength $g^2/a_1$. According to Eq.~\eqref{eq:parameters},   one can increase  the Hubbard interactions via a Feshbach resonance, or decrease the spin-flip tunnellings,  until  the desired point of  parameter space  $(ma_1,\xi_2,g^2/a_1)$ is reached.
Once this groundstate is approximately prepared, let us discuss possible characterization techniques in the experiment.

Note that the scalar condensate~\eqref{eq:atoms_condensates} can be inferred by spin-resolved in-situ imaging, where the trapped atoms are illuminated by  an incoming  laser and cast a shadow on a CCD-camera that is used to extract the integrated, so-called columnar, density of the atoms~\cite{cond-mat/9904034,0801.2500}.  Absorptive ~\cite{Partridge503} and dispersive~\cite{PhysRevLett.97.030401} techniques can be applied in subsequent shots  with  lasers addressing each of the internal states separately, such that  $\rho_\sigma(\boldsymbol{x})=\langle\Phi^{\dagger}_\sigma(\boldsymbol{x})\Phi_\sigma^{\phantom{\dagger}}(\boldsymbol{x})\rangle$ can be reconstructed from the spin-resolved columnar densities. To get lattice-site resolution for $\langle f^{{\dagger}}_{\boldsymbol{n},\sigma}f^{\phantom{\dagger}}_{\boldsymbol{n},\sigma}\rangle$, one can exploit the so-called quantum gas microscopes~\cite{10.1093/nsr/nww023,Bakr2009,Sherson2010,Haller2015,PhysRevLett.116.235301,doi:10.1126/science.aad9041}. To achieve spin resolution, one may    separate different spin
components spatially prior to the  microscope imaging~\cite{Weitenberg2011,doi:10.1126/science.aag1635}. Alternatively,  one can  remove the atoms with a specific  spin state by shining resonant light before the imaging~\cite{doi:10.1126/science.aag1430,doi:10.1126/science.aag3349}. Finally, in order to infer the values of the $\Pi_j$ condensates, one would need to apply an additional microwave/Raman term that drives a $\pi/2$ rotation of the spins on  the Bloch sphere in Fig.~\ref{fig:bloch_sphere} {\bf (a)}. Controlling the phase of this rotation, provided that it is locked to the rotating frame of the original Raman beams,  one can map the differential spin population to the $x$ or $y$ axis, such that the subsequent quantum-gas-microscope imaging gives the desired information about the Lorentz-breaking condensates~\eqref{eq:atoms_condensates}.

Density imaging  can also be performed after switching off the confining potential, which leads to the so-called time-of-flight (TOF) imaging.   If the density is imaged  after a sudden release~\cite{Greiner2002}, one gains  information about the coherence properties of the system  in momentum space~\cite{Bloch_2008}.    Instead of  the sudden release, one can adiabatically ramp-down  the lattice potential, which gives access to the quasi-momentum atomic distribution  through the so-called band-mapping technique~\cite{PhysRevLett.87.160405,PhysRevLett.94.080403}. In reference~\cite{Jotzu2014}, this band-mapping technique was used to measure the differential drift of the atom cloud when subjected to two opposite gradients, which allows one to distinguish trivial from non-trivial Berry curvatures in  Haldane's honeycomb model of the QAH effect. This type of measurements could be used to infer the value of the Chern number~\cite{PhysRevA.85.033620,PhysRevLett.111.135302}, as has been demonstrated in cold-atom experiments of the integer QHE~\cite{Aidelsburger_2014}.

Coming back to the QAH effect, we note  that the  spin-resolved TOF densities have been measured  in quasi-momentum space for the simulated Qi-Wu-Zhang model  with the $^{87}$Sr gas~\cite{liang2021realization}. Remarkably, exploiting a symmetry that corresponds exactly  to our inversion symmetry~\eqref{eq:inv_symm},   the measured differential spin densities at four high-symmetry points of the Brillouin zone give experimental access to the Chern number~\cite{PhysRevLett.111.120402}. These points correspond to the center $\Gamma$, edge centers $M$ and corner $R$ where the fermion doublers reside, as discussed below Eq.~\eqref{eq:WIlson_masses}. According to our discussion of the static self-energy and the topological Hamiltonian in Sec.~\ref{sec:self_energy_th}, we expect that these spin densities will get renormalized by the non-zero scalar condensate, but still serve to characterise the topological invariant in the presence of interactions in an analogous fashion to our Eq.~\eqref{topoconst}. It would be interesting to combine this observable, together with the in-situ symmetry-breaking $\Pi_1$ ($\Pi_2$) condensates, to explore the full phase diagram of the model and the topological and symmetry-breaking phase transitions.  We note that, in this case, the condensates would change into AFM$_y$ (AFM$_x$) N\'eel orders due to the alternation in Eq.~\eqref{eq:atoms_condensates}, and the $\sigma^x\to\sigma^y$ change of the different implementation of the Raman lattice.

Before closing this section, we emphasise that recovering  the continuum limit of the lattice field theory~\eqref{eq:total_H} implemented by the cold-atom QS does not imply a drastic  modification of  the laser wavelength $\lambda\to 0$. Instead, it requires  setting the microscopic parameters $(t_j,\tilde{t}_{j},\delta,U_{\uparrow\downarrow})$ to certain values, such that the bare couplings $(ma_1, \xi_2,r_j, g^2/a_1)$ lie in the vicinity of a critical point. Here,   the energy gap is much smaller than the tunnellings $\Delta\epsilon\ll \tilde{t}_j$, and the relevant length scale   $\xi_l\gg a_1,a_2$ leads to a continuum QFT. The question that could be addressed by the cold-atom QS is to explore this region and determine the critical scaling shedding light on the nature of the strongly-coupled fixed points of the continuum four-Fermi QFT, addressing questions that might otherwise require large-scale LFT simulations. Ultimately, the goal would be to explore different fillings and real-time dynamics in these models, going beyond the capabilities of classical simulations.

\section{\bf Conclusion and outlook}
 
 We have shown that Wilson-type discretizations of four-Fermi QFTs in (2+1) dimensions~\eqref{continuum} with irreducible representations of the gamma matrices~\eqref{eq:gammas} yield a neat playground to address interesting questions in both high-energy physics and condensed matter. Although explicitly lacking chiral symmetry, these regularised QFTs present the analogue of dynamical mass generation which, in contrast  to other spacetime dimensions, occurs both at a non-zero coupling strength within a renormalizable QFT with a strongly-coupled fixed point. In contrast to chiral-invariant theories, these four-Fermi-Wilson model can also host  fermion condensates that break inversion symmetry spontaneously and, thus, Lorentz invariance in the continuum limit. From a condensed-matter perspective, these lattice models host QAH phases with non-zero Chern numbers, and the four-Fermi terms can be used to explore the role of interactions as one enters  strongly-correlated regimes. Both of these topics are actively investigated in these two fields.
 
 We have presented a multidisciplinary approach that combines tools from these communities to advance our understanding of these four-Fermi-Wilson models. In the strong-coupling and single-flavour limit, we have shown that the condensed-matter concept of super-exchange interactions can be used to find an effective description in terms of a 90$^0$ quantum compass model in a transverse field. Analyzing this model with variational mean-field and tensor-network techniques, we have identified two possible symmetry-breaking channels that connect to two versions of the aforementioned Lorentz-breaking condensates. This has allowed us to formulate a large-$N$ limit of this field theory in terms of auxiliary fields, which has been used to predict the whole extent of the condensates  away from the strong-coupling limit by solving a set of non-linear gap equations. By comparing a continuum-time Hamiltonian formalism with a discrete-time Euclidean approach, we have been able to identify additive renormalizations of the bare parameters that must be carefully considered when one explores the phases of the model using the discrete-time Euclidean approach common to the lattice field theory community. 
 
 Moreover, using the Euclidean approach, we have calculated the effective potential resumming the  leading-order Feynman diagrams  for $N\to\infty$. This has allowed us to unveil a new type of radiative corrections that give rise to novel effects in comparison to chiral-invariant theories, such as an interesting crossover between first- and second-order phase transitions. This effective potential has also allowed us to explore regions of parameter space where the inversion symmetry  remains intact, extracting the large-$N$ contributions to the self-energy, and using those to calculate the many-body Chern numbers that characterise the groundstate and the QAH in the presence of correlations. This leads to a non-perturbative characterization of a rich phase diagram, which contains  large-$N$ Chern insulators, trivial band insulators, and Lorentz-breaking fermion condensates separated by various critical lines, around which one can recover  continuum QFTs and explore the nature of the corresponding strongly-coupled fixed points. 
 
 Finally, we have shown that quantum simulators based on ultra-cold alkali-earth atoms trapped in optical lattices and subjected to synthetic spin-orbit coupling yield a very promising avenue to realise these four-Fermi-Wilson models in experiments. In particular, we have argued that recent experiments with a $^{87}$Sr gas subjected to Raman lattice potentials~\cite{liang2021realization} is related to the non-interacting limit of our four-Fermi-Wilson model, and gives a unique opportunity to explore a correlated QAH effect in the laboratory. Such an implementation would benefit from the microscopic tunability of cold-atom quantum gases, which would allow to infer all of the relevant observables such as the symmetry-breaking order parameters related to the fermion condensates or the many-body Chern numbers. Moreover, these quantum simulators would open the route to the study of real-time dynamics and finite-fermion densities, overcoming current limitations of numerical studies based on classical hardware. 
 
 As an outlook, we believe that exploiting the current multi-disciplinary view will be very interesting to explore effects that have not been covered by this article. An interesting open question is to determine the role of anisotropic Wilson parameters in the nature of the Lorentz-breaking fermion condensates, exploring if new phases can appear as one increases the interactions. We also believe that further studies of the isotropic regime $a_1=a_2$ will be very interesting. As noted in this manuscript, there is an emerging $O(2)$ symmetry in the $\boldsymbol{\pi}(x)$ field, which can modify substantially the dynamics of the theory due to an additional Chern-Simons term that is generated by quantum corrections. In analogy to quantum electrodynamics in (2+1) dimensions, where photons acquire a mass due to a Chern-Simons term~\cite{PhysRevD.29.2366}, the low-lying excitations about the condensate phase that breaks the $O(2)$ symmetry spontaneously can also acquire a non-zero mass. It will  be very interesting to explore this phenomenon in the presence of boundaries, since there can be an interplay with the QAH effect and the topological edge states. Finally, we finish by mentioning that the cold-atom quantum simulators would open a new route to the study of real-time dynamics and finite-fermion densities, overcoming current limitations of numerical studies based on classical hardware. In the future, it will be very interesting to exploit this multidisciplinary view and to identify new interaction-induced topological phases, such as fractional Chern insulators.


\section*{Acknowledgements}

The ICFO group acknowledges support from ERC AdG NOQIA, State Research Agency AEI (“Severo Ochoa” Center of Excellence CEX2019-000910-S) Plan National FIDEUA PID2019-106901GB-I00 project funded by MCIN/ AEI /10.13039/501100011033, FPI, QUANTERA MAQS PCI2019-111828-2 project funded by MCIN/AEI /10.13039/501100011033, Proyectos de I+D+I “Retos Colaboración” RTC2019-007196-7 project funded by MCIN/AEI /10.13039/501100011033, Fundació Privada Cellex, Fundació Mir-Puig, Generalitat de Catalunya (AGAUR Grant No. 2017 SGR 1341, CERCA program, QuantumCAT \ U16-011424, co-funded by ERDF Operational Program of Catalonia 2014-2020), EU Horizon 2020 FET-OPEN OPTOLogic (Grant No 899794), and the National Science Centre, Poland (Symfonia Grant No. 2016/20/W/ST4/00314), Marie Sk\l odowska-Curie grant STREDCH No 101029393, “La Caixa” Junior Leaders fellowships (ID100010434),  and EU Horizon 2020 under Marie Sk\l odowska-Curie grant agreement No 847648 (LCF/BQ/PI19/11690013, LCF/BQ/PI20/11760031,  LCF/BQ/PR20/11770012).).   A.B. acknowledges support from the Ram\'on y Cajal program RYC-2016-20066,  CAM/FEDER Project S2018/TCS- 4342 (QUITEMADCM),  and PGC2018-099169-B-I00 (MCIU/AEI/FEDER, UE). S.J.H. acknowledges the support of STFC grant ST/T000813/1.

\appendix

\section{Absence of  two simultaneous $\pi$ condensates}
\label{app:two_condensates}

In this Appendix, we give a detailed account of the gap equations obtained by using a  Hubbard-Stratonovich transformation to rewrite the action~\eqref{rawaction2} in terms of two  $\pi$ fields, and show that there cannot be a simultaneous condensation for general anisotropies.

To find the effective action in  terms of the $\sigma(x)$ and $\boldsymbol{\pi}(x)=(\pi_1(x),\pi_2(x))$ auxiliary fields, we note that in the single-flavour limit ${N=1}$ one can exactly rewrite the  quartic interaction   as the combination  $(\overline{\Psi}(\boldsymbol{x})\Psi(\boldsymbol{x}))^2\to
\frac{1}{3}((\overline{\Psi}(\boldsymbol{x})\gamma^1\Psi(\boldsymbol{x}))^2+ (\overline{\Psi}(\boldsymbol{x})\gamma^2\Psi(\boldsymbol{x}))^2+(\overline{\Psi}(\boldsymbol{x})\Psi(\boldsymbol{x}))^2)
$,  such that the auxiliary fields can be introduced symmetrically. We proceed by assuming once more  that the corresponding condensates are homogeneous, i.e. ${\Sigma(\boldsymbol{x})=\Sigma}$, $(\Pi_1(\boldsymbol{x}),\Pi_2(\boldsymbol{x}))=(\Pi_1,\Pi_2)=\boldsymbol{\Pi}$, such that the effective action is now
\beq
\label{eq:free_action_aux_app}
\begin{split}
S_E\!= \!\!\!\int_{0}^{\beta}\!\!\!\!{\rm d}\tau \bigg( &\frac{3NA_{s}}{2g^2}\left(\Sigma^2+\boldsymbol{\Pi}^2 \right)\\+&\sum_{\boldsymbol{k}\in {\rm BZ}}\!\!\! \Psi^\star_{\boldsymbol{k}}(\tau)\big(\partial_\tau+{h}_{\boldsymbol{k}}(m+\Sigma,\boldsymbol{\Pi})\big)\Psi_{\boldsymbol{k}}(\tau) \!\!\!\bigg).
\end{split}
\eeq
Here, in analogy to the derivation  for a single $\pi$ channel presented in Sec.~\ref{sec:H_ft_gap_eqs}, we find that   the  single-particle Hamiltonian~\eqref{freehamiltonian} gets modified to \beq
{h}_{\boldsymbol{k}}(m)\to{h}_{\boldsymbol{k}}(m+\Sigma,\boldsymbol{\Pi})= {\boldsymbol{d}}_{\boldsymbol{k}}(m+\Sigma,\boldsymbol{\Pi})\cdot (\mathbb{I}_N\otimes \boldsymbol{\sigma}).
\eeq
where the vector~\eqref{freehamiltonian_d_vector} also gets modified $\boldsymbol{d}_{\boldsymbol{k}}(m)\to\boldsymbol{d}_{\boldsymbol{k}}(m+\Sigma,\boldsymbol{\Pi})$ due to the presence of the auxiliary  fields 
\beq
\boldsymbol{d}_{\boldsymbol{k}}(m+\Sigma,\boldsymbol{\Pi})=\boldsymbol{d}_{\boldsymbol{k}}(m)+\left(\Pi_1,\Pi_2,\Sigma\right).
\label{freehamiltonian_d_vector_aux_app}
\eeq
Therefore, the only differences with respect to Sec.~\ref{sec:H_ft_gap_eqs} are that the couplings strength $g^2\to2g^2/3$, and that this vector is simultaneously shifted by both $\Pi_j$ condensates.

The rest of the derivation follows exactly the steps described in Sec.~\ref{sec:H_ft_gap_eqs}, and leads to a saddle point of the action that is now determined by three non-linear gap equations
\begin{align}
{\bf I}:\,\,\, \frac{\Sigma}{g^2}&=\frac{1}{3}\int_{\boldsymbol{k}} \, \frac{m+\Sigma+{m}_W(\boldsymbol{k})}{\epsilon_{\boldsymbol{k}}(m+\Sigma,\boldsymbol{\Pi})}\label{firstequation_app}\\
{\bf II}: \,\,\,\frac{\Pi_1}{g^2}&=\frac{1}{3}\int_{\boldsymbol{k}}  \, \frac{\frac{1}{a_1}\sin( k_1a_1)+\Pi_1}{\epsilon_{\boldsymbol{k}}({m}+\Sigma,\boldsymbol{\Pi})}\label{secondequation_app}\\
{\bf III}: \,\,\,\frac{\Pi_2}{g^2}&=\frac{1}{3}\int_{\boldsymbol{k}}  \, \frac{\frac{1}{a_2}\sin(k_2 a_2)+\Pi_2}{\epsilon_{\boldsymbol{k}}({m}+\Sigma,\boldsymbol{\Pi})},\label{thirdequation_app}
\end{align}

Let us note that from {\bf II}~\eqref{secondequation_app} and {\bf III}~\eqref{thirdequation_app}, it follows that
\begin{align}
\int_{\boldsymbol{k}}\bigg(\frac{\frac{1}{\Pi_1 a_1}\sin(a_1 k_1)}{\epsilon_{\boldsymbol{k}}({m}+\Sigma,\boldsymbol{\Pi})}	- \frac{\frac{1}{\Pi_2 a_2}\sin(k_2 a_2)}{\epsilon_{\boldsymbol{k}}({m}+\Sigma,\boldsymbol{\Pi})}\!\bigg)\!\!=0. \label{twothree}
\end{align}
The left hand side  does not vanish for {any} point ($\sigma$, $\Pi_1$, $\Pi_2$) in the anisotropic case ${a_1 \neq a_2}$, meaning that {\bf II} and {\bf III} cannot be satisfied simultaneously. For ${a_1=a_2}$, gap equations {\bf II} and {\bf III} are equal and, around this regime,  Eq.~(\ref{twothree}) changes from being  negative to  positive. As a consequence, there is no  condensation with two simultaneous non-zero vacuum expectation values of the $\pi$ fields for generic anisotropy, meaning that we must consider either the $\Pi_1$ or the $\Pi_2$ symmetry-breaking channels individually. This is in line with the prediction of the variational mean-field where, in the language of an effective spin model, the ground state is either a $x$- {or} $y$-ferromagnet unless $a_1=a_2$. This discussion implies that we can improve on the auxiliary-field description by considering  two independent sets of gap equations for $\Sigma, \Pi_1$ and $\Sigma, \Pi_2$ separately.
 
\section{Continuum-time limit and time doublers}
 \label{app:add_renorm}
 
 In this Appendix, we present a detailed derivation of the additive renormalization of the bare parameters in the time-continuum limit~\eqref{eq:rescalings_parameters} caused by the spurious time doublers~\eqref{eq:continuum_theory_euclidean}. To get an explicit expression for this renormalization, we   start from the action for dimensional fields~\eqref{rawaction3}, using a subscript $E$ in the bare parameters and condensates to distinguish this   discrete-time Euclidean-lattice approach from the Hamiltonian one in Eqs.~\eqref{firstequation}-\eqref{secondequation}. 
 
 After repeating similar steps as those described in Sec.~\ref{sec:H_ft_gap_eqs}, we obtain an effective action for the auxiliary fields, and arrive at a pair of gap equations 
\begin{align}
\frac{\Sigma_E }{g_E^2}&=\frac{1}{Q}\sum_{\boldsymbol{k}} \frac{m_E+\Sigma_E+m_E(\boldsymbol{k})}{\mathsf{s}^2_{\boldsymbol{k}}({m}_E+{\Sigma_E},{\Pi}_{2,E})},\label{eq:diemnsional_gap_1}\\
\frac{\Pi_{2,E} }{g_E^2}&=\frac{1}{Q}\sum_{\boldsymbol{k}} \frac{\frac{1}{a_2} \sin(k_2 a_2)+\Pi_{2,E}}{\mathsf{s}^2_{\boldsymbol{k}}({m}_E+{\Sigma_E},{\Pi}_{2,E})},\label{eq:diemnsional_gap_2}
\end{align}
where  the Euclidean   contribution to the mass~\eqref{Euclidean_wilson_contribution} due to the Wilson-term  reads, for dimensional couplings, as follows
\beq
\label{eq:dimensional_wmass}
m_E(\boldsymbol{k})=\sum_{\alpha}\frac{1-\cos(k_\alpha a_\alpha)}{a_\alpha}.
\eeq
In the above gap equations, we have also introduced the  analogue of Eq.~\eqref{eq:action_momentum_aux} for dimensional couplings
\beq
\begin{split}
\mathsf{s}^2_{\boldsymbol{k}}({m}_E+{\Sigma}_E,{\Pi}_{2,E})&=\left(m_E+\Sigma_E+m_E(\boldsymbol{k})\right)^2\\
&+\sum_{\alpha}\frac{1}{a^2_\alpha}\big(\sin(k_\alpha a_\alpha)+\Pi_{\alpha,E}\delta_{\alpha,2}\big)^2.
\end{split}
\eeq

With these expressions at hand, let us perform a long-wavelength approximation $k_0=K_0+{\delta k_0}$ around $K_0\in\{0,\pi/a_0\}$, such that we can identify   the contributions of the physical and spurious doublers. After a  Taylor expansion, and using contour techniques for the integrals along $\delta k_0$, we find that the second gap equation can be expressed as
\begin{align}
\frac{\Pi_{2,E}}{g_{E}^2}\!=\! \frac{1}{2}\!\!\int_{\boldsymbol{k}}\!\!\bigg( \!\!\frac{\frac{1}{a_2}\sin(k_2 a_2)+\Pi_{2,E}}{\epsilon_{\boldsymbol{k}}\left({m}_E\!+\!\Sigma_E\!,\!\Pi_{2,E}\boldsymbol{e}_2\!\right)}\! +\frac{\frac{1}{a_2}\sin(k_2 a_2)+\Pi_{2,E}}{ \epsilon_{\boldsymbol{k}}(m_E\!+\!\Sigma_E\!\!+\!\frac{2}{a_0}\!,\!\Pi_{2,E}\boldsymbol{e}_2\!)}\!\! \bigg)\!\!,
\end{align}
where $\boldsymbol{k}=(k_1,k_2)^t$, and  we have made use of the single-particle energy defined in Eq.~\eqref{singleparticleenergy}.
We can now let  $a_0\rightarrow0$, noticing that the  mass term of the spurious time doublers, proportional to $1/a_0$,  dominates in the denominator of the second term, such that the integral vanishes linearly with $a_0$ in this limit. Accordingly, we recover  the continuum gap equation~\eqref{secondequation} for $j=2$, which is the desired single $\pi$-channel  gap equation.
Let us now recall that, in our numerical solution of the gap equations, we start by fixing the value of the coupling strength $g^2$ and  $M=m+\Sigma$, after which we   solve Eq.~\eqref{eq:diemnsional_gap_2} to get the $\pi$ condensate $\Pi_2$. As we have just shown that the Euclidean-lattice  gap equation~\eqref{eq:diemnsional_gap_2}  yields the same gap equation~\eqref{secondequation} in the continuum-time limit, we can thus conclude that
\beq
\label{eq:coupling_pi_cond_equiv}
g_E^2=g^2,\hspace{2ex}\Pi_{2,E}=\Pi_{2}.
\eeq

The situation changes for the first gap equation~\eqref{eq:diemnsional_gap_1}, since the Wilson-term contribution~\eqref{eq:dimensional_wmass} for the time doublers  leads to a term that scales with $1/a_0$ also in the numerator 
\begin{align}
\begin{split}
\frac{ \Sigma_E }{g_E^2}\!=\! \frac{1}{2}\!\!\int_{\boldsymbol{k}} \!\!\bigg( \frac{m_E+\Sigma_E+{m}_W(\boldsymbol{k})}{\epsilon_{\boldsymbol{k}}(m_E+\Sigma_E,{\Pi}_{2,E}\boldsymbol{e}_2)}+\frac{m_E+\Sigma_E+\frac{2}{a_0}+{m}_W(\boldsymbol{k})}{\epsilon_{\boldsymbol{k}}(m_E\!+\!\Sigma_E\!+\!\frac{2}{a_0}\!,\!{\Pi}_{2,E}\boldsymbol{e}_2)\!}\bigg)\!\!. 
\end{split}
\end{align}
In the limit $a_0\to 0$, the second term no longer vanishes, but yields instead
\begin{align}
\label{eq:sigma_gap_eq_continuum}
\frac{\Sigma_E}{g_E^2}=\frac{1}{2} \int_{\boldsymbol{k}}\frac{m_E+\Sigma_E+m_W(\boldsymbol{k})}{\epsilon_{\boldsymbol{k}}(m_E+\Sigma_E,\Pi_{2,E}\boldsymbol{e}_2)}+\frac{1}{2\,a_1 \,a_2}.
\end{align} 
One can readily check that this gap equation differs from the Hamiltonian one~\eqref{firstequation}, which was obtained by working directly in the time-continuum limit,  by a constant additive term, where  we consider the single $\pi$-channel. 
Going back to our numerical solution of the gap equations, and the discussion above Eq.~\eqref{eq:coupling_pi_cond_equiv}, once we have solved the second gap equation and know $g_E^2$ and $\Pi_{2,E}$, we can solve Eq.~\eqref{eq:diemnsional_gap_2} to obtain the value of the scalar condensate $\Sigma_E$, and finally infer the corresponding bare mass $m_E=M-\Sigma_E$. According to the shift in  equation~\eqref{eq:sigma_gap_eq_continuum}, we can readily infer that  the Euclidean-lattice formalism gives the same value of the scalar condensate, but renormalizes additively the bare mass 
\beq
\Sigma_E=\Sigma,\hspace{2ex}m_E=m-\frac{g^2}{2a_1a_2}.
\eeq
When this additive renormalization is taken into account, together with the required rescalings of the adimensional formulation~\eqref{eq:diemsnionless_fields}, we finally arrive to Eq.~\eqref{eq:rescalings_parameters}.

\section{Derivation of the effective potential}
 \label{effective}

In this Appendix, we give a detailed account of the derivation of the resummation for  the effective potential~\eqref{veffbasic} to all orders of the couplings strength $g^2$. Let us start from the most general expression with an even/odd number of auxiliary lines
\begin{align}
\begin{split}
V_{\rm eff}(\tilde{\Pi}_2)= \frac{N \tilde{\Pi}_2^2}{2\tilde{g}^2}+N \sum_{n=1}^{\infty} \frac{1}{n} \int_{\boldsymbol{p}}{\rm Tr}\left(-\ii \tilde{\gamma}_2 \frac{\tilde{\Pi}_2}{\ii \slashed{p}+\tilde{m}}\right)^{\hspace{-0.65ex}n},
\label{veffbasic_app}
\end{split}
\end{align}
where we recall that $\int_{\boldsymbol{p}}=\sum_{\boldsymbol{k}}$, $\slashed{p}=\tilde{\gamma}^\alpha p_{\alpha}$ with $p_{\alpha}=2\kappa_\alpha\sin k_\alpha$, and our choice for the Euclidean gamma matrices is given in Eq.~\eqref{eq:gamma_euc}. Although these are the specific for the anisotropic Wilson-type regularization of the four-Fermi QFT, we note that the following derivations are completely generic, and can be readily applied to the continuum case and to other representations of gamma matrices. For instance, if one aims at implementing chiral symmetry using the Euclidean version of Eq.~\eqref{eq:gammas_4}, the expressions below need only be modified by letting $N\to 2N$ due to the doubled dimension of the gammas.

The trace in Eq.~\eqref{veffbasic_app} can be calculated analytically via  an inductive method, where
\begin{equation}
{\rm Tr}\left(-\ii\tilde{\gamma}_2 \frac{\tilde{\Pi}_2}{i \slashed{p}+\tilde{m}}\right)^{\hspace{-0.65ex}n}= \frac{\tilde{\Pi}_2^n}{(p^2+\tilde{m}^2)^n}{\rm Tr}(I_n)
\end{equation}
and we have defined
\begin{align}
I_n=(-\ii\tilde{\gamma}_2(-\ii\slashed p+\tilde{m}))^n.
\end{align}
 One can derive a recursive relation by noting that 
\begin{align}
I_n&=-\tilde{\gamma}_2(-\ii\slashed{p}+\tilde{m})\tilde{\gamma}_2(-\ii\slashed{p}+\tilde{m})I_{n-2} \\
&=-(p^2+\tilde{m}^2) I_{n-2}-2 p_2 I_{n-1}.
\end{align}
In the last step, we use the anti-commutation rules $\{\tilde{\gamma}_\mu,\tilde{\gamma}_\nu\}=2\delta_{\mu,\nu}$, and the identity $\tilde{\gamma_2}\slashed{p}\tilde{\gamma}_2=-\slashed{p}-2p_2\tilde{\gamma}_2$. The recurrence relation starts with the first two terms $I_0=\mathbb{I}_2$ and $I_1=-\ii\tilde{\gamma}_2(-\ii\slashed{p}+\tilde{m})$ which, considering our choice in Eq.~\eqref{eq:gamma_euc} and the properties of the Pauli matrices, have the  traces
\begin{align}
{\rm Tr}(I_0)={\rm Tr}(\mathbb{I}_2)=2,\hspace{2ex} {\rm Tr}(I_1)={\rm Tr}(-p_2\mathbb{I}_2)=-2 p_2.
\end{align}
For higher-dimensional representations, such as Eq.~\eqref{eq:gammas_4}, these traces will have an overall multiplicative factor that shall carry onto the expressions of the effective potential.

With the aid of these first two terms, and the above recurrence relation, one can develop a general expression for arbitrary $n$ with   the general structure 
\begin{equation}
\label{eq:expansion_app}
{\rm Tr}(I_n)= 2 \sum_{k=1}^n A_{nk} p_2^k (p^2+\tilde{m}^2)^{\frac{n-k}{2}}, 
\end{equation}
where the first few values of the coefficients $A_{nk}$ are given in Table~\ref{coftabelle}. Note that we only present the coefficients for even integers $k=0,2,4,...,n$, and even $n$. Although the odd ones $k=1,3,5,...,n$, for odd $n$, can also be non-zero, they lead to mode sums, or integrals in the continuum, which vanish~\eqref{veffbasic_app}, as the function is odd in a symmetric interval. In the standard calculation of the effective potential for continuum four-Fermi QFTs~\cite{coleman_1985}, where the $\sigma$ field couples to $(\overline{\Psi}\Psi)$ for the chiral symmetry-breaking channel; the odd terms vanish directly since they are proportional to the  traces of the gamma matrices, all of which vanish. For the current $(\overline{\Psi}\tilde{\gamma}_1\Psi)$ channel, although the traces do not vanish, once we integrate over the Euclidean Brillouin zone~\eqref{veffbasic_app}, the odd terms do not contribute either. It is for this reason that the large-$N$ Feynman diagrams depicted in Fig.  only contain an even number of auxiliary $\pi$ lines, and lead to Eq.~\eqref{veffbasic_app}.

\begin{table}[t]
\caption{Coefficients ${A}_{nk}$ in the expansion of Eq.~(\ref{eq:expansion_app}). The lower row also gives the values of the coefficients $\mathcal{A}_k$ in Eq.~\eqref{eq:coeffs_Ak}.}
\label{coftabelle}
	\centering
         \begin{tabular}{|r|r| r| r| r| r| r|r|}
		\hline
		\hline
		$n/k$ & 0 & 2 & 4 & 6 & 8 & 10 &12 \\
		\hline
		2 & $-1$ & 2& & & & &\\
		4&1&$-8$&8& & & &\\
		6&$-1$&18&$-48$& 32& & &\\
		8 & 1 & $-32$ & 160 &$ -256$ & 128 & &\\
		10 & $-1$ & $50$ & $-400$ &$ 1120$ & $-1280$ & 512 &\\
		12 & $1$ & $-72$ & $840$ &$ -3584$ & $6912$ & $-6144$ &2048\\
		\hline
		\hline
		$\mathcal{A}_{k}$ & & $-2$ &$\frac{2}{3}$ & $-\frac{4}{45}$ & $\frac{2}{315}$ & $-\frac{4}{14175}$ & $\frac{4}{467775}$\\
		\hline
		\hline
\end{tabular}
\end{table}

In order to perform the resummation, we need to express these coefficients $A_{nk}$ in closed from. We have found that the exact expression is 
\begin{align}
A_{nk}\!=\! (-1)^{\frac{n}{2}}\! \mathcal{A}_k\! \left(\frac{n}{2}\right)^{\!\!\!2}\!\!\left(\!\!\!\left(\frac{n}{2}\right)^{\!\!\!2}-1^2 \right)...\left(\!\!\!\left(\frac{n}{2}\right)^{\!\!2}-\left(\frac{k}{2}-1\right)^{\!\!\!\!2}\!\right)\!\!,
\label{coefficients}
\end{align}
where we have introduced 
\begin{align}
\label{eq:coeffs_Ak}
\mathcal{A}_k= \frac{(-1)^{\frac{k}{2}} 2^{k-1}}{\left(\frac{k}{2}\right)^2\left(\left(\frac{k}{2}\right)^2-1^2\right)...\left(\left(\frac{k}{2}\right)^2-\left(\frac{k}{2}-1\right)^2\right)}.
\end{align}

The first values of $\mathcal{A}_k$  shown in table \ref{coftabelle} can be readily checked to follow this general expression. This expression~\eqref{coefficients} must be supplemented with the $k=0$ term $A_{n0}= (-1)^{\frac{n}{2}}$. Relabelling $n,k\to 2n,2k$ to only account for terms with an even number of external $\pi$ lines, we find
\begin{widetext}
\begin{align}
\label{eq:rad_terms_V_eff}
\frac{V_{\rm eff}(\tilde{\Pi}_2)}{N}= \frac{\tilde{\Pi}_2^2}{2 \tilde{g}^2}-\int_{\boldsymbol{p}} \log\left(1+\frac{\tilde{\Pi}_2^2}{p^2+\tilde{m}^2}\right) + \int_{\boldsymbol{p}} \sum_{n=1}^\infty\left(-\frac{\tilde{\Pi}_2^2}{p^2+\tilde{m}^2}\right)^{\!\!n} \sum_{k=1}^n \left(\frac{p_2^2}{p^2+\tilde{m}^2}\right)^{\!\!k}  \mathcal{A}_{2k} \,{n}(n^2-1^2)(n^2-2^2)...(n^2-(k-1)^2),
\end{align}
\end{widetext}
where the first radiative term comes from resumming the series in $n$ for the aforementioned $k=0$ term, where we have used  $\ln(1+z)=\sum_{n=1}^{\infty}(-z)^n/n$ for the parameter
\beq
\label{eq:z_parameter}
z=\frac{\tilde{\Pi}_2^2}{p^2+\tilde{m}^2},
\eeq 
such that convergence requires $|z|<1$. We note that  in the standard effective potential for continuum four-Fermi QFTs~\cite{coleman_1985}, where the $\sigma$ field couples to the $(\overline{\Psi}\Psi)$ bilinear, all the radiative quantum corrections to the classical potential are contained in a term that is completely analogous to this one
\beq
\delta V_{\rm q,1}=-N\int_{\boldsymbol{p}} \log\left(1+\frac{\tilde{\Pi}_2^2}{p^2+\tilde{m}^2}\right)
\eeq
 after making the substitution $\tilde{\Pi}_2\to\tilde{\sigma}$. Accordingly, the consequence of using a different $\pi$ channel is that there are additional quantum corrections contained in the remaining contributions for  $k\in\{1,2,3,\cdots\}$ of Eq.~\eqref{eq:rad_terms_V_eff}. Accounting for these new radiative corrections is crucial to find the correct phase diagram, identifying the regions that delimit the correlated QAH phase.

Let us now describe how to perform the resummation of these additional corrections. The idea is to focus on the different $k$ contributions separately, performing the sums over  $n$ to arbitrary orders of the coupling strength, here corresponding to arbitrary pairs of external lines,  by means of the following generating functions
\begin{equation*}
S_\ell(z)= \sum_n n^\ell (-z)^n= \left(z \frac{d}{dz}\right)^{\!\!\!{\ell}} \frac{1}{1+z}.
\end{equation*}
Note that the contributions in Eq.~\eqref{eq:rad_terms_V_eff} of order $\mathcal{O}(p_2^{2k})$, considering a fixed value of $k$, can be expressed in terms of a combination of  generating functions $\{S_{2r-1}(z)\}_{r=1}^k$, e.g. for $k=2$ one gets a contribution proportional to $S_{3}(z)-S_1(z)$.  Each of these generating functions is a rational function  $P_{2k-1}(z)/(1+z)^{2k}$, the numerator of which $P_{2k-1}(z)$ is a polynomial of order $2k-1$. In light of Eq.~\eqref{eq:z_parameter}, $z\propto\tilde{\Pi}_1^2$, and one would then expect that each $k$ term has a different  polynomial dependence on the $\pi$ condensate, such that resummation cannot be performed. Remarkably, we find that the prefactors that multiply  the generating functions inside each of the polynomials $P_{2k-1}(z)$ cancel all terms except for one scaling with $z^k$, e.g. for $k=2$, we get $P_3(z)=6z^2$. This leads to a scaling with the condensate of order $\mathcal{O}(\tilde{\Pi}_2^{2k})$ that can now be resummed.  The first few terms of these additional quantum-mechanical corrections to the effective potential read
\begin{widetext}
\begin{align}
\frac{\delta V_{\rm q,2}(\tilde{\Pi}_2)}{N}&\approx\frac{1}{2} \int_{\boldsymbol{p}} \left(\frac{4p_2^2 \tilde{\Pi}_2^2}{(p^2+\tilde{m}^2+\tilde{\Pi}_2^2)^2}\right.+\frac{1}{2}\left(\frac{4p_2^2 \tilde{\Pi}_2^2}{(p^2+\tilde{m}^2+\tilde{\Pi}_2^2)^2}\right)^2+\frac{1}{3}\left(\frac{4p_2^2 \tilde{\Pi}_2^2}{(p^2+\tilde{m}^2+\tilde{\Pi}_2^2)^2}\right)^3+ \frac{1}{4}\left.\left(\frac{4p_2^2\tilde{ \Pi}_2^2}{(p^2+\tilde{m}^2+\tilde{\Pi}_2^2)^2}\right)^4+\cdots\right).
\end{align}
\end{widetext}
The particular form of these last four terms suggests that a resummation is possible using again the the Taylor series of the logarithm series, such that 
\begin{equation*}
\delta V_{\rm q,2}(\tilde{\Pi}_2)=-\frac{N}{2}\int_{\boldsymbol{p}} \log\left(1-\frac{4p^2_2 \tilde{\Pi}_2^2}{(p^2+\tilde{m}^2+\tilde{\Pi}_2^2)^2}\right).
\end{equation*}

\bibliographystyle{apsrev4-1}
\bibliography{bibliography}

\end{document}